\documentclass[12pt]{article}
\usepackage[pass]{geometry}

\usepackage{graphicx} %Graphen
\usepackage{subfigure} %Abbildungen
\usepackage{lscape}	%Gedrehte Darstellungen
\usepackage{pst-plot, pstricks} %Graphen
\usepackage{fancybox,amssymb,color} %Fancy stuff (z.B. Umrahmungen), Formeln, Farben
\usepackage[english]{babel}
\usepackage{setspace} %Zeilenabstand
\usepackage{epstopdf}
\usepackage[capposition=top]{floatrow} 
\usepackage{float}
\usepackage{enumitem}
\usepackage{longtable}
\usepackage{tabularx,calc}
\usepackage{multirow}
\usepackage{booktabs}
\usepackage{acro}
\usepackage{setspace}
\usepackage{lscape}
\usepackage{mathtools }
 \usepackage{rotating}

\usepackage[authoryear]{natbib}
\usepackage{caption}

\usepackage{amsmath}
\usepackage{amsthm} 
\usepackage{amssymb}
\usepackage{amsbsy}
\usepackage{bm}

 \usepackage{xr}
\usepackage{changes}%added
\definechangesauthor[name={Sascha}, color=green]{Sascha}

\newtheorem{proposition}{Proposition}

\newtheorem{assumption}{Assumption} 
 
\title{Estimating Fiscal Multipliers by Combining Statistical Identification with Potentially Endogenous Proxies\thanks{We thank Juan Antolin Diaz, Jetro Anttonen, Regis Barnichon, Robin Braun, Martin Bruns, Ralf Brüggemann, Luca Fanelli, Francesco Furlanetto, Markku Lanne, Michael McCracken, Michael Owyang, Franck Portier, Morten Ravn, Malte Rieth, and Roland Winkler for their helpful comments. Our appreciation also extends to the participants of the BSE Summer Forum Workshop on Advances in Structural
Shocks Identification, Kings College London Workshop in Empirical and Theoretical Macroeconomics, IAAE 2023, and the RuhrMetrics Research Seminar for their comments and suggestions. Jan Prüser gratefully acknowledges the support of the German Research Foundation (DFG, 468814087). The opinions expressed in this article are the sole responsibility of the authors and should be interpreted as reflecting the views of Sveriges Riksbank.}}

\author{Sascha A. Keweloh\footnote{\scriptsize  TU Dortmund University, Department of Economics, D-44221 Dortmund, sascha.keweloh@tu-dortmund.de.} 
\hspace{3.5em} Mathias Klein\footnote{\scriptsize  Sveriges Riksbank, Monetary Policy Department - Research, SE-103 37 Stockholm, Sweden, mathias.klein@riksbank.se.} 
\hspace{5em} Jan Prüser\footnote{\scriptsize  TU Dortmund University, Department of Economics, D-44221 Dortmund,  prueser@statistik.tu-dortmund.de.}  
\\ 
\hspace{1.5em} TU Dortmund University 
\hspace{1.5em} Sveriges Riksbank 
\hspace{1em} TU Dortmund University   
}  
\date{\today}

\begin{document}
%\textbf{Alternative title: A unification of fiscal multiplier estimates in proxy VARs }
\clearpage\maketitle
\thispagestyle{empty}
\begin{abstract}

\noindent Different proxy variables used in fiscal policy SVARs lead to contradicting conclusions regarding the size of fiscal multipliers. Our analysis suggests that the conflicting results may stem from violations of the proxy exogeneity assumptions. We propose a novel approach to include proxy variables into a Bayesian non-Gaussian SVAR, tailored to accommodate potentially endogenous proxies. Using our model, we find that increasing government spending is more effective in stimulating the economy than reducing taxes.

%We construct new exogenous proxies that can be used in the traditional proxy VAR approach resulting in similar estimates compared to our proposed hybrid SVAR model. 

\vspace{0.75cm}
\end{abstract}

\clearpage
\pagenumbering{arabic}

\section{Introduction}

The proxy Structural Vector Autoregression (SVAR) has become a popular tool to identify fiscal policy shocks. Prominent examples are \citet{mertens2014reconciliation} who use a tax proxy and \citet{caldara2017analytics} who rely on non-fiscal proxy variables like a TFP measure. Notably, both studies reach contradictory conclusions about the size of fiscal multipliers, \citep{angelini2020fiscal}.  A potential explanation for these conflicting results are endogenous proxies.

Alternatively, statistical identification methods can achieve point identification by imposing stronger assumptions on the stochastic properties of the shocks, see \cite{matteson2017independent}, \cite{gourieroux2017statistical},  \cite{keweloh2020generalized}, \cite{lewis2021identifying},   \cite{guay2021identification}, \cite{jarocinski2024estimating}, \cite{crucil2024monetary}, \cite{pruser2024large}, or \cite{kociecki2025non}.
However,  even if the stronger stochastic assumptions are satisfied, inference based on statistical identification necessarily demands more from a finite sample than standard economic restriction- or proxy-based estimation approaches, see \cite{montiel2022svar}.

In this paper, we propose a hybrid approach combining proxies and statistical identification which is robust to endogenous proxies. We leverage prior knowledge of an exogenous proxy to reduce estimation uncertainty while maintaining the flexibility to downweight or discard the proxy if the data provide evidence against its exogeneity. This mitigates the bias that would otherwise result from endogenous proxies.  
Allowing for potentially endogenous proxies, we find that increasing government spending is a more effective tool to stimulate the economy than lowering taxes, and previous contradictions in the literature seem to be caused by endogenous proxies. 

Our econometric contribution consists of two parts. 
First, we combine proxies with a statistically identified SVAR using independent and non-Gaussian shocks. Crucially, in the statistically identified SVAR, we do not need to assume exogenous proxies. 
Instead, we estimate their correlation with non-target shocks and introduce a prior that shrinks towards exogeneity. 
We demonstrate that our shrinkage improves estimation precision when proxies are, in fact, exogenous. However, if the proxies are endogenous, the data can update the prior and reduce shrinkage, thereby mitigating the bias that endogenous proxies would otherwise introduce.  
 
We build on a growing literature that combines statistical identification and economically motivated restrictions, see  \cite{schlaak2023monetary}, \cite{drautzburg2023refining}, \cite{braun2021importance}, \cite{keweloh2023monetary}, \cite{HERWARTZ2023104630}, or \cite{carriero2024blended}. 
\cite{schlaak2023monetary} combine identification by heteroskedasticity and proxy variables. They show that combining statistical identification  with exogenous proxies improves efficiency, but performance deteriorates when the proxies are endogenous. 
Our approach advances this literature by shrinking the coefficients toward the exogenous proxy restriction instead of imposing these through a dogmatic prior.  
This makes sense if we believe the proxy is only weakly exogenous or are uncertain whether the proxy is exogenous at all. Importantly and in contrast to the hybrid literature cited above, our prior beliefs can still be updated by the data. 
This enables the model to inform us about endogenous proxies in a data-driven fashion, thereby mitigating the bias arising from their inclusion. But our aim extends beyond a mere test of proxy exogeneity. Instead, we try to leverage prior knowledge of an exogenous proxy (possible approximate) to improve the accuracy of the estimation and reduce estimation uncertainty. Overall, our framework preserves the key advantage of traditional hybrid approaches—namely, the ability to leverage exogenous proxies to improve estimation precision, while at the same time remaining robust to proxy endogeneity.

Our second econometric contribution is a novel approach to incorporate proxies into a Bayesian SVAR. In the frequentist setting, researchers can choose between two main strategies for estimating a proxy SVAR: (i) a moment-based approach, as in \cite{stock2012disentangling} and \cite{mertens2013dynamic}, which relies only on the assumption that the proxy is uncorrelated with non-target shocks; or (ii) the augmented proxy approach, as in \cite{angelini2019exogenous}, which additionally imposes a specific data-generating process for the proxy.
In contrast, Bayesian researchers currently lack this flexibility and must rely exclusively on the augmented proxy SVAR, as in \cite{caldara2019monetary}.
We fill this gap and introduce a new approach that incorporates moment conditions into the Bayesian SVAR, mirroring the frequentist moment-based proxy approach of \cite{stock2012disentangling} and \cite{mertens2013dynamic}, while preserving the advantages of Bayesian inference. 
Unlike the augmented proxy approach, and in line with moment-based frequentist proxy estimators, our proxy weighting method does not require specifying the functional form of the proxy. We show that this represents a key advantage in a non-Gaussian SVAR, where misspecifying the proxy's functional form can induce dependence among shocks and undermine the validity of non-Gaussian identification strategies.  

We use our model to estimate the effects of fiscal policy shocks and evaluate whether commonly used proxies are exogenous. We provide evidence that the shocks identified using the fiscal proxy SVAR from \cite{mertens2013dynamic} and using the non-fiscal proxy SVAR from \cite{caldara2017analytics} are non-Gaussian. Specifically, all shocks show heavy tails and the tax and output shocks are left skewed. This data feature, which is used for identification, also makes sense from an economic point of view because tax reductions are generally larger (in absolute terms) than tax increases, and GDP usually falls stronger during recessions than it rises during expansions. It turns out that the government spending multiplier is larger than the tax multiplier. On impact, the tax multiplier is close to zero and it remains below unity for the entire impulse response horizon of five years. The government spending multiplier, however, is above unity for the first year after the shock and then slowly converges back to its pre-shock level. Evidently, our results are in some contrast to previous studies. In particular,  \citet{mertens2014reconciliation}  estimate a tax multiplier around three. Our empirical findings indicate that the large differences between applying the \citet{mertens2014reconciliation} approach and our model are due to the tax proxy not being exogenous. Specifically, we provide evidence that the narrative tax measure is negatively correlated with structural output shocks. Intuitively, not accounting for this correlation leads to identified tax cuts that also include exogenous increases in output, which increases the size of the estimated tax multiplier. 

We emphasize that the reader should interpret our results with the understanding that they rely on strong assumptions about the error process, which may not necessarily hold. To address this concern, we carefully investigate the potential violation of these assumptions, but do not find any empirical evidence suggesting that they are invalid. Furthermore, we present historical evidence that validates our empirical findings based on the statistical identification approach. In particular, the narrative evidence indicates that the tax proxy shows a tendency to indicate an exogenous tax increase during periods of economic recessions and an exogenous tax decrease during episodes of economic expansion. Similarly, we also find evidence that the TFP proxy used by \citet{caldara2017analytics} is negatively correlated with exogenous government spending shocks. Therefore, assuming an exogenous TFP proxy identifies output shocks which include spending shocks of opposite sign. As a consequence, the estimated spending multiplier is biased upward. 

Several related papers use statistical identification to estimate the dynamic effects of fiscal shocks. For example, \citet{lewis2021identifying} exploits time-varying shock variances, finding government spending and tax multipliers of similar size, both below one. \citet{guay2021identification} relies on higher-order moments and shows that estimated multipliers depend heavily on additional assumptions. \citet{karamysheva2022we} use non-Gaussian errors and also find similar-sized multipliers.

The remainder of the paper is organized as follows: Section \ref{sec: SVARs} describes our novel proxy weighting approach and the combination of proxy variables with non-Gaussianity.   Section \ref{sec: Finite Sample Performance} provides a summary of the results of a range of Monte Carlo simulations.  Section \ref{sec: Application} uses the proposed model to analyze the effects of fiscal policy shocks.
Section \ref{sec: Conclusion} concludes.

  \section{SVARs}\label{sec: SVARs}
This section presents the basic notation. 
Afterwards, we introduce a novel approach to incorporate proxy variables into a Bayesian SVAR using moment conditions.
 Subsequently, we combine the proxy moment conditions with independent non-Gaussian shocks and propose priors which shrink to proxy exogeneity. 
 Finally, we discuss why it is important to use moment conditions to combine independent non-Gaussian shocks with proxy variables.

  \label{sec: Proxy VAR}
  An SVAR with $n$ variables can be written as
  \begin{align}
  \label{eq: SVAR}
  	\bm y_t  = \bm \nu +  \bm A_1 \bm y_{t-1} + ... +\bm  A_p y_{t-p}  +\bm u_t 
  \quad \text{ and } \quad 
  	\bm u_t =  \bm B_0 \bm \varepsilon_{t} ,
  \end{align}
  with 
  parameter matrices $\bm A_1,...,\bm A_p \in \mathbb{R}^{n \times n}$ and  $\text{det}(\bm I-\bm A_1c-...-\bm A_pc^p)\neq0$ for $|c|\leq 1$,
  an intercept  $\bm \nu$, 
  an invertible matrix $\bm B_0 \in \mathbb{B} := \{\bm B \in \mathbb{R}^{n \times n} | \text{det}(\bm B)\neq 0 \}$, 
  an $n$-dimensional vector of  time series  $\bm y_t=[\bm y_{1t} ,...,\bm y_{nt} ]'$,  
    an $n$-dimensional vector of reduced form shocks $\bm u_t=[ u_{1t} ,..., u_{nt} ]'$,
  and an $n$-dimensional vector of serially uncorrelated structural shocks $ \bm \varepsilon_t=[ \varepsilon_{1t},..., \varepsilon_{nt}]'$ with mean zero and unit variance.  
  
  We define $\bm \pi_0 = \text{vec}([\bm v,\bm A_1',...,\bm A_p']')$
  to collect the specific parameter matrices of the data-generating process and we use $\bm \pi $ to denote a vector of SVAR parameter matrices which can differ from the data-generating parameter matrices.
  Moreover, the innovation $e_{it}(\bm B,\bm \pi)$ denotes the $i$th component of $\bm e_t(\bm B,\bm \pi)=\bm B^{-1} (\bm y_t - \bm \nu -  \bm A_1 \bm y_{t-1} - ... - \bm A_p \bm y_{t-p})$.
Therefore, $\bm e_t(\bm B,\bm \pi)$ represents the shocks given $\bm \pi$  and $\bm B$. If $\bm \pi$  and $\bm B$ are equal to the corresponding values of the data-generating process, $\bm e_t(\bm B,\bm \pi)$ is equal to the structural shocks $\bm \epsilon_t$.
Without additional restrictions, the model is not identified.%, i.e. any orthogonal matrix $\bm Q$ 
%$\bm Q \in \mathcal{O}$ of the orthogonal group $\mathcal{O}(m)=\{\bm Q \in\mathbb{R}: \bm Q \bm Q'=\bm I_m \$ yields an observationally equivalent model $ \tilde{\bm B}=\bm B \bm Q$. 
We will use proxy variables and non-Gaussian shocks to archive identification.

 %\added{In addition we provide evidence based on simulated as well as real data to provide further evidence for the need of avoid specifying the proxy process when using independent non-Gaussian errors. den satz streichen?}

%, and misspecification of the proxy process in the augmented proxy SVAR can result in incorrect coverage rates, as discussed in the \textcolor{red}{Appendix}
%compare \cite{jentsch2019dynamic},   \cite{bruns2022alternative}, or \cite{budnikidentifying}.  
%To address these limitations, w

\subsection{A novel proxy weighting approach}\label{sec:weighting} 

A valid proxy is correlated with the target shock and uncorrelated with non-target shocks.
  \begin{assumption}
\label{assumption: proxy}
A variable $z_t$ is a valid proxy for the target shock $\epsilon_{it}$ if it satisfies:
\begin{align} 
\nonumber
    1. \textbf{ Relevance: }    E[z_t \varepsilon_{it}]\neq 0 
    \quad \quad \quad \quad \quad
    2. \textbf{ Exogeneity: } E[z_t \varepsilon_{jt}]= 0 \text{ for } j \neq i
\end{align}
\end{assumption}  
 For simplicity, we assume that the first shock $ \varepsilon_{1t}$ is the target shock.
 
We propose a Bayesian proxy approach using moment conditions based on the proxy exogeneity assumption to reweight the likelihood of the SVAR. 
In contrast to the augmented proxy SVAR proposed by \cite{caldara2019monetary}, our proposed method does not require  specifying the functional form of the proxy.
Section \ref{sec: Why do we use moment conditions for the estimation?}, compares both methods and discusses the implications of misspecifying the functional form of the proxy.

%In comparison, the Bayesian augmented proxy SVAR typically assumes $z_t = \beta_1 \varepsilon_{1t} + \hat{\eta}_t $, meaning the proxy is specified as a linear function of the target shock and a noise term which is assumed to be uncorrelated with all structural shocks, see, e.g. \cite{caldara2019monetary}. However, misspecification of the functional form of the proxy can lead to a failure of identification. To see this, consider a proxy generated by the process $z_t = \beta_1 \varepsilon_{1t} + \beta_2 \varepsilon_{1t}^2 + \eta_t $. Misspecification of the proxy process in the augmented proxy SVAR leads to noise $\hat{\eta}_t = z_t - \beta_1 \varepsilon_{1t} =  \beta_2 \varepsilon_{1t}^2 + \eta_t $. Due to misspecification, the target shock still affects the misspecified noise term $\hat{\eta}_t$. Consequently, the misspecified noise term is correlated with the target shock, which is a contradiction to the assumptions of the augmented proxy SVAR. 

%. The approach proposed in this section is similar to moment based proxy estimators in the frequentist literature and does not require specifying the functional form of the proxy.

Define the proxy exogeneity moment conditions measuring the correlation of proxy and non-target shocks $ e_{jt}(\bm B,\bm \pi)$ for $j=2,...,n$ with
\begin{align}
\label{eq: D }
    D_j( \bm z, \bm y, \bm B, \bm \pi) = \frac{1}{ \sqrt{T} } \sum_{t=1}^{T} z_t  e_{jt}(\bm B,\bm \pi) 
\end{align}
and define $D(\bm z) = [D_2 (\bm z, \bm y, \bm B, \bm \pi),..., D_n (\bm z, \bm y, \bm B, \bm \pi)]'$.
The joint density of the data $\bm y$ and the proxy exogeneity moment conditions is
\begin{equation}
\label{eq: joint like z D}
  p(\bm y, D(\bm z) |\bm B, \bm \pi)=   p(\bm y| \bm B, \bm \pi)  p( D(\bm z) | \bm y ,  \bm B, \bm \pi). 
\end{equation} 
The first term denotes the likelihood of the data $\bm y$ and
the second term is the conditional density of the proxy exogeneity moment conditions given the data. 
For an exogenous proxy variable, we assume that 
\begin{align}
\label{eq: Dz dist}
    D(\bm z) | \bm y, \bm B, \bm \pi \sim \mathcal{N}(\bm 0, \bm \Sigma_z),
\end{align} 
where  $\bm \Sigma_z $ denotes the covariance matrix of $( z_t  e_{2t} ,... ,z_t  e_{nt} )'$, see section \ref{Dz}.

The moment conditions do 
not constitute a generative model,  and as such we view $p( D(\bm z) | \bm y ,  \bm B, \bm \pi)$ as a pseudo-likelihood following \cite{yin2009bayesian}.\footnote{A pseudo-likelihood function is also used for Bayesian inference in narrative SVARs \citep{antolin2018narrative} as well as in  local projections \citep{ferreira2023bayesian}. For a general discussion on pseudo-likelihoods for Bayesian inference, see \cite{ventura2016pseudo}.} \cite{yin2009bayesian} presents a Bayesian version of the Generalized Method of Moments (GMM) by defining a pseudo-likelihood
function as follows
\begin{align}
\tilde{L}(\bm z|  \bm y ,  \bm B, \bm \pi) \propto \text{exp}( -0.5 D(\bm z)' \bm \Sigma_z^{-1} D(\bm z)),
\end{align}
which is the kernel of a normal density.
\cite{yin2009bayesian} shows the validity of the posterior distribution resulting from this pseudo-likelihood. Specifically, \cite{yin2009bayesian} demonstrate that the pseudo-likelihood defined by GMM criteria leads to valid Bayesian inferences in finite samples using probability coverage of posterior sets as suggested by \cite{monahan1992proper}. \cite{kim2002limited} justifies the incorporation of the GMM  criterion in the  Bayesian framework asymptotically. Furthermore, \cite{chernozhukov2003mcmc} shows that a GMM criterion can be seen as the Laplace approximation of the negative true likelihood
evaluated around the mode. Hence, we view our pseudo-likelihood
as a convenient device that allows us to use the Bayesian toolkit when estimating our hyprid SVAR.

Throughout the paper, we refer to the model using a Gaussian likelihood $p(\bm y| \bm B, \bm \pi) $ and the conditional density based on an exogenous proxy variable in Equation (\ref{eq: Dz dist}) as the \textit{Gaussian proxy weighting approach}.  
 If we assume a Gaussian likelihood, we need to assume that the proxy variables are exogenous to archive identification.  Intuitively, the second term in Equation (\ref{eq: joint like z D}) re-weights the likelihood of the VAR data $\bm y$ by giving more weight to structural parameters that result in non-target shocks that are uncorrelated with the proxy.  
The following proposition shows that for a valid proxy, the joint likelihood in Equation (\ref{eq: joint like z D}) identifies the impact of the target shock.

\begin{proposition}
    \label{prop: identification proxy weighting}
    Consider an SVAR $u_t = \bm B_0 \varepsilon_t$ with a valid proxy variable $z_t$ for the target shock $\varepsilon_{1t}$. Let $\bm b_{1,0}$ be the impact of the target shock equal to the first column of $\bm B_0$.
    Assume that $z_t \epsilon_{lt}$ for $l=1,...,n$ is i.i.d. with finite mean and variance such that we can approximate the distribution of $\sqrt{T}\frac{1}{T}\sum_{t=1}^{T} z_t \epsilon_{lt}$ using the central limit theorem.\footnote{We need the i.i.d assumption to apply the central limit theorem. This assumption may be violated in practice if a proxy variables can be explained by its own past or the past of other variables. In this case we can clean the proxy variables using an extra regression step. In the Online Appendix we find that in our empirical application the results are robust to such an step.} 
    
    The joint likelihood in Equation (\ref{eq: joint like z D}) 
    identifies the target shock and its impact $\bm b_{1,0}$, i.e. for $\bm B= \bm B_0 \bm Q$ with an orthogonal matrix $\bm Q$ the distribution of $D(z)| \bm y, \bm B $ is asymptotically equal to the distribution of $D(z)| \bm y, \bm B_0 $ from Equation (\ref{eq: Dz dist})  if and only if the first column of $\bm B$ is equal to $\bm b_{1,0}$.
    
    Moreover,  the  conditional likelihood  $p( D(\bm z) | \bm y ,  \bm B, \bm \pi)$  converges in probability to zero if the first column of $\bm B$ is not equal to $\bm b_{1,0}$, i.e. for $\bm B= \bm B_0 \bm Q$ with an orthogonal matrix  $\bm Q$ it  holds that
$ 
      p(D(\bm z) | \bm y ,  \bm B)  \overset{p}{\rightarrow} 0 
$ 
if the first column of $\bm B$ is not equal to $\bm b_{1,0}$.
 
\end{proposition}
\begin{proof}
    See the appendix.
\end{proof}
The approach can be extended to the case of multiple proxies for multiple target shocks. A corresponding proposition can be found in the Online appendix.

The conditional density of the proxy exogeneity moment conditions downweights the likelihood of $\bm B$ and $\bm \pi$ values leading to non-target shocks which are correlated with the proxy, i.e. it downweights the solutions which lead to an endogenous proxy. However, the ability to downweight endogenous proxy solutions depends on the relevancy of the proxy.  The inclusion of an exogenous but irrelevant proxy asymptotically only scales the joint likelihood in Equation (\ref{eq: joint like z D}) but does not affect its shape or maxima.  
\begin{proposition}
    \label{prop: relevance proxy weighting}
    In an SVAR   $u_t = \bm B_0 \varepsilon_t$  with Gaussian shocks and an exogenous but irrelevant proxy variable $z_t$ for the target shock $\varepsilon_{1t}$, the conditional likelihood of the proxy exogeneity moment conditions is asymptotically flat, that is, we have $ D(\bm z) | \bm y, \bm B_0 \bm Q   \sim  \mathcal{N} (  0, \Sigma_z )$ for all orthogonal matrices $ \bm Q$ and $E[p(D(\bm z) | \bm y, \bm B_0 \bm Q_1 ) ] = E[p(D(\bm z) | \bm y, \bm B_0 \bm Q_2 ) ] $ for all orthogonal matrices $\bm Q_1$ and $\bm Q_2$.  
\end{proposition}
\begin{proof}
    See the  Online appendix.
\end{proof}

\subsection{Non-Gaussian SVARs}\label{sec: Bayes nG SVAR} 
This section explains how independent non-Gaussian structural shocks can be used to identify and estimate the simultaneous interaction in the SVAR. Technically, we impose the following assumptions:
\begin{assumption}
   \label{ass: shocks} 
\begin{enumerate}
    \item The components of the structural shocks $\varepsilon_t$ are a sequence of independent and identically distributed
random vectors with   zero mean and unit variance.
    \item The components of the structural shocks $\varepsilon_t$ are mutually independent and at most one component has a Gaussian marginal distribution.
\end{enumerate}
\end{assumption}

\cite{lanne2017identification} show that these assumptions are sufficient to identify the SVAR up to sign and permutation of the shocks. Assumption \ref{ass: shocks} imposes independence and non-Gaussianity of the structural shocks. Both conditions have been relaxed in the literature: \cite{mesters2024non}, \cite{keweloh2023uncertain}, and \cite{herwartz2023identification} explore relaxations of the independence assumption, while \cite{maxand2020identification} addresses the relaxation of non-Gaussianity. For simplicity,  we adopt the standard assumption of independent and non-Gaussian shocks in Assumption \ref{ass: shocks}.

Importantly, the identified shocks must be labeled manually by the researcher. A labeling strategy based on economic reasoning can attach an economic interpretation to the shocks and at the same time prevent permutation switches. We follow \cite{bertsche2022identification} and label a shock as the target shock if it has the highest correlation with the corresponding proxy in absolute magnitude. We discuss labeling in more detail in the Online Appendix.

\cite{lanne2020identification}, \cite{anttonen2021statistically}, and \cite{braun2021importance} propose Bayesian non-Gaussian SVAR models. 
   We simplify \cite{anttonen2021statistically} and assume that each shock follows a skewed t-distribution such that the density of the $i$th shock is given by
\begin{align}
f_i(\varepsilon_{it}; \lambda_i,q_i)=\frac{\Gamma(0.5 + q_i)}{v (\pi q_i)^{0.5} \Gamma(q_i) ( \frac{|\varepsilon_{it}+m|^2}{q_i v^2 (\lambda \text{sign}(\varepsilon_{it}+m)+1)^2} 
)^{0.5+q_i}},
\end{align} 
with $|\lambda_i|<1$, $q_i>2$ which implies that the fourth moment of the $i$th  shock exists, and
  the normalization $m= \frac{2 v \lambda q_i^{0.5} \Gamma(q_i-0.5)}{\pi^{0.5}\Gamma(q_i+.5)}$, $v=q_i^{-0.5} \left[ (3 \lambda^2 + 1)(\frac{1}{2q_i-2}) - \frac{4\lambda^2}{\pi} (\frac{\Gamma(q_i-0.5)}{\Gamma(q_i)})^2 \right]^{-0.5}$ to mean zero and unit variance, which normalizes the  size of the shocks.
The likelihood of the data  $\bm y$  given $\bm y_{-p+1},...,\bm y_0 $  follows from \cite{lanne2017identification} and is equal to
\begin{align}
\label{eq: likeli svar}
    p(\bm y| \bm \pi, \bm B,\bm \lambda, \bm q)=|\text{det}(\bm B)|^{-T} \prod_{i=1}^{n} \prod_{t=1}^{T} f_i(  e_{it}(\bm B,\bm \pi); \lambda_i,q_i).
\end{align} 
Notably, by estimating $\bm \lambda$ and $\bm q$ our framework is flexible and the data can inform us over the degree of skewness and excess kurtosis, see \cite{anttonen2021statistically}.

\subsection{Combining proxy variables with non-Gaussianity}
\label{sec: nG proxy}

This section combines the re-weighting based on proxy exogeneity moment conditions with a non-Gaussian SVAR. Specifically, we use the non-Gaussian likelihood of Equation (\ref{eq: likeli svar}) in Equation (\ref{eq: joint like z D}), which ensures identification under Assumption \ref{ass: shocks}.
Therefore, the proxy is not required for identification. Consequently, in contrast to the existing proxy approaches, we are able to identify shocks even if the proxy is endogenous. If the data do not support the exogeneity of the proxy, the model can effectively ignore its information. However, if the proxy is exogenous and relevant, incorporating it can lead to improved estimation accuracy and reduced estimation uncertainty in the non-Gaussian SVAR framework.

We generalize the proxy moment condition in Equation (\ref{eq: D }) to 
\begin{align}
\label{eq: D 2}
    D_j( \bm z, \bm y, \bm B, \bm \pi, \mu) =   \frac{1}{\sqrt{T}} \sum_{t=1}^{T}( z_t  e_{jt}(\bm B,\bm \pi)-\mu_j) ,
\end{align}
and allow for a non-zero mean $\bm \mu=(\mu_2,\dots,\mu_{n})$ of the moment conditions, which corresponds to endogenous proxies. Our aim is to estimate the exogeneity of the proxy variable $\bm \mu$ instead of restricting the proxy variable to an exogenous proxy by imposing $\bm \mu=0$.

This leads to the joint density
\begin{equation}
\label{eq: joint like z D 2}
  p(\bm y, D(\bm z) |\bm B, \bm \pi, \bm  \mu,\bm \lambda, \bm q)=   p(\bm y| \bm B, \bm \pi,\bm \lambda, \bm q)  p( D(\bm z) | \bm y ,  \bm B, \bm \pi,\bm \mu) 
\end{equation} 
and for $\bm \mu =  [  E[z_t  e_{2t}],...,E[z_t  e_{nt}]]'$ we have 
\begin{align}
\label{eq: Dz dist 2}
    D(\bm z) | \bm y, \bm B, \bm \pi, \bm \mu \sim \mathcal{N}(\bm 0,\bm \Sigma_z).
\end{align}

For an exogenous proxy $\bm \mu$ contains only zeros.
We propose a prior distribution for $\bm \mu$ that reflects the belief in an exogenous proxy. Specifically, we shrink the solution towards an exogenous proxy variable, i.e. we shrink all coefficients $\mu_{ j}$ in $\bm \mu$ to zero with the prior
\begin{align} 
    \label{eq: prior mu}
	\mu_{ j}    \sim N(0, \sigma_{\mu j }^2), 
    \quad
	\sigma_{\mu j }^2  \sim IG(a,b).
    %& \quad \text{ and } 
	%\sigma_j^2   =\text{Var}(\bm z).
\end{align} 
We set $a=b=0$ which results in a flat prior, see \cite{tipping2001sparse}. The prior variance $\sigma_{\mu j }^2$  controls shrinkage of $\mu_{ j}$ to zero. If we have strong beliefs in the validity of the proxy, we would set $\sigma_{\mu j }^2$ to be small. However, such prior beliefs can be controversial. To avoid fixing $\sigma_{\mu j }^2$ at an inappropriate value, we use the data and estimate $\sigma_{\mu {j}}^2$ hierarchically using an inverse Gamma prior.
Hence, by shrinking $\mu_{ j}$ to zero, the model benefits from valid proxies, but if empirical warranted it can avoid such shrinkage and $\mu_{ j}$ can take on values different from zero to allow for endogenous proxy variables.

The joint posterior is proportional to
\begin{align}
 p( \bm \pi, \bm B,\bm \lambda, \bm q, \bm \mu, \bm \Sigma_{\boldsymbol{\mu}}|\bm y,  D(\bm z) )  \propto  p(\bm y| \bm B, \bm \pi,\bm \lambda, \bm q)  p( D(\bm z) | \bm y ,  \bm B, \bm \pi,\bm \mu)p(\bm \mu| \bm \Sigma_{\boldsymbol{\mu}}),
\end{align} 
with $p(\bm \mu| \bm \Sigma_{\boldsymbol{\mu}})=N(\bm \mu; \bm 0,\bm \Sigma_{\boldsymbol{\mu}}) $ and $ \bm \Sigma_{\boldsymbol{\mu}} = \text{diag}(  \sigma^2_{\mu_1},....,\sigma^2_{\mu_n})$.\footnote{We use flat priors for all model parameters. Alternatively, it would be possible to use a Minnesota type prior or a more flexible global local prior on $\bm \pi$, see e.g., \cite{huber2019adaptive}, \cite{cross2020macroeconomic} and \cite{pruser2023data}.} 
The Markov Chain Monte Carlo (MCMC) algorithm to sample from the posterior is in the Online Appendix.\footnote{We rely on the algorithm proposed by \citet{ter2008differential} to draw from the posterior distribution, which makes our approach more computationally intensive than alternative proxy methods, e.g. \cite{arias2021inference}. However, we find that the sampler performs well across a wide range of Monte Carlo experiments and robustness checks. Our experience is consistent with the findings reported in \citet{anttonen2021statistically}, who document similarly reliable performance using this algorithm.}

Throughout the paper, we refer to the model using the non-Gaussian likelihood $p(\bm y| \bm B, \bm \pi, \bm \lambda, \bm q) $, the conditional likelihood $p( D(\bm z) | \bm y ,  \bm B, \bm \pi, \bm  \mu)$ based on the proxy moment conditions in Equation (\ref{eq: Dz dist 2}), and the proxy exogeneity shrinkage prior in Equation (\ref{eq: prior mu}) as the \textit{non-Gaussian proxy weighting approach}.

The conditional density $p( D(\bm z) | \bm y ,  \bm B, \bm \pi, \mu)$ still re-weights the SVAR likelihood and thereby utilizes the information of the proxy.
However, by estimating $\bm \mu$ we allow for exogenous proxy  ($\bm \mu=0$) and an endogenous proxy with ($\bm \mu \neq 0$). If we were to estimate $\bm \mu$ without shrinking it toward the exogeneity restriction $\bm \mu = 0$ via the shrinkage prior in Equation (\ref{eq: prior mu}), $D(z)$ would simply contribute an equal number of moment conditions and new parameters in $\bm \mu$. In that case, one could choose a $\bm \mu \neq 0$ to perfectly satisfy the moment conditions, resulting in no impact on the estimation of $\bm B$; that is, neither efficiency gain nor bias would arise.  
Conversely, if we were to fix $\bm \mu = 0$ to enforce an exogenous proxy, we would arrive at a typical blended identification approach, combining statistical identification with an economic restriction, as in \cite{schlaak2023monetary} or \cite{carriero2024blended}. In that case, the proxy moment conditions $D(z)$ can only be satisfied by choosing $\bm B$ such that the proxy is exogenous. This leads to efficiency gains in the estimation of $\bm B$ if the proxy is truly exogenous, but induces a bias if the proxy is endogenous.\footnote{
Analytically proving that combining statistical identification with economic restrictions yields efficiency gains over using either approach alone is challenging. Hybrid approaches typically demonstrate such gains via Monte Carlo simulations; we do so as well in our application. Intuitively, our non-Gaussian estimator identifies the model from higher-moment information, which can result in volatile estimates, whereas an exogenous proxy relies on a simple covariance restriction. Combining both sources of information stabilizes estimation and thereby improves efficiency. To our knowledge, the only analytical proof that combining statistical identification with economically motivated short-run restrictions improves efficiency over an estimator relying solely on the short-run restriction is given in \cite{keweloh2025higher}.
 }
Instead, we estimate $\bm \mu$ and impose the shrinkage prior in Equation (\ref{eq: prior mu}) to pull $\bm \mu$ toward zero. By estimating the variances $\sigma_{\mu_j}^2$ of $\bm \mu$, the data determine the cost of deviating from an exogenous proxy. If the proxy is exogenous, the estimated variances $\sigma_{\mu_j}^2$ will be small, making deviations of $\bm \mu$ from zero costly. In this case, the estimator will keep $\bm \mu$ close to zero and instead adjust $\bm B$ to satisfy the proxy moment conditions $D(z)$, thereby achieving efficiency gains. Conversely, if the proxy is endogenous, the estimated variances $\sigma_{\mu_j}^2$ will be large, making deviations from zero inexpensive. In that case, the estimator can use $\bm \mu$ to satisfy the proxy moment conditions $D(z)$ without inducing bias in $\bm B$.  
Our model therefore exploits information from proxy variables to improve the precision and reduce the uncertainty of the non-Gaussian SVAR estimation, while also mitigating bias when the proxies are endogenous.

\subsection{The covariance matrix $\bm \Sigma_z$}\label{Dz}
Implementing the proxy weighting estimators in Sections  \ref{sec:weighting} and \ref{sec: nG proxy} requires specifying or estimating $\bm \Sigma_z$, the covariance matrix of $( z_t  e_{2t} ,... ,z_t  e_{nt} )'$. 

First, if the proxy is exogenous in a strong sense—meaning it is not only uncorrelated but also satisfies $E[z_t^2 \varepsilon_{jt} \varepsilon_{kt}]=E[z_t^2 ]  E[\varepsilon_{jt} \varepsilon_{kt}] $ with non non-target shocks $\varepsilon_{jt} \varepsilon_{kt}$ which for example follows from independence—then $\bm \Sigma_z = \text{diag}(\sigma_z^2, \dots, \sigma_z^2)$, where $\sigma_z$ denotes the standard deviation of the proxy variable.
Second, if the proxy is endogenous, the estimator stops shrinking $\mu_j$ toward zero, making the choice of $\bm \Sigma_z$ less critical. Specifically, when deviations of $\mu_j$ from zero entail little cost in Equation (\ref{eq: prior mu}), $\mu_j$ can be adjusted such that $D_j(\bm z, \bm y, \bm B, \bm \pi, \mu)$ in Equation (\ref{eq: D 2}) equals zero. In this case, the proxy becomes irrelevant to the estimation, and so does the precise specification of $\bm \Sigma_z$.
%Intuitively, by estimating $\bm \mu$ we can shift the distribution of the moment conditions away from zero in the presence of an endogenous proxy, making the choice of $\bm \Sigma_z$ less important. 

 In the remainder of the study, we use the simple assumption $\bm \Sigma_z=\text{diag}(\sigma_z^2,...,\sigma_z^2)$.
However, we consider two alternatives approaches. For the first, we follow \cite{yin2009bayesian} and set it to the empirical covariance matrix $ \bm \Sigma_z=  \frac{1}{T}\sum_{t=1}^T (z_t e_{2t}, \dots,z_t e_{nt})'(z_t e_{2t}, \dots,z_t e_{nt})-D(\bm z)D(\bm z)' $. For the second we estimate $\bm \Sigma_z$ and  shrink $\bm \Sigma_z$ to $\text{diag}(\sigma_z^2,...,\sigma_z^2)$. In particular, we assume that $\bm \Sigma_z (i,k)\sim N(m_{i,k},\sigma_{i,k}^2)$ with $m_{i,k}=\sigma_z^2$ if $i=k$ and otherwise we set $m_{i,k}=0$ and $\sigma_{i,k}^ 2\sim  IG(c,d)$ with $c=d=0$. In the Online Appendix we show the all the alternatives lead to similar result in our empirical application as well as in our Monte Carlo simulations.

 \subsection{Why do we use moment conditions for the estimation?  }\label{sec: Why do we use moment conditions for the estimation?} 
In the frequentist framework, there are two main approaches to incorporate proxy variables into an SVAR: (a) the moment-based proxy approach of \cite{stock2012disentangling} and \cite{mertens2013dynamic}, and (b) the augmented proxy SVAR approach of \cite{angelini2019exogenous}. In contrast, the Bayesian proxy SVAR literature relies exclusively on the augmented proxy SVAR, see  \cite{caldara2019monetary} or \cite{arias2021inference}.

This section compares the moment-based and augmented proxy SVAR approaches. We demonstrate that specifying the proxy’s data-generating process, as required in the augmented proxy SVAR, introduces a risk of misspecification, which can compromise identification—particularly in non-Gaussian models relying on independent shocks.

The augmented proxy SVAR approach adds an additional equation to the SVAR to model the proxy variable, such that the augmented system is equal to
 \begin{align}
  \label{eq: augmented proxy SVAR}
     	\begin{bmatrix}
     	\bm y_t \\    z_t
     	\end{bmatrix}  
     	=  	\begin{bmatrix}
     	\bm \nu  \\   \bm \nu_z
     	\end{bmatrix}
     	+ \sum_{i=1}^{p} \begin{bmatrix}
     	\bm A_i & 0 \\ \bm \Gamma_{1} & \bm \Gamma_{2}
     	\end{bmatrix} 	\begin{bmatrix}
     	\bm y_{t-i} \\    z_{t-i}
     	\end{bmatrix}   
     	+ \begin{bmatrix}
     	\bm B_0 & 0 \\ \bm \Phi & \bm \Sigma_{\eta}
     	\end{bmatrix}  \begin{bmatrix}
     	\bm \varepsilon_{t}  \\    \eta_t
     	\end{bmatrix}  ,
 \end{align} 
 with a measurement error $   \eta_t$ uncorrelated with the structural form shocks $\bm \varepsilon_{t}$. To simplify, let $\bm \nu_z=0$, $\bm \Sigma_{\eta}=1$, and $\bm \Gamma_{1} = \bm \Gamma_{2}=0$, which implies that the proxy is equal to a linear combination of structural shocks and measurement error, i.e., $ z_t =   \bm \Phi 	\bm\varepsilon_{ t} +  \eta_{t}$.   Imposing a valid proxy implies zero restrictions on the $\bm \Phi$ matrix, such that the proxy variable is a linear combination of the target-shock and the measurement error, i.e.  $z_t =   \Phi_i	\varepsilon_{it} +  \eta_{t}$.

%For Bayesian proxy SVARs, \cite{caldara2019monetary} propose to write the joint likelihood of the data $\bm y$ and the proxy $\bm z$ as
%\begin{equation}
 % p(\bm y, \bm z |\bm B, \bm \pi)=   p(\bm y| \bm B, \bm \pi)  p( \bm z | \bm y ,  \bm B, \bm \pi).
  %\label{eq:linearproxy}
%\end{equation} 
%The first term,   $p(\bm y| \bm B, \bm \pi)$,  denotes the Gaussian likelihood of the VAR data $\bm y$. 
%The second term, $p( \bm z | \bm y ,  \bm B, \bm \pi)$, is the conditional likelihood of the proxy given the data. The conditional likelihood is derived from the augmented proxy equation and the distribution of the proxy noise term $\eta_{t}$, see \cite{caldara2019monetary}. 
%Intuitively, the second term re-weights the likelihood of the VAR data $\bm y$ by putting more weight to the structural parameters that result in target shocks equal to a scaled version of the proxy.

The augmented proxy SVAR requires to specify the data generating process of the proxy.
Specifically, the proxy is defined as a linear function of the target shock and a measurement error. However,  many proxy variables may not follow a linear process, like for instance the tax proxy in our empirical analysis, which may rather follow a process like $ 
    z_t = \psi_t (  \Phi_i	\varepsilon_{it} +    \eta_{it}),
 $ 
where $\psi_t$ is a Bernoulli random variable, compare \cite{jentsch2019dynamic},   \cite{bruns2022alternative}, or \cite{budnikidentifying}. If the proxy process is misspecified, it can lead to dependent shocks, which renders identification approaches based on independent shocks invalid.

\begin{proposition}
   \label{prop augmented miss} 
   For a non-linear proxy variable generated by the process  
   $ z_t = \psi_t (  \tilde{\Phi}_i	\varepsilon_{it} +    \tilde{\eta}_{t}) $ where $\psi_t$ is a Bernoulli random variable,
   the misspecified linear augmented proxy SVAR with the linear proxy specification 
   $ z_t =  \Phi_i	\varepsilon_{it} +    \eta_{t} $
   leads to a measurement error 
   \begin{align}
      \eta_t = \begin{cases}
     (\tilde{\Phi}_i - \Phi_i)	\varepsilon_{it} +    \tilde{\eta}_{t} &,    \text{ if }  \psi_t=1
    \\
     -  \Phi_i	\varepsilon_{it} &, \text{ else }
    \end{cases}.
    \end{align}
    
\end{proposition}
\begin{proof}
    In the linear augmented proxy model, the noise term is equal to $ \eta_{t} = z_t - \Phi_i	\varepsilon_{it}$. Plugging in the data generating process of the proxy yields $ \eta_{t} = \psi_t (  \tilde{\Phi}_i	\varepsilon_{it} +    \tilde{\eta}_{t}) - \Phi_i	\varepsilon_{it}$.
\end{proof} 

In the misspecified augmented proxy SVAR,  the measurement error $\eta_t$ is a function of the target shock $\varepsilon_{it}$.  
Consequently, measurement error and structural shocks are not independent in the misspecified augmented proxy SVAR. Therefore, estimators based on independent shocks should not be used to estimate augmented proxy SVARs with a misspecified proxy process. 
Of course, a correctly specified non-linear augmented proxy SVAR is a possible solution. Nevertheless, it again requires to correctly specify the non-linear proxy process and misspecifications again leads to similar problems.

One might argue that imposing Equations (\ref{eq: Dz dist}) and (\ref{eq: Dz dist 2}) amounts to specifying the proxy’s DGP, similar to the augmented proxy SVAR. However, the distribution of $D(z)$ does not depend on any distributional assumption about $z$; it follows asymptotically from the central limit theorem. In other words, regardless of the DGP of $z$, as long as the central limit theorem’s conditions hold, the moment conditions $D(z)$ will be asymptotically normal, thereby justifying Equations (\ref{eq: Dz dist}) and (\ref{eq: Dz dist 2}).

The consequences of misspecification in the augmented non-Gaussian proxy SVAR can be observed in the following Monte Carlo simulation. We generate data using a non-Gaussian SVAR with three variables and one exogenous proxy
  \begin{align}
  \label{eq: MC one proxy}
  \begin{bmatrix}
  u_{g, t} \\
  u_{y, t} \\
   u_{\tau, t} \\ 
  \end{bmatrix} &=
  \begin{bmatrix}
   1 & 0  & 0        \\
    0.15 & 1 &  -0.5    \\
   0 & 1.5 & 1    
  \end{bmatrix}
  \begin{bmatrix}
  \varepsilon_{g, t} \\
  \varepsilon_{y, t} \\
  \varepsilon_{\tau ,t} \\
  \end{bmatrix} . 
  \end{align}   
In the first setup, the proxy is generated by a linear DGP, i.e. $z_t = \varepsilon_{\tau, t}  +	  \eta_t$. 
In the second simulation, the proxy is generated by a truncated linear DGP, i.e. $z^{new}_t = \psi_t z_t$ where $\psi_t$ is Bernoulli random variable such that on average $80$\% of the proxy observations are truncated to zero. The second case is relevant in our empirical work.
We estimate the augmented proxy SVAR (without any zero restrictions implied by a valid proxy) using the non-Gaussian likelihood from Section \ref{Table:MSE} . 
Table \ref{Table:non-Gaussian augmented}  shows the estimated effect of the target shock $\varepsilon_{\tau, t}$. We observe that proxy misspecification in the augmented proxy SVAR leads to a bias of the non-Gaussian SVAR estimator and large estimation error (compare with results in Table \ref{Table: Finite sample performance proxy process }).

\begin{table}
	\caption{Results for non-Gaussian augmented SVAR - mean and mse}
	\label{Table:non-Gaussian augmented} 
	\begin{tabular}{ c  | c    c          }
		&  proxy  1  &  proxy 2    \\  
			&$z_{\tau ,t}= \varepsilon_{\tau, t}  +	  \eta_t$
            &$z_{\tau ,t}= \psi_t (\varepsilon_{\tau, t}  +	 \eta_t) $ 
		\\ \hline 
		$T=800$&& 
  		 
		\\
		
non-Gaussian (augmented)  &
		$\begin{bmatrix} 
		\underset{(0.005)}{\hphantom{-}0.00}  &
		\underset{(0.005)}{-0.49} & 
		\underset{(0.011)}{\hphantom{-}0.99} 
		\end{bmatrix}'$
		
		&
		
		$\begin{bmatrix}           
		\underset{(0.049)}{-0.02} & 
		\underset{(0.069)}{-0.43} & 
		\underset{(0.180)}{\hphantom{-}0.88} 
		\end{bmatrix}'$

	\end{tabular}  
	\footnotesize{ $ $ \\ \textit{Note: The true impact of the shock $\varepsilon_{\tau, t}$ is $\begin{bmatrix}   
			0  & -0.5   & 1
			\end{bmatrix}'$. 
   The average and MSE of the Bayesian estimators are calculated based on the median of the posterior of B in each simulation.}}
\end{table}

   In summary, the augmented proxy SVAR requires to specify the DGP of the proxy and misspecification of the DGP can lead to dependencies and render a non-Gaussian estimation based on the independence assumption invalid. In contrast, a moment-based proxy estimation approach never specifies the DGP of the proxy and hence, cannot suffer from misspecification consequences.

\section{Main Monte Carlo Study} 
  \label{sec: Finite Sample Performance}

This section demonstrates the ability of our non-Gaussian proxy weighting approach in handling both exogenous and endogenous proxy variables.
Our model is able to use the information of the proxy, which leads to a better performance compared to a purely non-Gaussian model.
Furthermore, even when the proxy exhibits weak exogeneity, meaning it is only minimally affected by non-target shocks, our approach retains the ability to utilize the proxy to enhance efficiency, surpassing the performance of the non-Gaussian model alone.
Moreover, our approach can detect whether the data provide evidence against the exogeneity of the proxy and can neglect information from endogenous proxies.

We simulate a system containing a government spending shock $ \varepsilon_{g, t}$, an output shock $ \varepsilon_{y, t}$, and a tax shock $\varepsilon_{\tau, t}$ with
  \begin{align}
  \label{eq: MC one proxy}
  \begin{bmatrix}
  u_{g, t} \\
  u_{y, t} \\
   u_{\tau, t} \\ 
  \end{bmatrix} &=
  \begin{bmatrix}
   1 & 0  & 0        \\
    0.15 & 1 &  -0.5    \\
   0 & 1.5 & 1    
  \end{bmatrix}
  \begin{bmatrix}
  \varepsilon_{g, t} \\
  \varepsilon_{y, t} \\
  \varepsilon_{\tau ,t} \\
  \end{bmatrix} . 
  \end{align}  
  The structural shocks are drawn independently and identically from a Pearson distribution with mean zero, variance one, skewness $0.68$ and excess kurtosis $2.33$ and simulate $1000$ data sets of length $T=250$ and $T=800$.

  We construct a variable $z_{ t}$ as a proxy for the tax shock $\epsilon_{\tau, t}$.  To demonstrate the flexibility of our model, we consider three different scenarios summarized in Table \ref{Table: Finite sample performance proxy process }. In the first scenario, the proxy is exogenous, in the second scenario, the proxy is weakly endogenous, and in the third scenario the proxy is endogenous. In all simulations, the proxy noise $\eta_{t}$ is i.i.d. and drawn from the same distribution as the structural shocks.    

  \begin{table}
	\caption{ Data-generating process of the proxy in different scenarios.     }
	\label{Table: Finite sample performance proxy process } 
	\begin{tabular}{   c  |  c  | c       }
		  exogenous proxy    & weakly endogenous proxy   & endogenous  proxy  \\  \hline 
		 $z_{\tau, t}=	\varepsilon_{\tau, t} +\eta_t  $&$z_{\tau, t}= \varepsilon_{\tau, t}   -0.05 	\varepsilon_{y, t}+	\eta_t  $&$z_{\tau ,t}= \varepsilon_{\tau, t} -0.37 	\varepsilon_{y, t} +	\eta_t  $
	 
	\end{tabular}  
\end{table}

We consider four Bayesian estimation approaches.
\begin{enumerate} 
    \item \textit{Proxy (augmented)}:  We estimate the SVAR using the standard augmented proxy SVAR approach assuming Gaussian shocks and imposing exogeneity of the proxy.

    \item  \textit{Proxy (weighting)}: We estimate the SVAR assuming Gaussian shocks and impose the exogenous proxy moment conditions without the ability to update the proxy exogeneity assumption. Specifically, the model uses the exogenous proxy moment conditions 
    $D(\bm z)=[D_1(\bm z ), D_2(\bm z )]'$ with
        \begin{align*} 
        D_1(\bm z  )=   \frac{1}{\sqrt{T}} \sum_{t=1}^{T} z_{t} e_{g,t}(\bm B,\bm \pi)   
         \text{ and }
         D_2(\bm z  )= \frac{1}{\sqrt{T}} \sum_{t=1}^{T} z_{t} e_{y,t}(\bm B,\bm \pi).
     \end{align*} 

     \item \textit{Non-Gaussian}:  We estimate a  non-Gaussian SVAR  without the proxy variable.

     \item \textit{Non-Gaussian proxy weighting}:   We estimate   the  non-Gaussian SVAR  with the proxy moment conditions  $D(\bm z)=[D_1(\bm z, \mu_1 ), D_2(\bm z,\mu_2)]'$  and
    \begin{align*}
% \label{eq: propxy prior sim tau g}
         D_1(\bm z,  \mu_1 )=   \frac{1}{\sqrt{T}} \sum_{t=1}^{T} (z_{t} e_{g,t}(\bm B,\bm \pi)    -  \mu_1) 
        \text{ and }
     D_2(\bm z,  \mu_2 )= \frac{1}{\sqrt{T}} \sum_{t=1}^{T} (z_{t} e_{y,t}(\bm B,\bm \pi) - \mu_2)   
 \end{align*} 
 and the proxy exogeneity shrinkage prior.
     
\end{enumerate}  
 Both non-Gaussian models use the true impact matrix $\bm B_0$ from the DGP in Equation (\ref{eq: MC one proxy}) to label the shocks in the MCMC. Specifically, for each proposal $\bm B$ we calculate $\bm B_0^{-1} \bm B$    and accept the proposal if each diagonal element  of  $\bm B_0^{-1} \bm B$  is larger in absolute value than the upper-right elements of $\bm B_0^{-1} \bm B$ in the same row, see \cite{keweloh2023uncertain}.

Table \ref{Table:MSE} shows the average point estimates and the mean squared error (MSE) of the estimated impact of $\varepsilon_{\tau ,t}$. 
The non-Gaussian model does not use the proxy variable and its performance is not affected by the different scenarios.
Both Gaussian proxy approaches lead to very similar results and perform notably better than the non-Gaussian model if the proxy variable is exogenous, however, including an endogenous proxy leads to biased estimates. 
Combining statistical identification with proxy variables provides a balanced solution between these two extreme cases.
If the proxy variable is exogenous, adding the proxy to the non-Gaussian model leads to an increase of the performance of the model, i.e. the MSE is twice as small compared to the non-Gaussian model. Therefore, the model is able to utilize the information of a valid proxy.
However, in contrast to the Gaussian proxy models, our proposed combination approach is able to deal with endogenous proxy variables. 
If the proxy variable is only weakly endogenous, our non-Gaussian proxy weighting approach can still exploit the information of the proxy to deliver improved estimation accuracy in comparison to the non-Gaussian model. 
Moreover, if the proxy variable is endogenous,  our non-Gaussian proxy weighting approach is less biased compared to the pure proxy approaches, and with more data and more evidence against the validity of the proxy variable, the prior gets updated and the bias decreases.

\begin{table}
\begin{footnotesize}
	\caption{ Average point estimates and MSE for the impact of $\varepsilon_{\tau ,t}$.}
	\label{Table:MSE} 
	\begin{tabular}{ c  | c    c   c       }
		&  exogenous proxy  &  weakly endogenous proxy  &   endogenous proxy  \\  
		&$z_{\tau ,t}= \varepsilon_{\tau, t}  +	\eta_t$&
  $z_{\tau ,t}= \varepsilon_{\tau, t} -0.10 	\varepsilon_{y, t} +	\eta_t$&
  $z_{\tau ,t}= \varepsilon_{\tau, t} -0.37 	\varepsilon_{y, t} +	\eta_t$
		\\ \hline
		$T=250$&&&
		\\
proxy (augmented)  &
		$\begin{bmatrix} 
		\underset{(0.008)}{\hphantom{-}0.00}  &
		\underset{(0.009)}{-0.50} & 
		\underset{(0.022)}{\hphantom{-}1.00} 
		\end{bmatrix}'$
		&
		
		$\begin{bmatrix}
		\underset{(0.008)}{\hphantom{-}0.00}  &
		\underset{(0.018)}{-0.59} & 
		\underset{(0.046)}{\hphantom{-}0.86} 
		\end{bmatrix}'$
		&
		
		$\begin{bmatrix}  
		\underset{(0.008)}{\hphantom{-}0.00}  &
		\underset{(0.106)}{-0.81} & 
		\underset{(0.363)}{\hphantom{-}0.42} 
		\end{bmatrix}'$

		\\
proxy (weighting)  &
				$\begin{bmatrix} 
		\underset{(0.008)}{\hphantom{-}0.00}  &
		\underset{(0.009)}{-0.50} & 
		\underset{(0.022)}{\hphantom{-}1.00} 
		\end{bmatrix}'$
		&
		
		$\begin{bmatrix} 
		\underset{(0.008)}{\hphantom{-}0.00}  &
		\underset{(0.017)}{-0.59} & 
		\underset{(0.046)}{\hphantom{-}0.85} 
		\end{bmatrix}'$
		&
		
		$\begin{bmatrix}  
		\underset{(0.008)}{\hphantom{-}0.00}  &
		\underset{(0.106)}{-0.81} & 
		\underset{(0.372)}{\hphantom{-}0.42} 
		\end{bmatrix}'$

		\\    
non-Gaussian  &
				$\begin{bmatrix} 
		\underset{(0.017)}{\hphantom{-}0.00}  &
		\underset{(0.021)}{-0.49} & 
		\underset{(0.051)}{\hphantom{-}0.95} 
		\end{bmatrix}'$
		&
		
		$\begin{bmatrix}
		\underset{(0.018)}{\hphantom{-}0.00}  &
		\underset{(0.019)}{-0.48} & 
		\underset{(0.045)}{\hphantom{-}0.95} 
		\end{bmatrix}'$
		&
		
		$\begin{bmatrix}  
				\underset{(0.018)}{\hphantom{-}0.00}  &
		\underset{(0.019)}{-0.48} & 
		\underset{(0.046)}{\hphantom{-}0.95} 
		\end{bmatrix}'$

		\\   
\begin{tabular}{@{}c@{}}non-Gaussian  \\proxy weighting\end{tabular}    & 
				$\begin{bmatrix} 
		\underset{(0.008)}{\hphantom{-}0.00}  &
		\underset{(0.009)}{-0.50} & 
		\underset{(0.022)}{\hphantom{-}0.99} 
		\end{bmatrix}'$
		&
		
		$\begin{bmatrix}
		\underset{(0.008)}{\hphantom{-}0.00}  &
		\underset{(0.011)}{-0.54} & 
		\underset{(0.028)}{\hphantom{-}0.93} 
		\end{bmatrix}'$
		&
		
		$\begin{bmatrix}  
				\underset{(0.009)}{\hphantom{-}0.00}  &
		\underset{(0.026)}{-0.59} & 
		\underset{(0.081)}{\hphantom{-}0.82} 
		\end{bmatrix}'$

			\\ \hline
		$T=800$&&&
		\\
proxy (augmented)  &
			$\begin{bmatrix} 
		\underset{(0.002)}{\hphantom{-}0.00}  &
		\underset{(0.003)}{-0.50} & 
		\underset{(0.007)}{\hphantom{-}1.00} 
		\end{bmatrix}'$
		&
		
		$\begin{bmatrix}
		\underset{(0.002)}{\hphantom{-}0.00}  &
		\underset{(0.012)}{-0.60} & 
		\underset{(0.032)}{\hphantom{-}0.84} 
		\end{bmatrix}'$
		&
		
		$\begin{bmatrix}  
		\underset{(0.002)}{\hphantom{-}0.00}  &
		\underset{(0.103)}{-0.82} & 
		\underset{(0.355)}{\hphantom{-}0.41} 
		\end{bmatrix}'$

		\\
proxy (weighting)  &
				$\begin{bmatrix} 
		\underset{(0.002)}{\hphantom{-}0.00}  &
		\underset{(0.003)}{-0.50} & 
		\underset{(0.007)}{\hphantom{-}1.00} 
		\end{bmatrix}'$
		&
		
		$\begin{bmatrix}
		\underset{(0.002)}{\hphantom{-}0.00}  &
		\underset{(0.012)}{-0.60} & 
		\underset{(0.031)}{\hphantom{-}0.85} 
		\end{bmatrix}'$
		&
		
		$\begin{bmatrix}  
		\underset{(0.002)}{\hphantom{-}0.00}  &
		\underset{(0.102)}{-0.82} & 
		\underset{(0.353)}{\hphantom{-}0.41} 
		\end{bmatrix}'$

		\\    
non-Gaussian  &
				$\begin{bmatrix} 
		\underset{(0.005)}{\hphantom{-}0.00}  &
		\underset{(0.005)}{-0.49} & 
		\underset{(0.012)}{\hphantom{-}0.99} 
		\end{bmatrix}'$
		&
		
		$\begin{bmatrix}
		\underset{(0.005)}{\hphantom{-}0.00}  &
		\underset{(0.005)}{-0.49} & 
		\underset{(0.012)}{\hphantom{-}0.99} 
		\end{bmatrix}'$
		&
		
		$\begin{bmatrix}  
		\underset{(0.005)}{\hphantom{-}0.00}  &
		\underset{(0.005)}{-0.49} & 
		\underset{(0.012)}{\hphantom{-}0.99} 
		\end{bmatrix}'$

		\\   
\begin{tabular}{@{}c@{}}non-Gaussian  \\proxy weighting\end{tabular}    & 
				$\begin{bmatrix} 
		\underset{(0.002)}{\hphantom{-}0.00}  &
		\underset{(0.003)}{-0.50} & 
		\underset{(0.006)}{\hphantom{-}1.00} 
		\end{bmatrix}'$
		&
		
		$\begin{bmatrix}
		\underset{(0.002)}{\hphantom{-}0.00}  &
		\underset{(0.005)}{-0.53} & 
		\underset{(0.012)}{\hphantom{-}0.94} 
		\end{bmatrix}'$
		&
		
		$\begin{bmatrix}  
		\underset{(0.003)}{\hphantom{-}0.00}  &
		\underset{(0.006)}{-0.52} & 
		\underset{(0.017)}{\hphantom{-}0.96} 
		\end{bmatrix}'$

	\end{tabular} 
	\footnotesize{\textit{Note: The true impact of the shock $\varepsilon_{\tau, t}$ is $\begin{bmatrix}   
			0  & -0.5   & 1
			\end{bmatrix}'$. 
   The average and MSE of the Bayesian models are calculated based on the median of the posterior of B in each simulation.}}
   \end{footnotesize}
\end{table}

Next, we turn our attention to the finite sample properties of the $68$\% credible bands of the models.
Table \ref{Table:coverage} shows the coverage rate (defined as the proportion in which the credible bands contain the true value) and the average length of the credible bands. 
The non-Gaussian model ignores the proxy variable, performs similarly throughout the three specifications, and has correct coverage rates (the coverage rate is close to the probability chosen for the credible bands). If the prior belief of an exogenous proxy is correct, adding the proxy also leads to correct coverage and more informative credible bands in the sense that the bands are up to $50$\% smaller compared to the non-Gaussian model ignoring the proxy variable. 
Adding an endogenous proxy using to the non-Gaussian model via our weighting approach worsens the coverage rates. However, with an increasing sample size and more information against the prior belief of an exogenous proxy, the coverage rate improves and the difference of the error bands length between the model with and without prior vanishes. Importantly, adding a weakly endogenous proxy also helps in a small sample size to lower the lengths of the credible bands without distorting the coverage much. In the Online Appendix we present various additional Monte Carlo experiments to further demonstrate the flexibility of our approach.

\begin{table}
\begin{footnotesize}
	\caption{ Coverage  and average length of  $68$\% credible bands of the estimated impact of $\varepsilon_{\tau ,t}$.}
	\label{Table:coverage} 
		\begin{tabular}{ c  | c    c   c       }
		&  exogenous proxy  &  weakly endogenous proxy  &   endogenous proxy  \\  
		&$z_{\tau ,t}= \varepsilon_{\tau, t}  +	\eta_t$&
  $z_{\tau ,t}= \varepsilon_{\tau, t} -0.10 	\varepsilon_{y, t} +	\eta_t$&
  $z_{\tau ,t}= \varepsilon_{\tau, t} -0.37 	\varepsilon_{y, t} +	\eta_t$
		\\ \hline
		$T=250$&&&
		\\
proxy (augmented)  &
		$\begin{bmatrix} 
		\underset{(0.018)}{0.68} & 
		\underset{(0.019)}{0.68} & 
		\underset{(0.028)}{0.69} 
		\end{bmatrix}'$
		&
		
		$\begin{bmatrix}
		\underset{(0.018)}{0.67} & 
		\underset{(0.018)}{0.46} & 
		\underset{(0.030)}{0.48} 
		\end{bmatrix}'$
		&
		
		$\begin{bmatrix}  
		\underset{(0.017)}{0.68} & 
		\underset{(0.016)}{0.00} & 
		\underset{(0.031)}{0.01} 
		\end{bmatrix}'$

		\\
proxy (weighting)  &
				$\begin{bmatrix} 
		\underset{(0.018)}{0.68} & 
		\underset{(0.019)}{0.67} & 
		\underset{(0.029)}{0.70} 
		\end{bmatrix}'$
		&
		
		$\begin{bmatrix} 
		\underset{(0.018)}{0.68} & 
		\underset{(0.018)}{0.47} & 
		\underset{(0.030)}{0.48} 
		\end{bmatrix}'$
		&
		
		$\begin{bmatrix}  
		\underset{(0.017)}{0.56} & 
		\underset{(0.015)}{0.06} & 
		\underset{(0.030)}{0.07} 
		\end{bmatrix}'$

		\\    
non-Gaussian  &
				$\begin{bmatrix} 
		\underset{(0.026)}{0.73} & 
		\underset{(0.027)}{0.68} & 
		\underset{(0.041)}{0.69} 
		\end{bmatrix}'$
		&
		
		$\begin{bmatrix} 
		\underset{(0.026)}{0.70} & 
		\underset{(0.027)}{0.70} & 
		\underset{(0.040)}{0.71} 
		\end{bmatrix}'$
		&
		
		$\begin{bmatrix}  
		\underset{(0.026)}{0.69} & 
		\underset{(0.027)}{0.69} & 
		\underset{(0.040)}{0.70}  
		\end{bmatrix}'$

		\\   
\begin{tabular}{@{}c@{}}non-Gaussian  \\proxy weighting\end{tabular}    & 
				$\begin{bmatrix} 
		\underset{(0.017)}{0.70} & 
		\underset{(0.018)}{0.68} & 
		\underset{(0.028)}{0.70}  
		\end{bmatrix}'$
		&
		
		$\begin{bmatrix}
		\underset{(0.017)}{0.70} & 
		\underset{(0.018)}{0.64} & 
		\underset{(0.029)}{0.64} 
		\end{bmatrix}'$
		&
		
		$\begin{bmatrix}  
		\underset{(0.018)}{0.71} & 
		\underset{(0.024)}{0.52} & 
		\underset{(0.040)}{0.53} 
		\end{bmatrix}'$

			\\ \hline
		$T=800$&&&
		\\
proxy (augmented)  &
			$\begin{bmatrix} 
		\underset{(0.010)}{0.70} & 
		\underset{(0.010)}{0.65} & 
		\underset{(0.016)}{0.66} 
		\end{bmatrix}'$
		&
		
		$\begin{bmatrix}
		\underset{(0.010)}{0.70} & 
		\underset{(0.010)}{0.18} & 
		\underset{(0.016)}{0.17} 
		\end{bmatrix}'$
		&
		
		$\begin{bmatrix}  
		\underset{(0.010)}{0.71} & 
		\underset{(0.009)}{0.00} & 
		\underset{(0.017)}{0.00} 
		\end{bmatrix}'$

		\\
proxy (weighting)  &
				$\begin{bmatrix} 
		\underset{(0.010)}{0.70} & 
		\underset{(0.010)}{0.65} & 
		\underset{(0.016)}{0.66} 
		\end{bmatrix}'$
		&
		
		$\begin{bmatrix}
		\underset{(0.010)}{0.70} & 
		\underset{(0.010)}{0.19} & 
		\underset{(0.016)}{0.17} 
		\end{bmatrix}'$
		&
		
		$\begin{bmatrix}  
		\underset{(0.018)}{0.67} & 
		\underset{(0.016)}{0.01} & 
		\underset{(0.031)}{0.01} 
		\end{bmatrix}'$

		\\    
non-Gaussian  &
				$\begin{bmatrix} 
		\underset{(0.013)}{0.66} & 
		\underset{(0.013)}{0.66} & 
		\underset{(0.020)}{0.66} 
		\end{bmatrix}'$
		&
		
		$\begin{bmatrix}
		\underset{(0.013)}{0.67} & 
		\underset{(0.013)}{0.66} & 
		\underset{(0.020)}{0.66} 
		\end{bmatrix}'$
		&
		
		$\begin{bmatrix}  
		\underset{(0.013)}{0.67} & 
		\underset{(0.013)}{0.66} & 
		\underset{(0.020)}{0.66} 
		\end{bmatrix}'$

		\\   
\begin{tabular}{@{}c@{}}non-Gaussian  \\proxy weighting\end{tabular}    & 
				$\begin{bmatrix} 
		\underset{(0.009)}{0.68} & 
		\underset{(0.010)}{0.68} & 
		\underset{(0.015)}{0.66} 
		\end{bmatrix}'$
		&
		
		$\begin{bmatrix}
		\underset{(0.009)}{0.68} & 
		\underset{(0.010)}{0.54} & 
		\underset{(0.017)}{0.54} 
		\end{bmatrix}'$
		&
		
		$\begin{bmatrix}  
		\underset{(0.010)}{0.70} & 
		\underset{(0.014)}{0.64} & 
		\underset{(0.021)}{0.63} 
		\end{bmatrix}'$

	\end{tabular} 
   \end{footnotesize}
\end{table}

\section{Estimating fiscal multipliers}
\label{sec: Application} 
This section applies our proposed non-Gaussian proxy weighting approach to estimate the effects of exogenous changes in tax revenues and government spending. First, we describe the data and show that the time series feature a sizable degree of non-Gaussianity. Then our main findings are discussed. We find a larger government spending than tax multiplier. We provide evidence indicating that the fiscal and non-fiscal proxies used by \citet{mertens2014reconciliation} and \citet{caldara2017analytics}, respectively, do not fulfill the crucial exogeneity assumption, which biases the results of Gaussian proxy approaches.

\subsection{Data and specification}
To achieve comparability, we use the same trivariate VAR as adopted by \citet{mertens2014reconciliation}.\footnote{\citet{caldara2017analytics} consider a slightly larger VAR consisting of five endogenous variables as baseline model but also report results for the three variables specification we rely on. We show below that our main findings are robust to extending the baseline trivariate VAR by inflation and the interest rate as done by \citet{caldara2017analytics}.} The three endogenous variables are federal tax revenues $\tau_{t}$, federal government consumption and investment expenditures $g_{t}$, and output $y_{t}$, all in log real per capita terms and for the sample 1950Q2 to 2006Q4.\footnote{The data are downloaded from Karel Mertens' website.} The VAR has four lags and includes a constant, linear, and quadratic trends, and a dummy for 1975Q2 all contained in $\bm X_t$. The SVAR is given by 
\begin{align}
\label{eq: SVAR application}
    \begin{bmatrix}
\tau_t   
\\
g_t
\\
y_t
\end{bmatrix}
=
\bm\gamma \bm X_t
+ 
\sum_{i=1}^{4}
\bm A_i
\begin{bmatrix}
\tau_{t-i}   
\\
g_{t-i}
\\
y_{t-i}
\end{bmatrix}
+ 
\begin{bmatrix}
u_{\tau,t   }
\\
u_{g,t   }
\\
u_{y,t   }
\end{bmatrix}
\text{ and }
\begin{bmatrix}
u_{\tau,t   }
\\
u_{g,t   }
\\
u_{y,t   }
\end{bmatrix}=
\begin{bmatrix}
b_{11} & b_{12} & b_{13}  \\
b_{21} & b_{22} & b_{23}  \\
b_{31} & b_{32} & b_{33}  \\ 
\end{bmatrix}
\begin{bmatrix}
\varepsilon_{\tau,t}    
\\
\varepsilon_{g,t} 
\\
\varepsilon_{y,t} 
\end{bmatrix},
\end{align}   
with tax shocks $\varepsilon_{\tau,t}    $, government spending shocks $\varepsilon_{g,t} $, and output shocks $\varepsilon_{y,t}$. For comparison, we map the interaction into the notation  used by \cite{mertens2014reconciliation} 
\begin{align}
    u_{\tau,t   } &= \Theta_G \sigma_G \varepsilon_{g,t}     
                    + \Theta_Y u_{y,t   } + \sigma_{\tau} \varepsilon_{\tau,t}    \\
    u_{g,t   } &= \gamma_{\tau} \sigma_{\tau} \varepsilon_{\tau,t}     
                    + \gamma_Y u_{y,t   } + \sigma_{G} \varepsilon_{g,t}  \\
    u_{y,t   } &=  \eta_{\tau}  u_{\tau,t   }    
                    + \eta_G u_{g,t   } + \sigma_{Y} \varepsilon_{y,t} .  
\end{align}

 \citet{mertens2014reconciliation}  use a tax proxy $z_{\tau,t}$ and a zero restriction imposing that government spending does not respond contemporaneously to changes in economic activity.  
The tax proxy relies on a series of possibly unanticipated tax shocks, a subset of the \citet{Romer2010} tax shocks identified by studying narrative records of tax policy decisions. As argued, the tax proxy measures changes in the tax system that are not related to the state of the economy and thereby it should offer a valid proxy for tax shocks. Therefore, \citet{mertens2014reconciliation} impose the following exogeneity assumptions
\begin{align}
    E[z_{\tau,t} \varepsilon_{g,t} ] = 0
    \quad
    \text{ and }
    \quad
    E[z_{\tau,t} \varepsilon_{y,t} ] = 0.
\end{align}

 Instead of using a fiscal proxy, \citet{caldara2017analytics} rely on the non-fiscal \citet{Fernald2012} TFP measure as a proxy $z_{y,t}$ for output shocks and additionally assume that government spending does not respond contemporaneously to tax shocks.  The \citet{Fernald2012} technology series measures total factor productivity adjusted for changes in factor utilization. \citet{Fernald2012} carefully eliminates these sources of endogenous movements such that his resulting purified TFP series can be understood as solely reflecting exogenous technology variations which motivates the two exogeneity assumptions 
\begin{align}
    E[z_{y,t} \varepsilon_{\tau,t} ] = 0
    \quad
    \text{ and }
    \quad
    E[z_{y,t} \varepsilon_{g,t} ] = 0 
\end{align}
used by \citet{caldara2017analytics}. In what follows, we rely on the original TFP measure used by \citet{caldara2017analytics}.

Both approaches start with carefully motivated identifying assumptions, yet both approaches lead to different conclusions regarding the effects of tax and spending shocks. In particular, the fiscal proxy leads to a large tax multiplier while the non-fiscal proxy leads to a large spending multiplier. This difference indicates that at least one of the identifying proxy exogeneity assumptions is invalid.\footnote{
In principal, the differences could be driven by the different zero restrictions. However, computing the tax shocks based on the tax proxy and the output shocks based on the TFP proxy leads to correlated shocks. This correlation is not affected by the zero restrictions and indicates that the different results are indeed driven by invalid proxies.} 
However, both approaches rely on the exogeneity assumptions to identify the SVAR and hence, cannot detect endogenous proxies. Our model fills this gap since it allows to evaluate the empirical support for the exogeneity assumptions and thus helps in understanding diverging findings regarding the size of fiscal multipliers between the fiscal and non-fiscal approach. 

Besides the tax and TFP proxies, we use the growth rate of military spending per head of population as a proxy $z_{g,t}$ for government spending shocks.  \cite{Hall2009,Barro2011,Miyamoto2019}, amongst others, use military spending to identify exogenous government spending shocks. Changes in military spending are often large and regularly respond to foreign policy developments, suggesting that these changes are exogenous in the sense that they are less likely to be driven by domestic cyclical forces.  

We estimate the SVAR in Equation (\ref{eq: SVAR application}) using the non-Gaussian proxy weighting approach without any zero restrictions on the response of government spending. We normalize all proxies to mean zero and unit variance and use the  proxy weighting moment conditions   
\begin{align*}
D_1(\bm z_{\tau},\mu_{\tau g} )=   \frac{1}{\sqrt{T}} \sum_{t=1}^{T} (z_{\tau,t} e_{g,t}(\bm B,\bm \pi)    -  \mu_{\tau g}) 
&\text{, } \quad
D_2(\bm z_{\tau}, \mu_{\tau y} )=   \frac{1}{\sqrt{T}} \sum_{t=1}^{T} (z_{\tau,t} e_{y,t}(\bm B,\bm \pi)    -  \mu_{\tau y}),
\\
D_3(\bm z_{y}, \mu_{y \tau} )=   \frac{1}{\sqrt{T}} \sum_{t=1}^{T} (z_{y,t}  e_{\tau,t}(\bm B,\bm \pi)    -  \mu_{y \tau})
&\text{, } \quad
D_4(\bm z_{y},\mu_{y g} )=   \frac{1}{\sqrt{T}} \sum_{t=1}^{T} (z_{y,t}  e_{g,t}(\bm B,\bm \pi)    - \mu_{y g}),
\\
D_5(\bm z_{g}, \mu_{g \tau} )=   \frac{1}{\sqrt{T}} \sum_{t=1}^{T} (z_{g,t}  e_{\tau,t}(\bm B,\bm \pi)    -  \mu_{y \tau})
&\text{, } \quad
D_6(\bm z_{g},\mu_{g y} )=   \frac{1}{\sqrt{T}} \sum_{t=1}^{T} (z_{g,t}  e_{y,t}(\bm B,\bm \pi)    - \mu_{y g}),
 \end{align*}
 with $D(z)=[D_1(\bm z_{\tau}, \mu_{\tau g} ), D_2(\bm z_{\tau}, \mu_{\tau y}),D_3(\bm z_{y},\mu_{y \tau} ),D_4(\bm z_{y},\mu_{y g}), D_5(\bm z_{g}, \mu_{g \tau} ),D_6(\bm z_{g},\mu_{g y} ) ]'$ 
 and the proxy exogeneity shrinkage prior from Equation (\ref{eq: prior mu}).
Therefore, we start with the prior that all proxies are valid and shrink towards exogenous proxies. However, the data can update the prior and deviate from a given exogeneity assumption. Moreover, we use the proxy variables to label the shocks similar to \cite{bertsche2022identification}. Specifically, the shock exhibiting the highest correlation in absolute terms with the TFP proxy is labeled as the output shock. Among the two remaining shocks, the one displaying the strongest correlation in absolute terms with the tax proxy is identified as the tax shock, while the remaining shock is labeled as the government spending shock.\footnote{An alternative labeling approach relying on a first step estimator is shown in the Appendix and leads to similar results.}

For comparison, we also estimate the \cite{mertens2014reconciliation} fiscal proxy SVAR with the Gaussian proxy weighting approach, the government spending restriction $b_{23}=0$, and the tax proxy weighting moment conditions $D_1(\bm z_{\tau},\mu_{\tau g} )$ and $D_2(\bm z_{\tau}, \mu_{\tau y})$  with $\mu_{\tau g}=\mu_{\tau y}=0$   and    the  \cite{caldara2017analytics} non-fiscal proxy SVAR with the Gaussian proxy-weighting approach, the government spending restriction $b_{13}=0$, and the TFP proxy weighting moment conditions $D_3(\bm z_{y},\mu_{y \tau} )$and $D_4(\bm z_{y},\mu_{y g})$  with $\mu_{y \tau}=\mu_{y g}=0$. Therefore, both Gaussian models impose proxy exogeneity without the ability to deviate.\footnote{Note that the two Gaussian proxy weighting models yield impulse responses akin to those acquired through the conventional moment-based frequentist proxy estimator, as shown in the Appendix.}

\begin{figure}[h]
	\centering
	\includegraphics[width=0.99\textwidth]{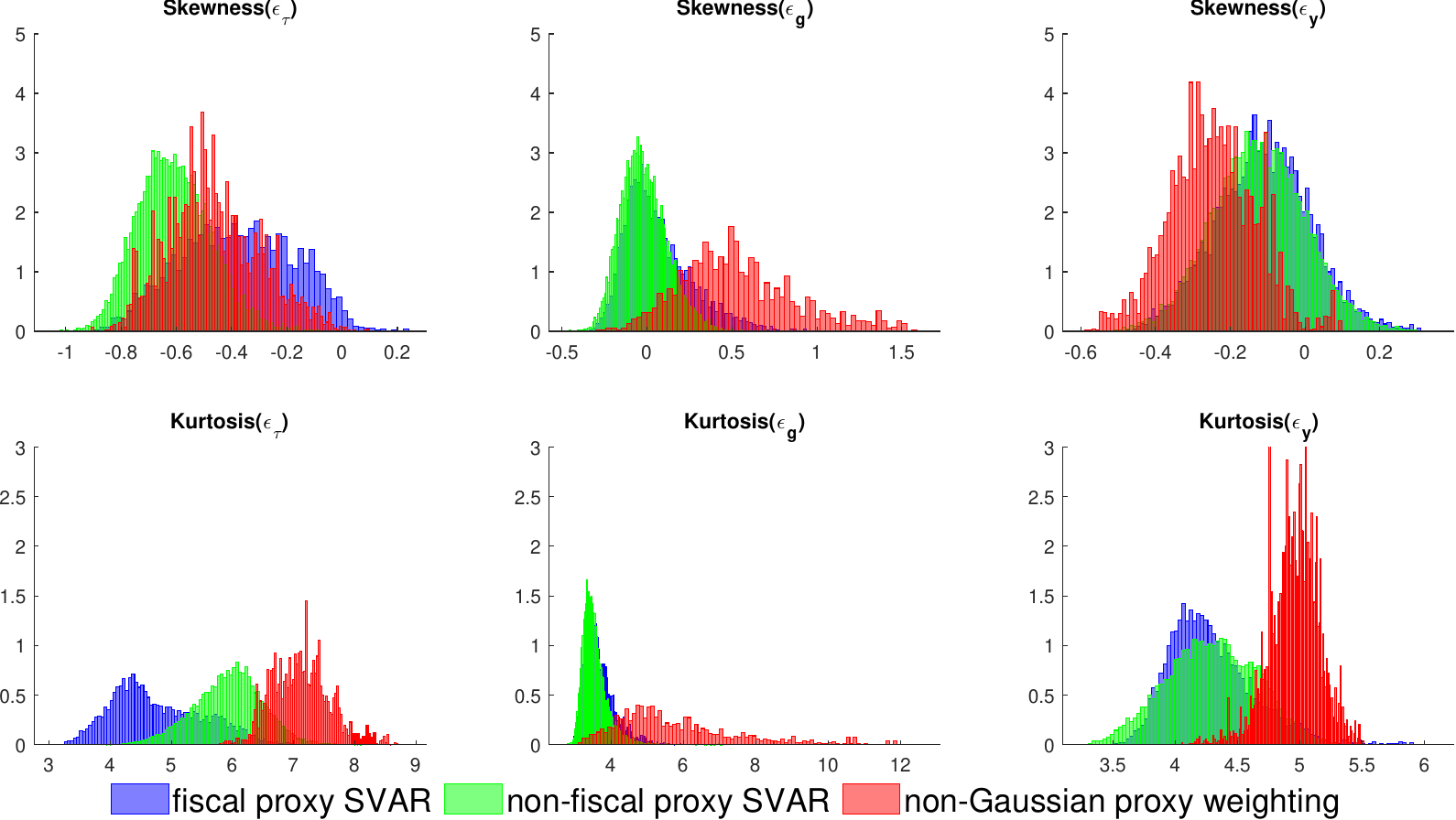} 
 \caption{Evidence for non-Gaussianity } 	\label{non-gaussianit}
 \floatfoot{The figure shows the posterior distributions of the skewness as well as the kurtosis of the structural shocks. We show results for our non-Gaussian proxy weighting SVAR, the non-fiscal proxy SVAR proposed by \citet{caldara2017analytics} as well as the fiscal proxy SVAR from \cite{mertens2014reconciliation}. }
\end{figure}

A requirement for updating proxy exogeneity by the data is that we work with non-Gaussian structural shocks. Figure \ref{non-gaussianit} shows the posterior of the skewness and kurtosis of the estimated structural shocks in the non-Gaussian proxy weighting SVAR and the two Gaussian proxy weighting SVARs. All three models show a sizeable degree of non-Gaussianity in the estimated structural shocks. In particular, the skewness of the tax and output shock is centered around non-zero values and the kurtosis shows positive values above three. Across all models, the tax and output shocks are left skewed, which is economically reasonable because tax cuts tend to be larger than tax hikes and output falls stronger during recessions than it rises during expansions. In addition, there is some evidence that the government spending shock is right skewed, indicating that spending stimuli are larger than spending consolidations.

 \begin{figure}[h]
	\centering
	\includegraphics[width=0.95\textwidth]{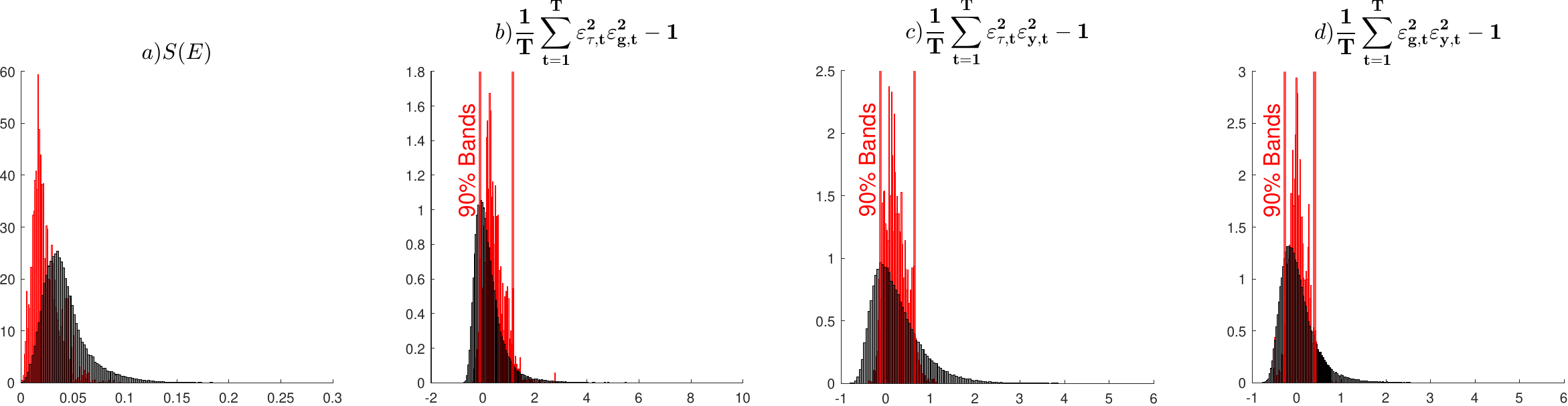} 
 \caption{Evidence for independent shocks } 	\label{figure-independence}
 \floatfoot{The figure shows in red the posterior of the test statistic proposed by \cite{montiel2022svar} and the posteriors of the cross-moments of squared shocks. The black posterior is the corresponding simulated posterior under the null hypothesis of independent shocks.}
\end{figure}

The second requirement for our non-Gaussian identification approach are independent structural shocks. The common critique to the independence assumption is a potentially shared volatility process, see \cite{montiel2022svar}.
Figure \ref{figure-independence} displays the posterior of the test statistic $S(E)=\sqrt{ \frac{1}{n(n-1)} (Corr(\varepsilon_{\tau,t}^2, \varepsilon_{g,t}^2)^2 +Corr(\varepsilon_{\tau,t}^2, \varepsilon_{y,t}^2)^2 +Corr(\varepsilon_{g,t}^2, \varepsilon_{y,t}^2)^2 )}$ proposed by \cite{montiel2022svar} to measure the common volatility of all shocks and the posterior of the three possible cross-moments of squared shocks together with the corresponding distributions approximated by bootstrap under the null of independent shocks by randomly permutating the shocks similar to the approach in \cite{braun2021importance}. 
In all cases, we find that the posterior with the posterior simulated under the null of independent shocks overlap to a large extend, suggesting no evidence against mutual independence.
Moreover, the $90$\% bands of the squared cross-moments contain zero, which is the value of the squared cross-moments $E[\varepsilon_{i,t}^2\varepsilon_{j,t}^2-1]$ of mutually independent shocks $\varepsilon_{i,t}$ and $\varepsilon_{j,t}$.

\begin{figure} 
	\centering
	\caption{Posterior of proxy relevancy moment conditions}\label{fig:relevance}
	\includegraphics[width=0.95\textwidth]{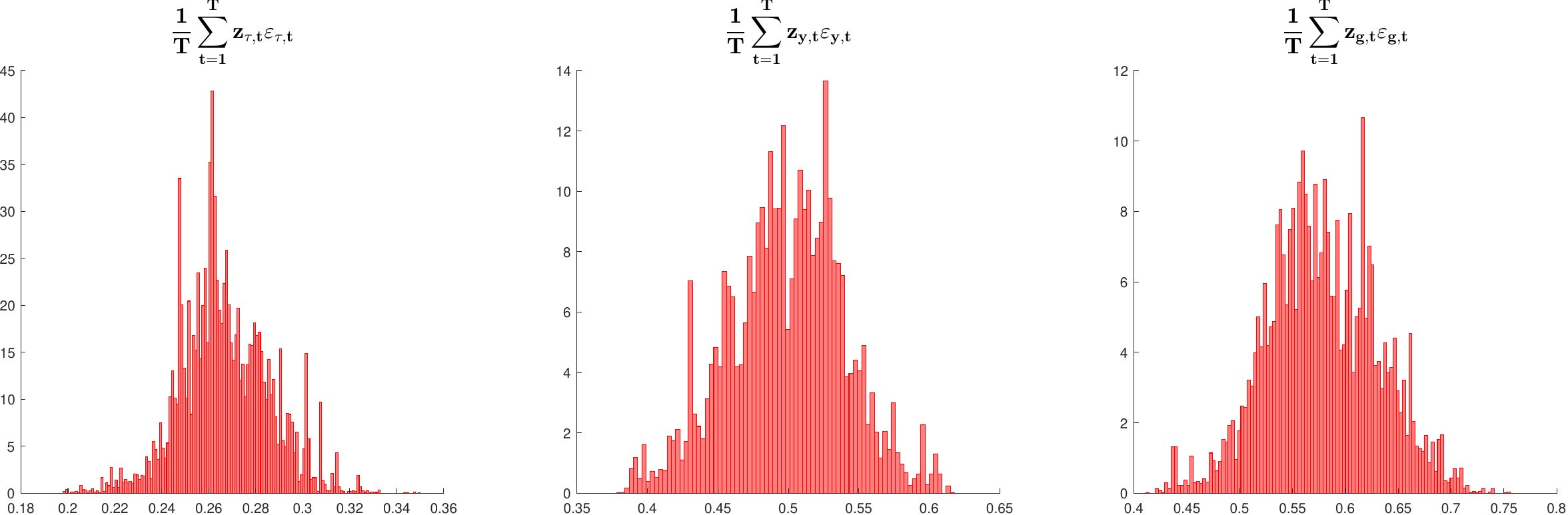}  
\end{figure}

Figure \ref{fig:relevance} plots results on the relevance of the proxies in our non-Gaussian proxy weighting model. The left column shows the posterior of the correlation between the tax proxy and tax shock, the middle column presents the posterior of the correlation between the TFP proxy and the output shock, and the right column shows the posterior of the correlation of the military spending proxy and the government spending shock. All distributions are centered around a positive mean with no support of values close to zero. Thus, we find that all proxies are strong in the sense that they satisfy the relevance condition.\footnote{Note that, we plot the correlation between the instrument and the structural shock, whereas \citet{mertens2014reconciliation} and \citet{caldara2017analytics} have to focus on the correlation between the instrument and the reduced form shock.}

\subsection{Baseline results}
\begin{figure}[t]
	\centering
	\caption{Comparison of estimated output multipliers between the different models}\label{fig:multipliers comp}
	\includegraphics[width=1\textwidth]{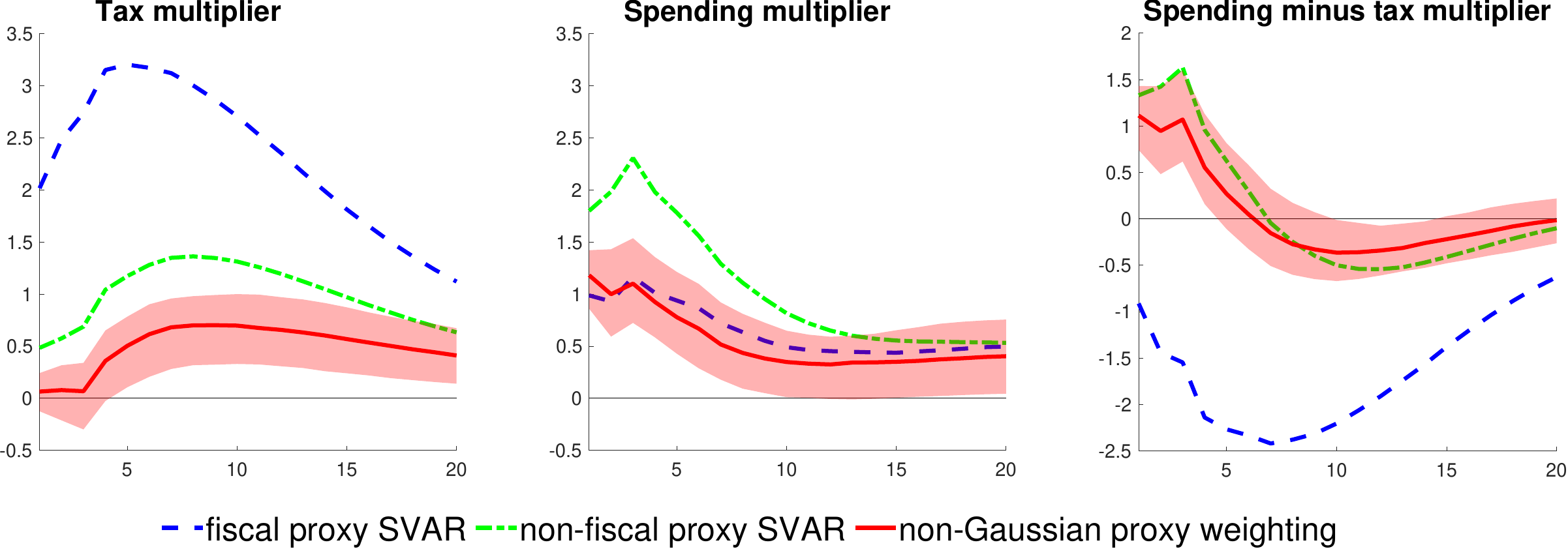} %comparisonMedian.pdf
 \floatfoot{The figure compares the output responses between our non-Gaussian proxy weighting SVAR with $68$\% credible bands to the median responses in the non-fiscal proxy SVAR proposed by \citet{caldara2017analytics} as well as the fiscal proxy SVAR from \cite{mertens2014reconciliation}. }
\end{figure}

Figure \ref{fig:multipliers comp} presents the estimated tax and spending multipliers, where the first column reports the tax multiplier, the second column shows the spending multiplier, and the third column presents the estimated difference between the tax and spending multiplier.\footnote{Similar to \citet{mertens2014reconciliation}, we calculate tax multipliers by dividing the output response of a tax revenue shock of minus one percent by the average ratio of federal tax revenues to GDP in the sample of 17.5\%. Government spending multipliers are calculated by dividing the output response of a public spending shock of one percent by the average ratio of federal spending to GDP in the sample of 9.1\%. Equivalently, the numbers reflect the present response to a tax cut (government spending increase) that lowers (increases) tax revenues (government spending) by one percentage point of GDP.} The solid lines report the estimates of the non-Gaussian proxy weighting approach and shaded areas indicate 68\% credible bands. As a comparison with the existing approaches, dashed lines show the estimates of the \cite{mertens2014reconciliation} fiscal proxy SVAR and dashed-dotted lines report the estimates of the \cite{caldara2017analytics} non-fiscal proxy SVAR. 
%We show the posterior median as well as 68\% credible bands.

We first discuss the estimates of our proposed non-Gaussian proxy weithing approach. The tax multiplier reported in the first column of Figure \ref{fig:multipliers comp} is estimated to be close to zero for the impact period. Only around one year after the shock, the multiplier increases. The multiplier peaks at a value of $0.7$ two years after the shock materialized. Thereafter, the response slowly converges back to its pre-shock level. The estimated government spending multiplier presented in the second column shows some stark differences. For the impact period, the spending multiplier is estimated to be different from zero taking on a value above but close to unity.  The third column showing the difference between the government spending and tax multiplier clearly depicts the divergent dynamics of both multipliers. For the first year after the shock, the spending multiplier clearly exceeds the tax multiplier. Over the medium term, the difference becomes negative implying that the tax multiplier is larger than the government spending multiplier, although the point estimate for the difference is small. Thus, our baseline model suggests that positive government spending shocks have a larger stimulating impact on economic activity than exogenous tax cuts.

 %\begin{figure} 
%	\centering
%		\caption{Estimated output multipliers for the non-Gaussian proxy weighting SVAR }\label{fig:multipliers}
%	\includegraphics[width=1\textwidth]{img/shocks_diff_DD.pdf}
%  \floatfoot{The figure shows the posterior median as well as $68$\% credible bands for the impulse responses of output to a tax shock and a government spending shock. The right subfigure shows the posterior median as well as 68\% credible bands for the difference of the two output responses. }
%\end{figure}   

\begin{figure}
	\centering
	\caption{Comparison of estimated output multipliers with non-Gaussian estimates}\label{fig:multipliers comp NG}
	\includegraphics[width=1\textwidth]{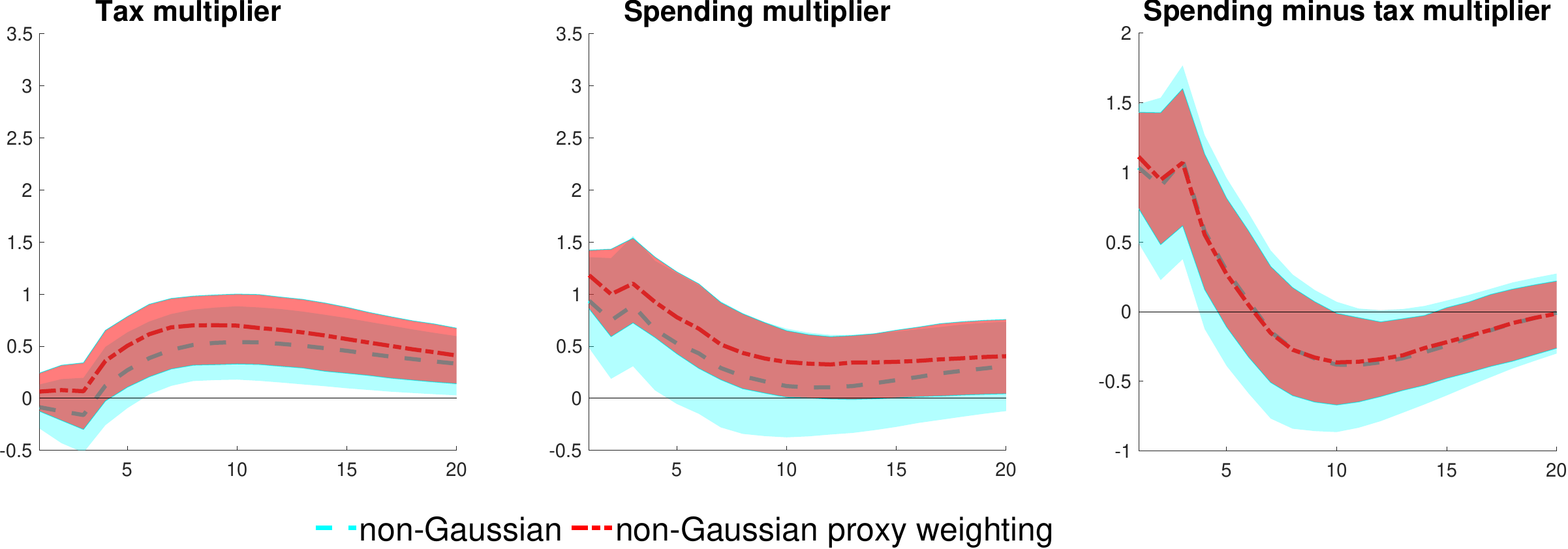} 
 \floatfoot{The figure compares the output responses between our non-Gaussian proxy weighting to a non-Gaussian estimator not including any proxy variables. Both estimators are shown with $68$\% credible bands. }
\end{figure}

Figure \ref{fig:multipliers comp NG} compares our baseline estimates of the non-Gaussian proxy weighting SVAR with the estimates obtained when only relying on non-Gaussianity. The non-Gaussian model does not use any information included in the proxy variables. While the point estimates show only small differences to our baseline model, estimation uncertainty is much larger in the model without proxy variables. Thus, in line with our Monte Carlo simulation results presented above, the fiscal multiplier application shows that potentially endogenous proxies still provide important information that increases estimation precision. Identification through non-Gaussianity misses this information, such that error bands become wider.

As shown in Figure \ref{fig:multipliers comp}, our estimates show strong differences to the fiscal proxy SVAR of \cite{mertens2014reconciliation}.   
The fiscal proxy SVAR leads to a much larger tax multiplier compared to our baseline estimates. In particular, the fiscal proxy SVAR delivers an on-impact tax multiplier around two and the multiplier further increases in the subsequent periods reaching a peak value close to three. Given the much larger tax multiplier, the fiscal proxy SVAR implies that exogenous tax cuts are a more powerful tool to stimulate the economy compared to exogenous increases in government spending, which is the opposite of our baseline results.
Additionally, the non-fiscal proxy SVAR of \citet{caldara2017analytics} leads to a larger government spending multiplier compared to our baseline estimates. In particular, the spending multiplier peaks at a value above two, whereas our baseline estimate shows a maximum value above but close to unity. Similarly to our non-Gaussian proxy weighting model, applying the non-fiscal proxy SVAR also results in a positive difference between the government spending and tax multiplier. In the Appendix, we show that the differences in the estimated multipliers can be explained by different estimates on the cyclical elasticities of tax revenues ($\Theta_{y}$) and government spending ($\gamma_{y}$).

Our main finding that the government spending multiplier is larger than the tax multiplier is highly robust to modifications of the baseline empirical specification. We report a battery of robustness checks in the Appendix. Specifically, our results are robust to including additional endogenous variables in the VAR, controlling for fiscal foresight, splitting and extending the sample. Moreover, the results are not significantly affected when separately estimating the models with a single proxy compared to the baseline model that uses all proxies. 
 
%To summarize, our non-Gaussian proxy weighting model produces a tax multiplier that is smaller than the one found when relying the on the fiscal proxy SVAR. Furthermore, we find a considerably smaller government spending multiplier compared to the non-fiscal proxy SVAR. We will show below that the differences in the estimated multipliers across models is due to the fact that the proxies used in the fiscal and non-fiscal SVAR do not fulfill the crucial exogeneity restrictions. When relying on newly constructed exogenous proxies, the fiscal and non-fiscal SVAR lead to estimated multipliers that are very similar to the ones of our proposed non-Gaussian proxy weighting SVAR. 

%\textcolor{red}{Figure on FEVD (Figure 5 in \citet{lewis2021identifying}}

\subsection{Understanding the differences}
 
The key advantage of our non-Gaussian proxy weighting approach is that we can update the proxy exogeneity priors. That is we shrink towards exogenous proxies, however, if the data provide evidence against a given exogeneity assumption, our model can update the prior and stop to shrink towards exogenous proxies.
In contrast, when applying the Gaussian fiscal proxy SVAR from \citet{mertens2014reconciliation} or the Gaussian non-fiscal proxy SVAR from \citet{caldara2017analytics}, it has to be assumed that the respective proxy is exogenous and this prior cannot be updated. In the following, we show that the data provide evidence against the exogeneity of both proxies, which further helps in understanding the different multiplier estimates across identification strategies.  
 
\begin{figure} [t]
	\centering
	\caption{Posterior of exogeneity moment conditions}\label{fig:exogeneity}
	\includegraphics[width=1\textwidth]{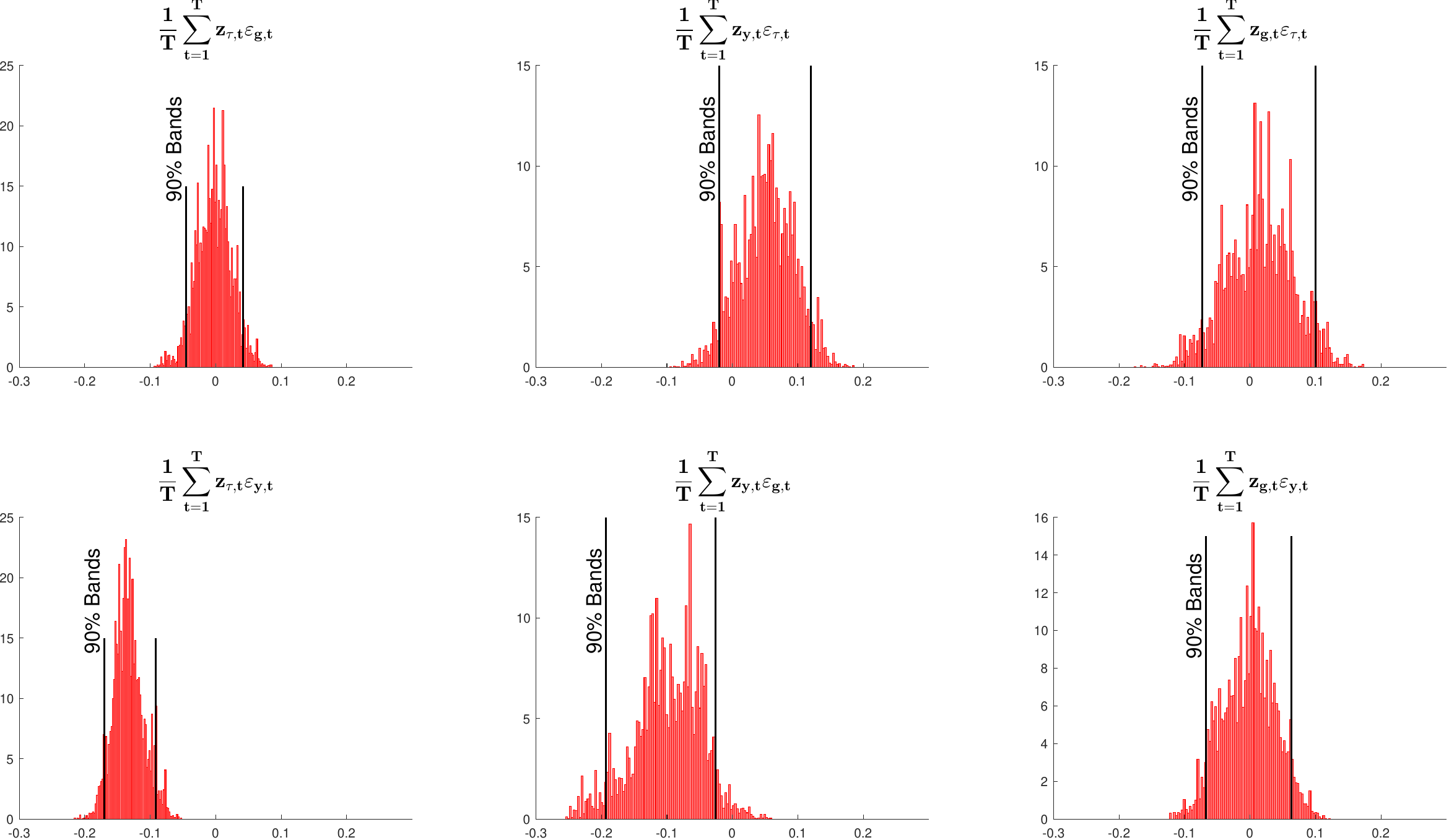} 
 %\floatfoot{The figure shows  the posterior of the proxy exogeneity moments.}
\end{figure}

Figure \ref{fig:exogeneity} provides evidence on the exogeneity assumptions obtained from our non-Gaussian model. The two graphs in the first column show the posterior distributions of the correlation between the tax proxy and the structural government spending and output shock, respectively. The second column presents the posterior distributions of the correlation between the TFP proxy and the structural tax and government spending shock, respectively. The third column shows the posterior distributions of the correlation between the military spending proxy and the structural tax and output shock, respectively. With an exogenous proxy variable, meaning $\mu_j=0$, the proxy prior shrinks toward shocks without systematic correlation with the proxy variables.
However, the data provide evidence against the exogeneity assumptions and the model prevents shrinkage towards some exogeneity assumptions.
For example, the tax proxy has a clear negative correlation with the output shock. Put differently, positive (negative) output shocks coincide with negative (positive) values for the tax instrument. Intuitively, not accounting for this correlation leads to identified tax cuts that also include exogenous increases in output, and vice versa, which increases the size of the estimated tax multiplier. 
Concerning the TFP proxy, we find evidence that the instrument is negatively correlated with exogenous government spending shocks. Therefore, the TFP proxy identifies positive (negative) output shocks that also include negative (positive) government spending shocks, which reduces the fraction of GDP movements explained by the identified output shocks. As a result, the estimated government spending elasticity is reduced, which as shown in \cite{caldara2017analytics}, increases the size of the estimated government spending multiplier. Finally, the military spending proxy seems to fulfill the exogeneity assumptions. The correlations with the structural tax shock and the output shock, respectively, are centered around zero.

Our findings on the endogeneity of the tax proxy align with \cite{bruns2024testing}, who develop a proxy exogeneity test based on the idea that if a proxy $z_t$ is exogenous, any function of it, i.e.   $z_t^2$, should also be exogenous. This yields an overidentified system, allowing for a J-test of proxy exogeneity. Applying this test, \cite{bruns2024testing} find evidence of endogeneity in the tax proxy, but no such issues for the government spending proxy, consistent with our results. However, because the output proxy lacks skewness, $z_t^2$ becomes irrelevant, rendering the J-test powerless to detect its endogeneity. In contrast, our identification strategy based on the assumption of independent shocks does not suffer from this limitation. Indeed, we find indications of endogeneity in the output proxy as well, though the evidence is substantially weaker than for the tax proxy.

 To further highlight the implied differences across identification strategies, in the Online Appendix we show the times series for the estimated structural shocks. Although the estimated shocks share a similar pattern for most periods, there are some discrepancies worth mentioning. In general, we find that the fiscal proxy VAR approach shows the tendency to interpret positive output shocks as negative tax shocks, and vice versa. As a consequence, by taking this shortcoming of the fiscal proxy into account, our non-Gaussian proxy weighting approach leads to a much smaller tax multiplier compared to the standard fiscal proxy identification strategy. We further highlight some episodes in which the non-fiscal proxy model and the non-Gaussian proxy weighting model differ in the sequence identified output and government shocks. We explain in detail why we think that the identified shock of our proposed non-Gaussian proxy weighting model fits better to the historical narrative.

 Furthermore, we go one step further and construct new proxy measures that are orthogonal to the disturbances of the model. To get new proxy measures we proceed as follows. For the narrative tax proxy, we regress the proxy on a constant and the median structural government spending and output shocks obtained from applying the non-Gaussian proxy weighting approach. Similarly, for the TFP measure, we regress the proxy on a constant and the median structural tax and government spending shocks. The estimated residuals of these regressions capture movements in the original proxy measures that are not related to other disturbances of the model.\footnote{One should keep in mind that the construction of the new proxies is model-specific. In particular, for our application, we use the simple three variables baseline VAR which is a common model used in the fiscal policy literature. In addition, using the newly constructed proxies as observable data might suffer from a generated regressor problem.} The estimated fiscal multipliers using the new proxies using a Gaussian proxy approach are very similar to the ones of our baseline non-Gaussian proxy weighting approach. The new proxies and estimation results are discussed in the Appendix.

  \section{Conclusion}
  \label{sec: Conclusion} 
This paper discusses the challenge of measuring the effects of fiscal policy and the recent use of the proxy VAR approach as a tool to identify fiscal policy shocks. We propose a new non-Gaussian proxy weighting approach that combines non-Gaussian identification with proxy variables. The method shrinks towards exogenous proxy variables but allows for updating the prior and stopping the shrinkage if empirically warranted. We use our model to provide evidence that the contradicting results of \citet{mertens2014reconciliation} and \citet{caldara2017analytics} may be due to the use of invalid instruments. Finally, we find that increasing government spending is more effective in stimulating the economy than lowering taxes.% Finally, we utilize our estimation results to construct new proxy variables and show that these new proxies imply very similar fiscal multiplier estimates when used in traditional proxy SVARs. 

%\vspace{1.5cm}

 %\bibliographystyle{chicago}
 %\bibliography{literatur}

    % references

 %   \begin{thebibliography}{}

  %  \bibitem[\protect\citeauthoryear{Aitchison}{Aitchison}{1962}]{Aitchison}
   % Aitchison, J. (1962).
    %\newblock Large-sample restricted parametric tests.
    %\newblock {\em Journal of the Royal Statistical Society,\/} Series B {\em 69}, 234--50.

 %   \end{thebibliography}

    % appendix (put online supplement in a separate file)

   % \section*{Appendix A: Proofs of Results}
   % \renewcommand{\theequation}{A.\arabic{equation}}
   % \renewcommand{\thesection}{A}
   % \setcounter{equation}{0}

 %\bibliographystyle{chicago}
 %\bibliography{literatur}
% \small{\setstretch{1}
%\addcontentsline{toc}{section}{References} 
%\bibliography{literatur}}\clearpage

\small{\setstretch{0.85}
\addcontentsline{toc}{section}{References}
\bibliographystyle{frbcle.bst}
\bibliography{literatur}

\clearpage
\setcounter{section}{0}
\setcounter{figure}{0}
\setcounter{table}{0}
\setcounter{equation}{0}
\setcounter{footnote}{0}
\setcounter{page}{1}
\renewcommand{\thepage}{\roman{page}}
\renewcommand\thefigure{\thesection.\arabic{figure}} 
\renewcommand\thetable{\thesection.\arabic{table}} 
\renewcommand\theequation{\thesection.\arabic{equation}} 

\appendix  
\section*{Online Appendix}

  \begin{abstract}

The Online Appendix contains proofs, provides a description about the Gibbs sampler, presents additional empirical results, provides details about a wide range of Monte Carlo simulations, discusses labeling, provides results for corrected proxies, and presents narrative evidence.

    \end{abstract}

    % main body

\section{Proofs} 

  %\medskip

    \textbf{Proof of Proposition 1:}

%\begin{proof}[Proof of Proposition 1]
For Gaussian shocks,  the likelihood 
$   p(\bm y| \bm B )  $ is identified up to orthogonal rotations $\bm Q$ of  $\bm B_0$, which is the well known identification problem.   
However, re-weighting the likelihood conditional likelihood  $p( D(\bm z) | \bm y ,  \bm B, \bm \pi)$ breaks the symmetry.
For an orthogonal matrix $\bm Q$  it follows that
    \begin{align}\label{ejt q}
         e_{t}(\bm B_0 \bm Q) = \bm Q' \bm B_0^{-1} u_t = Q' \varepsilon_t
    \quad \text{and} \quad 
        e_{jt}(\bm B_0 \bm Q)= q_{1j} \varepsilon_{1t} + ... + q_{nj} \varepsilon_{nt},
    \end{align}
    where the  $\ q$'s display the corresponding elements of $\bm Q$.
    Therefore, the proxy exogeneity moment condition is equal to
    \begin{equation}
        \sqrt{T} \frac{1}{T} \sum_{t=1}^{T} z_t  e_{jt}(\bm B_0 \bm Q ) =  q_{1j} \sqrt{T} \frac{1}{T} \sum_{t=1}^{T} z_t \varepsilon_{1t} + \sum_{l=2}^{n}   q_{lj} \sqrt{T} \frac{1}{T} \sum_{t=1}^{T} z_t \varepsilon_{lt}.
    \end{equation} 
     For a valid proxy and with $\sigma_{z \epsilon_{l} }^2=Var(z_t  \epsilon_{lt})< \infty $   the CLT implies
     $ 
         \sqrt{T} \frac{1}{T} \sum_{t=1}^{T} z_t \varepsilon_{1t} \sim  \mathcal{N}( \sqrt{T} E[z_t \varepsilon_{1t}] , \sigma_{z \epsilon_{1} }^2 ) $
       and 
         $\sqrt{T} \frac{1}{T} \sum_{t=1}^{T} z_t \varepsilon_{lt} \sim \mathcal{N}( 0 , \sigma_{z \epsilon_{l} }^2)$ for $l=2,...,n$.
     Consequently, for a single moment condition $D_j(z)$ and a valid proxy,  the distribution of $D(\bm z)$ at $\bm B_0 \bm Q $   is equal to
   \begin{align}
   \label{eq: proof Dz B dist}
        D_j(\bm z) | \bm y, \bm B_0 \bm Q   \sim   
           \mathcal{N} (  q_{1j} \sqrt{T} E[z_t \varepsilon_{1t}],  q_{1j}  \sigma_{z \epsilon_{1} }^2 + \sum_{l=2}^{n}q_{lj}     \sigma_{z \epsilon_{l} }^2 ).
    \end{align} 

    However, for a single moment condition and a valid proxy Equation (\ref{eq: Dz dist}) states  that at $\bm B_0$, the distribution of $D_j(\bm z)$ is equal to
    \begin{align}
    \label{eq: proof CLT}
        D_j(\bm z) | \bm y, \bm B_0 \sim \mathcal{N}(\bm 0, \bm \sigma_{z \epsilon_{j} }^2), 
    \end{align}

     Therefore, the distribution of $ D_j(\bm z) | \bm y, \bm B_0 \bm Q $ is asymptotically equal to the   distribution of $D(z)| \bm y, \bm B_0 $  only if $q_{1j}=0$.
    This holds for all moment conditions $D_j(\bm z)$ and implies that  the distribution of $ D(\bm z) | \bm y, \bm B_0 \bm Q   $ containing all moment conditions is asymptotically equal to to the   distribution of $D(z)| \bm y, \bm B_0 $  if and only if $q_{12}=...=q_{1n}=0$.   
     Orthogonality of $\bm Q$ implies  $q_{11}=1$ and $q_{21}=....=q_{n1}=0$, which proofs the first statement.

  Let $p( . )$ denote the likelihood of  $   D(\bm z) | \bm y, \bm B_0, \bm \pi_0 \sim \mathcal{N}(\bm 0, \bm \Sigma_z)$,
    Equation (\ref{eq: proof Dz B dist}) implies that if  $q_{1j}\neq 0$, the distribution of $D(\bm z) | \bm y, \bm B_0 \bm Q $ has a finite variance, however,  the mean converges to infinity. 
    Consequently, for all $\epsilon>0$ it holds that
    \begin{align}
        \lim_{T \to \infty}   P( |  p(D(\bm z) | \bm y ,  \bm B) | > \epsilon)=0,
    \end{align} 
    if $q_{12}=...=q_{1n}=0$.  Orthogonality of $\bm Q$ implies  $q_{11}=1$ or $q_{11}=-1$ and $q_{21}=....=q_{n1}=0$, which proofs the second statement.  We normalize the shocks such that only $q_{11}=1$ is possible, see the section about labeling in the main paper.

%\end{proof}

    \hfill$\square$\\

    \textbf{Proof of Proposition 2:}

%\begin{proof}[Proof of Proposition 2]
    Analogous to the proof of Proposition \ref{prop: identification proxy weighting} it follows that  
$ D(\bm z) | \bm y, \bm B_0 \bm Q   \sim  \mathcal{N} (  q_{1j} \sqrt{T} E[z_t \varepsilon_{1t}], \sigma_z )$.
An irrelevance of the proxy implies $E[z_t  \varepsilon_{1t}] = 0$ and thus,
$ D(\bm z) | \bm y, \bm B_0 \bm Q   \sim  \mathcal{N} (  0, \sigma_z )$ for all orthogonal $ \bm Q$. 
    Therefore,  $E[p(D(\bm z) | \bm y, \bm B_0 \bm Q_1 ) ] = E[p(D(\bm z) | \bm y, \bm B_0 \bm Q_2 ) ] $ for all orthogonal matrices $\bm Q_1$ and $\bm Q_2$.  
%\end{proof}
    \hfill$\square$\\

\vspace{1.5cm}

\section{Identification with multiple proxies} 

Consider two proxy variables $z_{1t}$ and $z_{2t}$ for the first two structural shocks $\varepsilon_{1t}$ and $\varepsilon_{2t}$ with the relevance conditions $E[z_{1t} \varepsilon_{1t}] \neq 0$, $E[z_{1t} \varepsilon_{2t}] \neq 0$, $E[z_{2t} \varepsilon_{1t}] \neq 0$, and $E[z_{2t} \varepsilon_{2t}] \neq 0$ and the exogeneity conditions  $E[z_{1t} \varepsilon_{jt}] = 0$ and $E[z_{2t} \varepsilon_{jt}] = 0$ for $j=3,...,n$.  
For the two proxy variables, define the proxy exogeneity moment conditions
\begin{align}
\label{eq: D1 }
    D_{1j}( \bm z_1, \bm y, \bm B, \bm \pi) = \frac{1}{ \sqrt{T} } \sum_{t=1}^{T} z_{1t}  e_{jt}(\bm B,\bm \pi) 
\end{align}
and
\begin{align}
\label{eq: D2 }
    D_{2j}( \bm z_2, \bm y, \bm B, \bm \pi) = \frac{1}{ \sqrt{T} } \sum_{t=1}^{T} z_{2t}  e_{jt}(\bm B,\bm \pi) 
\end{align}
and $D(\bm z_1, \bm z_2) = [D_{13} (\bm z_1, \bm y, \bm B, \bm \pi),..., D_{1n} (\bm z_1, \bm y, \bm B, \bm \pi),D_{23} (\bm z_2, \bm y, \bm B, \bm \pi),..., D_{2n} (\bm z_2, \bm y, \bm B, \bm \pi)]'$.
The joint likelihood of the data $\bm y$ and the  proxy exogeneity moment conditions $D(\bm z_1 , \bm z_2 )$ can be written as 
\begin{equation}
\label{eq: joint like z D1 D2}
  p(\bm y, D(\bm z_1 , \bm z_2 ) |\bm B, \bm \pi)=   p(\bm y| \bm B, \bm \pi)  p( D(\bm z_1 , \bm z_2 ) | \bm y ,  \bm B, \bm \pi). 
\end{equation}  For an exogenous proxy variable, the conditional likelihood can be approximated using  the central limit theorem (CLT), i.e.
\begin{align}
\label{eq: Dz dist 2prox}
    D(\bm z_1 , \bm z_2 ) | \bm y, \bm B_0, \bm \pi_0 \sim \mathcal{N}(\bm 0, \bm \Sigma_z),
\end{align} 
where  $\bm \Sigma_z $ denotes the covariance matrix   of $( z_{1t}  e_{3t} ,... ,z_{1t}  e_{nt},z_{2t}  e_{3t} ,... ,z_{2t}  e_{nt} )'$. 

The following proposition shows that for two valid proxy variables, the  the joint likelihood in Equation (\ref{eq: joint like z D1 D2}) identifies the impact of the target shocks.

\begin{proposition}
    \label{prop: identification proxy weighting}
    Consider an SVAR  $u_t = \bm B_0 \varepsilon_t$ with   Gaussian shocks  and two valid proxy variables $z_{1t}$ and $z_{2t}$ for the two target shock $\varepsilon_{1t}$ and $\varepsilon_{2t}$. %Let $\bm b_{1,0}$ be the impact of the first target shock  equal to the first column of $\bm B_0$ and let $\bm b_{2,0}$ be the impact of the second target shock  equal to the second column of $\bm B_0$ .

    The    joint  likelihood  in Equation (\ref{eq: joint like z D1 D2}) 
    identifies the target shocks and their impact up to an orthogonal rotation, i.e. for $\bm B= \bm B_0 \bm Q$ with an orthogonal matrix $\bm Q$ the distribution of 
    $D(\bm z_1, \bm z_2)| \bm y, \bm B $ is asymptotically equal to the distribution of $D(z)| \bm y, \bm B_0 $ from Equation (\ref{eq: Dz dist 2prox})  if and only if the first two  columns of $\bm B$ are equal to an orthogonal rotation of the first two columns of $B_0$.

\end{proposition}
\begin{proof}
Analogous to the proof of Proposition 2, one can show that 
the distribution of $ D(\bm z_1, \bm z_2) | \bm y, \bm B_0 \bm Q $ for an orthogonal matrix $\bm Q$ is asymptotically equal to the   distribution of $D(\bm z_1, \bm z_2)| \bm y, \bm B_0 $  if and only if $q_{13}=...=q_{1n}=0$ and $q_{23}=...=q_{2n}=0$.  
 Orthogonality of $\bm Q$ implies  $\begin{bmatrix}
     q_{11} & q_{12}\\q_{21}& q_{22}
 \end{bmatrix}$ is orthogonal, $q_{31}=....=q_{n1}=0$, and $q_{32}=....=q_{n2}=0$, which proofs the second statement.
\end{proof}

\section{Labeling}
\label{subsec: labeling}
Estimating $\bm \mu$ and thus allowing endogenous proxies is possible due to the fact that the proxy is not used for identification. Instead, identification of the SVAR relies on the assumption of independent and non-Gaussian shocks. However, identification based on independent and non-Gaussian shocks only identifies the SVAR up to sign and permutations. The MCMC algorithm may sample from different sign-permutations, such that posterior draws from the response of one variable to one shock do not come from a unique shock, but rather from a combination of different shocks resulting in invalid inference, \cite{anttonen2021statistically}.
Furthermore, the shocks are statistical identified but do not carry a structural label. Hence, the statistical identified shocks must be labeled manually by the researcher based on economic assumptions. These assumptions do not constrain the possible values of the identified parameters, which are derived purely from statistical information. Instead, the economic assumptions guide the selection among the statistically identified shocks, a desirable feature when the restrictions are considered approximate but not strictly valid (see \cite{lewis2024identification}).

Preventing sign permutations of the shocks in the MCMC simply requires imposing an a priori rule, which also allows to attach structural labels to the shocks in each MCMC draw. To illustrate, consider a bivariate SVAR without lags $y_t = \bm B_0 \varepsilon_t$  in prices and quantities contained in $y_t$ and with a supply and demand shock contained in $ \varepsilon_t$. Changing the order and/or signs of the supply and demand shocks while applying a corresponding sign-permutation to $\bm B_0$ leads to the exact same prices and quantities, i.e. $y_t = \tilde{\bm B}_0 \tilde{\varepsilon}_t$ with $\tilde{\bm B}_0= \bm B_0 \bm P^{-1}$, $\tilde{\varepsilon}_t= \bm P \varepsilon_t$ for any sign-permutation matrix $\bm P$. There simply is no inherent order or sign of supply and demand shocks in the SVAR, instead, we always need a rule to determine the order of the shocks. 

For example, a recursive SVAR imposes the rule that the second shock, i.e. the supply shock, has no impact on the first variable, which labels the second shock as a supply and the first shock as a demand shock for any allowed proposal $\bm B$. 
The rule can be relaxed to the assumption that the first shock has the largest simultaneous impact, in absolute magnitude, on the first variable, compare \cite{lanne2017identification}. This still allows to detect if a given proposal $\bm B$ has the correct order of supply and demand shocks, however, it is problematic if we expect that supply and demand shocks have a similar effect on the first variable, see \cite{keweloh2023uncertain}. In this case, a labeled first step estimator $ \bar{\bm B}$  can be used to re-center the \cite{lanne2017identification} labeling approach, see \cite{keweloh2023uncertain}.

Similarly, in a proxy SVAR, the shocks are labeled based on the proxy variable. Specifically,   proxy SVARs typically impose the restriction that the proxy is exogenous and relevant, that is, only correlated with the target shock, which automatically assigns a label to the target shock for any allowed proposal $\bm B$ and also prevents permutations.
The proxy labeling rule can also be relaxed analogously to the recursive SVAR labeling rule. Specifically, we impose the assumption that the proxy variable has a larger correlation with the proxy in absolute magnitude than with the other shock to label the target shock for any proposal $\bm B$. This is desirable if proxy restrictions are considered approximate but not strictly valid. As a consequence we still allow the proxy to be correlated with non-target shocks.
\footnote{We have also used the algorithm from \cite{keweloh2023uncertain} as an robustness check to address the issue of permutations and labeling.}   

The implementation is straightforward; for each proposal $\bm B$   and $\bm \pi$ in the MCMC, we calculate the corresponding innovations $e_t(\bm B, \bm \pi)$ and the correlation of innovations and proxy. If the correlation of the proxy and the target shock is larger in absolute value than the correlation of the proxy with all non-target shocks, we proceed with the MCMC iteration. If not, we reject the proposal and continue with the next MCMC iteration.

\section{Gibbs sampler with a metropolis hasting step}
%p( \bm \pi, \bm B,\bm \lambda, \bm q, \bm \mu, \bm \Sigma_{\boldsymbol{\mu}}|\bm y,  D(\bm z) )  \propto  p(\bm y| \bm B, \bm \pi,\bm \lambda, \bm q)  p( D(\bm z) | \bm y ,  \bm B, \bm \pi,\bm \mu)p(\bm \mu| \bm \Sigma_{\boldsymbol{\mu}}),

 Estimation of our model can be carried out using MCMC techniques.
 We use a Gibbs sampler with a Metropolis Hasting step. A sample of the joint posterior $ p( \bm \theta,  \bm \mu, \bm \Sigma_{\boldsymbol{\mu}}|\bm y,D(\bm z)) $, where $\bm \theta=(\bm B, \bm \pi ,\bm \lambda, \bm q)$, can be obtained by sequentially drawing from these two conditional posterior distributions after an initial burn-in phase:

\begin{enumerate}
\item $p(\bm \Sigma_{\mu}|\bm \mu )$;
\item $p(\bm  \theta, \bm \mu|\bm y,D(\bm z), \bm \Sigma_{\mu})$.
\end{enumerate}

In the fist step, we draw each $\sigma_{\mu_{ij}}^2$ conditioning on $\mu_{ij}$ independently from each other.
Since the Gaussian and inverse-Gamma distribution are conjugate distributions, it is possible to derive the conditional posterior of $\sigma_{\mu_{j}}^2$
\begin{align}
\label{eq: prior sigma mu}
\sigma_{\mu j }^2 | \mu_{j} \sim IG(0.5+a,0.5 \mu_{j}^2 +b). 
\end{align}
 
 In a second step conditioning on $\Sigma_{\boldsymbol{\mu}}$ and the data we jointly draw $\bm \theta$ and $\bm \mu$ from their conditional posterior distribution using a Metropolis Hasting step as proposed by \cite{ter2008differential}. 

 \cite{ter2008differential} propose a type of Differential Evolution Monte Carlo Chain (DE-MC) algorithm. It consists of running $N$ different MCMC chains in parallel. Let the current states of chain $i$ be stored in matrix $\boldsymbol{X}$ of dimensions $N \times d$, with $d$ as the number of elements contained in $\bm \Xi=( \bm \theta,\bm  \mu)$. Moreover, define $\boldsymbol{Z}$ as a $M \times d$ matrix containing the current and past states of the chains. To update $\boldsymbol{\Xi}_i$ follow:
\begin{equation}
	\boldsymbol{\Xi}^\star =\boldsymbol{\Xi}_i + \gamma(\boldsymbol{x}_{R1} - \boldsymbol{x}_{R2}) +\bm e,
\end{equation} 
where $\boldsymbol{x}_{R1}$ and $\boldsymbol{x}_{R2}$ are row vectors randomly selected without replacement from $\boldsymbol{Z}$, $\gamma$ is an arbitrary scalar, and $e$ is a draw from a symmetric distribution with small variance and unbounded support, i.e. $\boldsymbol{e} \sim N(0,\bm I_d c)$ with $c$ relatively small. Following \cite{ter2008differential} and \cite{anttonen2021statistically}, we set $N=2$, $\gamma = 2.38/2\sqrt{d}$ and $c=0.0001^2$. We find that based on these values the sampler mixes well without any further tuning. Next, using $\boldsymbol{\Xi}^\star$ the Metropolis ratio $r$ is calculated as:
\begin{equation}
	\label{eq_mratio}
	r = \frac{ p(\bm y| \bm \theta^\star) p(D(\bm z)|\bm y, \bm \theta^\star,     \bm \mu^\star,)p(\bm \mu^\star| \bm \Sigma_{\mu})}{ p(\bm y| \bm \theta) p(D(\bm z)|\bm y, \bm \theta, \bm \mu)p(\bm \mu| \bm \Sigma_{\mu})}.
\end{equation}
Then based on a random draw $u\sim U(0,1)$, the new state is accepted or rejected with probability $\min(1,r)$, thus updating $\boldsymbol{X}$. This process of updating $\boldsymbol{X}$ is repeated for all chains, and then $\boldsymbol{X}$ is appended to $\boldsymbol{Z}$ such that $M$ increases with $N$ rows. To initialize the algorithm, we simply need $M_0$ initial values of $\boldsymbol{\Xi}$, which we sample from a normal distribution with mean obtained by numerical maximizing the likelihood function of the model and variance set to be small. Once obtained, these are copied to $\boldsymbol{Z}$ and $\boldsymbol{X}$, such that $M = M_0$. \\

After these steps have been repeated a certain number of times, all chains converge to the posterior distribution independently of the others, thus after discarding a burn-in phase in $\boldsymbol{Z}$, the remaining rows can be considered as draws from the posterior distribution of $\boldsymbol{\Xi}$. Each chain will possess the characteristics of:
\begin{align*}
	\text{E}(x_i - x_j) &= 0,\\
	\text{cov}(x_i - x_j)&=2\text{cov}(\bm \Xi),
\end{align*}
where $i$ and $j$ correspond to the indices of each chain.

\newpage

    % For a Gaussian likelihood, we show with simulated and real data that both methods are equivalent. But if we use a non-Gaussian likelihood and assuming independent shocks the two approaches lead to different results if the proxy process is missspecified.

 \section{Additional Simulation Results}
In this section, we present results of additional Monte Carlos simulations.
To further demonstrate the flexibility of our approach, we consider a range of different plausible scenarios an applied researcher may face. In many applications, the relevance of the proxy may be weak. In this case, we show that pure proxy approaches lead to biased estimates. In contrast, our combined approach still works and can exploit the proxy information to improve the pure statistical identification approach. In addition, alternative scenarios include a simulation with $t$-distributed shocks, with Gaussian shocks, a proxy with missing values, and a simulation with two proxy variables. Finally, we consider an alternative DGP based on the estimation results of \cite{mertens2014reconciliation}. The results show that our model is highly flexible and can adapt to different scenarios.

 If not otherwise stated, we assume the same DGP as in the main part of the paper. We simulate a system containing a government spending shock $ \varepsilon_{g, t}$, an output shock $ \varepsilon_{y, t}$, and a tax shock $\varepsilon_{\tau, t}$ with
  \begin{align}
 % \label{eq: MC one proxy}
  \begin{bmatrix}
  u_{g, t} \\
  u_{y, t} \\
   u_{\tau, t} \\ 
  \end{bmatrix} &=
  \begin{bmatrix}
   1 & 0  & 0        \\
    0.15 & 1 &  -0.5    \\
   0 & 1.5 & 1    
  \end{bmatrix}
  \begin{bmatrix}
  \varepsilon_{g, t} \\
  \varepsilon_{y, t} \\
  \varepsilon_{\tau ,t} \\
  \end{bmatrix} . 
  \end{align}  

\subsection{Weak proxy variables}
In some applications the correlation of the proxy variable with the target shock may be low. Therefore, we next assume that the proxy variables is only weakly relevant. Furthermore, the structural shocks are independently and identically   drawn from  a Pearson distribution with mean zero, variance one, skewness $0.68$ and excess kurtosis $2.33$ and we assume that the proxy variables is exogenous. Table \ref{Table:weakMSE} and \ref{Table:Weakcoverage} show the results. While the performance of the pure proxy approaches worsens considerably our combined approach still works fine and still exploits the information of the proxy to improve on the pure statistical identification approach.

\begin{table}
	\caption{Weakly relevant - mean and mse}
	\label{Table:weakMSE} 
	\begin{tabular}{ c  | c    c   c       }
		&  proxy  1  &  proxy 2  &   proxy 3  \\  
		&$z_{\tau ,t}= \varepsilon_{\tau, t}  +	2 \eta_t$&$z_{\tau ,t}= \varepsilon_{\tau, t}  +	3\eta_t $&$z_{\tau ,t}= \varepsilon_{\tau, t}  +	4\eta_t  $
		\\ \hline
		$T=250$&&&

		\\
proxy (augmented)  &
		$\begin{bmatrix}  
		\underset{(0.019)}{\hphantom{-}0.00}  &
		\underset{(0.021)}{-0.48} & 
		\underset{(0.048)}{\hphantom{-}0.98} 
		\end{bmatrix}'$
		&
		
		$\begin{bmatrix}
	\underset{(0.032)}{\hphantom{-}0.00}  &
		\underset{(0.033)}{-0.47} & 
		\underset{(0.075)}{\hphantom{-}0.95} 
		\end{bmatrix}'$
		&
		
		$\begin{bmatrix}  
		\underset{(0.039)}{\hphantom{-}0.00}  &
		\underset{(0.045)}{-0.45} & 
		\underset{(0.106)}{\hphantom{-}0.91} 
		\end{bmatrix}'$
		
		\\    
non-Gaussian  &
		$\begin{bmatrix} 
		\underset{(0.018)}{\hphantom{-}0.00}  &
		\underset{(0.020)}{-0.47} & 
		\underset{(0.045)}{\hphantom{-}0.96} 
		\end{bmatrix}'$
		
		&
		
		$\begin{bmatrix}          
		\underset{(0.017)}{\hphantom{-}0.00}  &
		\underset{(0.020)}{-0.48} & 
		\underset{(0.042)}{\hphantom{-}0.96} 
		\end{bmatrix}'$
		
		&
		
		$\begin{bmatrix}  
		\underset{(0.017)}{-0.01} & 
		\underset{(0.018)}{-0.48} & 
		\underset{(0.041)}{\hphantom{-}0.95} 
		\end{bmatrix}'$
		
		\\   
\begin{tabular}{@{}c@{}}non-Gaussian  \\proxy weighting\end{tabular}    & 
		$\begin{bmatrix} 
		\underset{(0.012)}{\hphantom{-}0.00}  &
		\underset{(0.013)}{-0.49} & 
		\underset{(0.029)}{\hphantom{-}0.99} 
		\end{bmatrix}'$
		&
		
		$\begin{bmatrix} 
		\underset{(0.013)}{\hphantom{-}0.00}  &
		\underset{(0.015)}{-0.49} & 
		\underset{(0.030)}{\hphantom{-}0.99} 
		\end{bmatrix}'$
		&
		
		$\begin{bmatrix}   
		\underset{(0.015)}{-0.01} & 
		\underset{(0.016)}{-0.49} & 
		\underset{(0.033)}{\hphantom{-}0.98} 
		\end{bmatrix}'$
		
			\\ \hline
		$T=800$&&&

		\\
proxy (bayes)  &
		$\begin{bmatrix}  
		\underset{(0.006)}{-0.01} & 
		\underset{(0.007)}{-0.50} & 
		\underset{(0.017)}{\hphantom{-}0.99} 
		\end{bmatrix}'$
		&
		
		$\begin{bmatrix}
	\underset{(0.012)}{\hphantom{-}0.00}  &
		\underset{(0.014)}{-0.49} & 
		\underset{(0.031)}{\hphantom{-}0.98} 
		\end{bmatrix}'$
		&
		
		$\begin{bmatrix}  
		\underset{(0.019)}{\hphantom{-}0.00}  &
		\underset{(0.023)}{-0.49} & 
		\underset{(0.052)}{\hphantom{-}0.95} 
		\end{bmatrix}'$
		
		\\    
non-Gaussian  &
		$\begin{bmatrix} 
		\underset{(0.005)}{\hphantom{-}0.00}  &
		\underset{(0.005)}{-0.50} & 
		\underset{(0.012)}{\hphantom{-}0.99} 
		\end{bmatrix}'$
		
		&
		
		$\begin{bmatrix}          
		\underset{(0.005)}{\hphantom{-}0.00}  &
		\underset{(0.005)}{-0.50} & 
		\underset{(0.011)}{\hphantom{-}0.99} 
		\end{bmatrix}'$
		
		&
		
		$\begin{bmatrix}  
		\underset{(0.005)}{\hphantom{-}0.00}  &
		\underset{(0.005)}{-0.50} & 
		\underset{(0.011)}{\hphantom{-}0.99} 
		\end{bmatrix}'$
		
		\\   
\begin{tabular}{@{}c@{}}non-Gaussian  \\proxy weighting\end{tabular}    & 
		$\begin{bmatrix} 
		\underset{(0.003)}{\hphantom{-}0.00}  &
		\underset{(0.004)}{-0.50} & 
		\underset{(0.008)}{\hphantom{-}0.99} 
		\end{bmatrix}'$
		&
		
		$\begin{bmatrix} 
		\underset{(0.004)}{\hphantom{-}0.00}  &
		\underset{(0.004)}{-0.50} & 
		\underset{(0.009)}{\hphantom{-}1.00} 
		\end{bmatrix}'$
		&
		
		$\begin{bmatrix}   
		\underset{(0.004)}{\hphantom{-}0.00}  &
		\underset{(0.004)}{-0.50} & 
		\underset{(0.010)}{\hphantom{-}0.99} 
		\end{bmatrix}'$
		 
	\end{tabular} 
	\footnotesize{\textit{Note: The true impact of the shock $\varepsilon_{\tau, t}$ is $\begin{bmatrix}   
			0  & -0.5   & 1
			\end{bmatrix}'$. 
   The average and MSE of the Bayesian estimators are calculated based on the median of the posterior of B in each simulation.}}
\end{table}

\begin{table}
	\caption{Weakly relevant - Coverage  and average length of  $68$\% credible bands of the estimated impact of $\varepsilon_{\tau ,t}$.}
	\label{Table:Weakcoverage} 
	\begin{tabular}{ c  | c    c   c       }
		&  proxy  1  &  proxy 2  &   proxy 3  \\  
		&$z_{\tau ,t}= \varepsilon_{\tau, t}  +	2 \eta_t$&$z_{\tau ,t}= \varepsilon_{\tau, t}  +	3\eta_t $&$z_{\tau ,t}= \varepsilon_{\tau, t}  +	4\eta_t  $
		\\ \hline
		$T=250$&&&
  		\\
		
proxy (augmented)  &
		$\begin{bmatrix} 
		\underset{(0.028)}{0.70} & 
		\underset{(0.029)}{0.68} & 
		\underset{(0.043)}{0.68} 
		\end{bmatrix}'$
		
		&
		
		$\begin{bmatrix}          
		\underset{(0.038)}{0.71} & 
		\underset{(0.039)}{0.73} & 
		\underset{(0.058)}{0.73} 
		\end{bmatrix}'$
		
		&
		
		$\begin{bmatrix} 
		\underset{(0.045)}{0.73} & 
		\underset{(0.047)}{0.73} & 
		\underset{(0.071)}{0.72} 
		\end{bmatrix}'$
  
		\\
		
non-Gaussian  &
		$\begin{bmatrix} 
		\underset{(0.026)}{0.71} & 
		\underset{(0.027)}{0.69} & 
		\underset{(0.041)}{0.70} 
		\end{bmatrix}'$
		
		&
		
		$\begin{bmatrix}          
			\underset{(0.026)}{0.72} & 
		\underset{(0.027)}{0.71} & 
		\underset{(0.041)}{0.73} 
		\end{bmatrix}'$
		
		&
		
		$\begin{bmatrix} 
		\underset{(0.026)}{0.71} & 
		\underset{(0.027)}{0.73} & 
		\underset{(0.042)}{0.74} 
		\end{bmatrix}'$
		
		\\   
\begin{tabular}{@{}c@{}}non-Gaussian  \\proxy weighting\end{tabular}    & 
		$\begin{bmatrix}
		\underset{(0.020)}{0.68} & 
		\underset{(0.021)}{0.67} & 
		\underset{(0.033)}{0.69} 
		\end{bmatrix}'$
		&
		
		$\begin{bmatrix}  
		\underset{(0.022)}{0.69} & 
		\underset{(0.023)}{0.68} & 
		\underset{(0.035)}{0.70} 
		\end{bmatrix}'$
		&
		
		$\begin{bmatrix}   
		\underset{(0.023)}{0.69} & 
		\underset{(0.024)}{0.71} & 
		\underset{(0.037)}{0.72} 
		\end{bmatrix}'$
		
 		\\ \hline
		$T=800$&&&
  		\\
		
proxy (augmented)  &
		$\begin{bmatrix} 
		\underset{(0.016)}{0.70} & 
		\underset{(0.016)}{0.66} & 
		\underset{(0.024)}{0.63} 
		\end{bmatrix}'$
		
		&
		
		$\begin{bmatrix}          
		\underset{(0.022)}{0.68} & 
		\underset{(0.022)}{0.65} & 
		\underset{(0.033)}{0.63} 
		\end{bmatrix}'$
		
		&
		
		$\begin{bmatrix} 
		\underset{(0.028)}{0.68} & 
		\underset{(0.028)}{0.66} & 
		\underset{(0.043)}{0.63} 
		\end{bmatrix}'$
  
		\\
		
non-Gaussian  &
		$\begin{bmatrix} 
	\underset{(0.013)}{0.70} & 
		\underset{(0.013)}{0.68} & 
		\underset{(0.020)}{0.68} 
		\end{bmatrix}'$
		
		&
		
		$\begin{bmatrix}          
		\underset{(0.013)}{0.66} & 
		\underset{(0.013)}{0.65} & 
		\underset{(0.020)}{0.70} 
		\end{bmatrix}'$
		
		&
		
		$\begin{bmatrix} 
		\underset{(0.013)}{0.68} & 
		\underset{(0.013)}{0.66} & 
		\underset{(0.020)}{0.70} 
		\end{bmatrix}'$
		
		\\   
\begin{tabular}{@{}c@{}}non-Gaussian  \\proxy weighting\end{tabular}    & 
		$\begin{bmatrix}
		\underset{(0.011)}{0.69} & 
		\underset{(0.011)}{0.67} & 
		\underset{(0.017)}{0.65} 
		\end{bmatrix}'$
		&
		
		$\begin{bmatrix}  
		\underset{(0.011)}{0.65} & 
		\underset{(0.012)}{0.65} & 
		\underset{(0.018)}{0.67} 
		\end{bmatrix}'$
		&
		
		$\begin{bmatrix}   
		\underset{(0.012)}{0.66} & 
		\underset{(0.012)}{0.65} & 
		\underset{(0.019)}{0.68} 
		\end{bmatrix}'$
	\end{tabular}  
 	\floatfoot{
% 	\textit{Note:}    ......
           }
\end{table}

%\begin{table} 
%	\caption{Gaussian with exogenous proxy  ($T=250 $).}
%	\label{Table: Finite sample performance -Gaussian} 
%	\begin{tabular}{ c  | c  |   c    } 
%		&  Median and MSE &  Coverage and length of    \\
%		
%		&    &     $68$\% credible bands 
%		\\ \hline
%		
%		proxy (augmented)   &
%		
%		
%		$\begin{bmatrix}
%		\underset{(0.008)}{\hphantom{-}0.00}  &
%		\underset{(0.009)}{-0.50} & 
%		\underset{(0.021)}{\hphantom{-}1.00} 
 %
%		\end{bmatrix}'$&
%		
%		
%		$\begin{bmatrix}
%		\underset{(0.018)}{0.69} & 
%		\underset{(0.019)}{0.67} & 
%		\underset{(0.028)}{0.67} 
%		\end{bmatrix}'$

%		\\    
%		proxy (weighting)   &

%		$\begin{bmatrix}          
%		\underset{(0.008)}{\hphantom{-}0.00}  &
%		\underset{(0.009)}{-0.50} & 
%		\underset{(0.021)}{\hphantom{-}1.00} 
%		\end{bmatrix}'$&

%		$\begin{bmatrix}          
%		\underset{(0.018)}{0.69} & 
%		\underset{(0.019)}{0.68} & 
%		\underset{(0.029)}{0.68} 
%		\end{bmatrix}'$

%		\\   
%\begin{tabular}{@{}c@{}}non-Gaussian  \\proxy weighting\end{tabular}    & 
%		
%		
%		$\begin{bmatrix} 
%		\underset{(0.016)}{\hphantom{-}0.00}  &
%		\underset{(0.020)}{-0.49} & 
%		\underset{(0.045)}{\hphantom{-}0.98} 
%		\end{bmatrix}'$ & 
%		
%		
%		$\begin{bmatrix} 
%		\underset{(0.034)}{0.83} & 
%		\underset{(0.035)}{0.81} & 
%		\underset{(0.053)}{0.82} 
%		\end{bmatrix}'$
%		
%	\end{tabular} 
%\end{table}  
%
%
%

\subsection{Further simulations }

Next we summarize the results of four additional Monte Carlo simulations to provide further evidence of the ability of our model to adapt to a range of different plausible scenarios. We also present results for the proxy (frequentist) estimator using $\hat{b}_{ij}= \frac{\sum_{t=1}^{T} (z_t u_{jt})}{\sum_{t=1}^{T}(z_t u_{it})}$ and additionally scaling the estimator to shocks with unit variance. 

First, we assume structural shocks are drawn from a Gaussian distribution and assume that we have valid proxies with different strength. We compare the augmented proxy VAR with our proposed proxy weighting approach for the case that we assume a Gaussian likelihood. 
Results in table \ref{Table: MSEnormal} and \ref{Table: Coveragenormal} show that both approaches lead to the same results.

Second, we assume that the shocks are now $t$-distributed shocks with seven degrees of freedom. 
Table \ref{Table: Finite sample performance - t  } and \ref{Table: Finite sample performance coverage -t}  show that the results are in line with the results from the case where the structural shocks are drawn from a Pearson distribution. Thus, our model performance does not deteriorate
 when using $t$-distributed shocks.

Third, we assume the practical relevant case that the proxy contains missing values. In particular, the proxy is generated by  $z^{new}_t = \psi_t z_t$ where $\psi_t$ is Bernoulli random variable such that $50$\% of proxy observations are set to zero. Results can be found in \ref{Table: Finite sample performance - missing} and \ref{Table: Finite sample performance coverage - missing}. It turns out that our model can deal well with proxy variables that contain missing values and therefore does not need to rely on a linear relation between the proxy and the structural shock of interest.

Fourth, we present results of different choices of $\bm \Sigma_z$, see Section 2.4 in the main paper, in table \ref{TableSigma}. The results do not differ across different choices. 

Finally, we add a second proxy for the shock of interest. In particular, the second proxy is generated by  $z^{y}_t = \varepsilon_{y ,t}  +	\eta_{y,t}  $. The proxy is valid in all cases. Results can be found in \ref{Table: Finite sample performance - 2proxy} and \ref{Table: Finite sample performance coverage - 2proxy} and show that our model can deal with multiple proxy variables.

\section{Alternative DGP with parameters calibrated to the data}

In this section, we consider a different DGP. The DGP is based on the estimation results from \cite{mertens2014reconciliation}:
\begin{align}
     \begin{bmatrix}
\tau_t   
\\
g_t
\\
y_t
\end{bmatrix}
= 
\sum_{i=1}^{4}
\bm A_i
\begin{bmatrix}
\tau_{t-i}   
\\
g_{t-i}
\\
y_{t-i}
\end{bmatrix}
+  
B_0
\begin{bmatrix}
\varepsilon_{\tau,t}    
\\
\varepsilon_{g,t} 
\\
\varepsilon_{y,t} 
\end{bmatrix},
\end{align}
with
\begin{align}
    A_1 &= 
    \begin{bmatrix}
      0.6964   &-0.1360 &   0.9441\\
    0.0886  &  1.1426 &  -0.2646\\
    0.0060  & -0.0268 &   1.2505
    \end{bmatrix}
    &
    A_2 &= 
    \begin{bmatrix}
       0.2014  &  0.0140 &  -0.5731\\
   -0.0302 &  -0.1210  &  0.3116\\
   -0.0124  &  0.0526  & -0.2179
    \end{bmatrix}
    \\
    A_3 &= 
    \begin{bmatrix}
      -0.0665  &  0.2171 &  -0.2626\\
   -0.0098 &   0.0543  & -0.2679\\
   -0.0535  & -0.0469  & -0.0346
    \end{bmatrix}
    &
    A_4 &= 
    \begin{bmatrix}
       -0.0352&   -0.1452 &   0.1000\\
   -0.1388   &-0.1793  &  0.2856\\
    0.0297  &  0.0168 &  -0.0151
    \end{bmatrix},
\end{align}
and
\begin{align}
    B_0 &= 
    \begin{bmatrix}
         0.0121 &   0.0011   & 0.0227\\
    0.0014 &   0.0235       &  0\\
   -0.0042&    0.0019   & 0.0073 
    \end{bmatrix}.
\end{align}

The results in table \ref{TableMSErealstic} and \ref{TableCoveragerealstic} are in line with the results from the more simple DGP.

%%%%%%%%%%%%%%%%%%%%%%%%%%%%%%%%%%%%%%%%%%%%%%%%%%%%

%%%%%%%%%%%%%%%%%%%%%%%%%%%%%%%%%%%%%%%%%%%%%%%%%

\begin{table}
	\caption{Comparing the proxy approaches - mean and mse}
	\label{Table: MSEnormal} 
	\begin{tabular}{ c  | c    c   c       }
		&  proxy  1  &  proxy 2  &   proxy 3  \\  
		&$z_{\tau ,t}= \varepsilon_{\tau, t}  +	1 \eta_t$&$z_{\tau ,t}= \varepsilon_{\tau, t}  +	3\eta_t $&$z_{\tau ,t}= \varepsilon_{\tau, t}  +	4\eta_t  $
		\\ \hline
		$T=250$&&&
		 \\
proxy (augmented)  &
		$\begin{bmatrix}  
		\underset{(0.008)}{\hphantom{-}0.00}  &
		\underset{(0.009)}{-0.50} & 
		\underset{(0.021)}{\hphantom{-}1.00} 
		\end{bmatrix}'$
		&
		
		$\begin{bmatrix}
	 \underset{(0.033)}{\hphantom{-}0.00}  &
		\underset{(0.037)}{-0.46} & 
		\underset{(0.081)}{\hphantom{-}0.96} 
		\end{bmatrix}'$
		&
		
		$\begin{bmatrix}  
		\underset{(0.042)}{\hphantom{-}0.00}  &
		\underset{(0.047)}{-0.45} & 
		\underset{(0.109)}{\hphantom{-}0.92} 
		\end{bmatrix}'$

		\\   
proxy (weighting)  &
		$\begin{bmatrix} 
		\underset{(0.008)}{\hphantom{-}0.00}  &
		\underset{(0.009)}{-0.50} & 
		\underset{(0.021)}{\hphantom{-}1.00} 
		\end{bmatrix}'$
		&
		
		$\begin{bmatrix} 
		\underset{(0.032)}{\hphantom{-}0.00}  &
		\underset{(0.037)}{-0.46} & 
		\underset{(0.081)}{\hphantom{-}0.96} 
		\end{bmatrix}'$
		&
		
		$\begin{bmatrix}   
		\underset{(0.040)}{\hphantom{-}0.00}  &
		\underset{(0.047)}{-0.45} & 
		\underset{(0.107)}{\hphantom{-}0.92} 
		\end{bmatrix}'$

	\end{tabular} 
	\footnotesize{\textit{Note: The true impact of the shock $\varepsilon_{\tau, t}$ is $\begin{bmatrix}   
			0  & -0.5   & 1
			\end{bmatrix}'$. 
   The average and MSE of the Bayesian estimators are calculated based on the median of the posterior of B in each simulation.}}
\end{table}

\begin{table}
	\caption{Comparing the proxy approaches - Coverage  and average length of  $68$\% credible bands of the estimated impact of $\varepsilon_{\tau ,t}$.}
	\label{Table: Coveragenormal} 
	\begin{tabular}{ c  | c    c   c       }
		&  proxy  1  &  proxy 2  &   proxy 3  \\  
			&$z_{\tau ,t}= \varepsilon_{\tau, t}  +	1 \eta_t$&$z_{\tau ,t}= \varepsilon_{\tau, t}  +	3\eta_t $&$z_{\tau ,t}= \varepsilon_{\tau, t}  +	4\eta_t  $
		\\ \hline 
		$T=250$&&&
  		\\
		
proxy (augmented)  &
		$\begin{bmatrix} 
		\underset{(0.018)}{0.69} & 
		\underset{(0.019)}{0.67} & 
		\underset{(0.028)}{0.67} 
		\end{bmatrix}'$
		
		&
		
		$\begin{bmatrix}          
		\underset{(0.037)}{0.68} & 
		\underset{(0.039)}{0.68} & 
		\underset{(0.057)}{0.68} 
		\end{bmatrix}'$
		
		&
		
		$\begin{bmatrix} 
		\underset{(0.046)}{0.71} & 
		\underset{(0.047)}{0.71} & 
		\underset{(0.070)}{0.71} 
		\end{bmatrix}'$
  
		\\
		
proxy (weighting)  &
		$\begin{bmatrix} 
		\underset{(0.018)}{0.69} & 
		\underset{(0.019)}{0.68} & 
		\underset{(0.029)}{0.68} 
		\end{bmatrix}'$
		
		&
		
		$\begin{bmatrix}           
		\underset{(0.038)}{0.69} & 
		\underset{(0.039)}{0.68} & 
		\underset{(0.058)}{0.68} 
		\end{bmatrix}'$
		
		&
		
		$\begin{bmatrix} 
		\underset{(0.046)}{0.72} & 
		\underset{(0.047)}{0.71} & 
		\underset{(0.071)}{0.71} 
 
		\end{bmatrix}'$

	\end{tabular}  
 	\floatfoot{
% 	\textit{Note:}    ......
           }
\end{table}

\begin{table}
	\caption{$t$-distributed shocks -  Average point estimates and MSE for the impact of $\varepsilon_{\tau ,t}$.     }
	\label{Table: Finite sample performance - t  } 
	\begin{tabular}{ c  |      c   c       }
	    & exogenous proxy   & endogenous  proxy  \\  
		 &$z_{\tau, t}= \varepsilon_{\tau, t}  +	\eta_t  $&$z_{\tau ,t}= \varepsilon_{\tau, t} -0.37 	\varepsilon_{y, t} +	\eta_t  $
		\\ \hline
		$T=250$&& 
		\\
		proxy (frequentist)  &

		$\begin{bmatrix}  
		\underset{(0.008)}{\hphantom{-}0.00}  &
		\underset{(0.009)}{-0.50} & 
		\underset{(0.023)}{\hphantom{-}0.99} 
		\end{bmatrix}'$&

		$\begin{bmatrix}  
		\underset{(0.008)}{\hphantom{-}0.00}  &
		\underset{(0.107)}{-0.82} & 
		\underset{(0.386)}{\hphantom{-}0.41} 
		\end{bmatrix}'$
		
		\\    
non-Gaussian  &
		$\begin{bmatrix} 
		\underset{(0.022)}{\hphantom{-}0.00}  &
		\underset{(0.026)}{-0.46} & 
		\underset{(0.063)}{\hphantom{-}0.92} 
		\end{bmatrix}'$
		
		&
		
		$\begin{bmatrix}          
		\underset{(0.022)}{\hphantom{-}0.00}  &
		\underset{(0.027)}{-0.46} & 
		\underset{(0.069)}{\hphantom{-}0.91} 
		\end{bmatrix}'$

		\\   
\begin{tabular}{@{}c@{}}non-Gaussian  \\proxy weighting\end{tabular}    &

			$\begin{bmatrix}   
		\underset{(0.009)}{\hphantom{-}0.00}  &
		\underset{(0.010)}{-0.50} & 
		\underset{(0.025)}{\hphantom{-}0.99} 
		\end{bmatrix}'$
		&
		
		$\begin{bmatrix}   
		\underset{(0.010)}{\hphantom{-}0.00}  &
		\underset{(0.040)}{-0.65} & 
		\underset{(0.142)}{\hphantom{-}0.71} 
		\end{bmatrix}'$
		
		\\ \hline
		$T=800$&& 
		\\
		
		proxy (frequentist)  &

		$\begin{bmatrix}
		\underset{(0.003)}{\hphantom{-}0.00}  &
		\underset{(0.003)}{-0.50} & 
		\underset{(0.007)}{\hphantom{-}1.00} 
		\end{bmatrix}'$
		&
		
		$\begin{bmatrix}  
		\underset{(0.003)}{\hphantom{-}0.00}  &
		\underset{(0.103)}{-0.82} & 
		\underset{(0.355)}{\hphantom{-}0.41} 
		\end{bmatrix}'$
		
		\\
non-Gaussian  &

		$\begin{bmatrix}      
		\underset{(0.008)}{\hphantom{-}0.00}  &
		\underset{(0.008)}{-0.49} & 
		\underset{(0.018)}{\hphantom{-}0.99} 
		\end{bmatrix}'$
		
		&
		
		$\begin{bmatrix}  
		\underset{(0.008)}{\hphantom{-}0.00}  &
		\underset{(0.008)}{-0.50} & 
		\underset{(0.018)}{\hphantom{-}0.99} 
		\end{bmatrix}'$
		
		\\
\begin{tabular}{@{}c@{}}non-Gaussian  \\proxy weighting\end{tabular}    &

		$\begin{bmatrix}  
		\underset{(0.003)}{\hphantom{-}0.00}  &
		\underset{(0.003)}{-0.50} & 
		\underset{(0.008)}{\hphantom{-}1.00} 
		\end{bmatrix}'$
		&
		$\begin{bmatrix} 
		\underset{(0.004)}{\hphantom{-}0.00}  &
		\underset{(0.012)}{-0.54} & 
		\underset{(0.033)}{\hphantom{-}0.92} 
		\end{bmatrix}'$
	\end{tabular}   
\end{table}

\begin{table}
	\caption{MC with $t$-distributed shocks - Coverage and average length of  $68$\% credible bands.}
	\label{Table: Finite sample performance coverage -t} 
	\begin{tabular}{ c  | c    c   c       }
		    & exogenous proxy   & endogenous  proxy  \\  
		 &$z_{\tau, t}= \varepsilon_{\tau, t}  +	\eta_t  $&$z_{\tau ,t}= \varepsilon_{\tau, t} -0.37 	\varepsilon_{y, t} +	\eta_t  $
		\\ \hline
		$T=250$&& 
		\\
		
non-Gaussian  &

		$\begin{bmatrix}          
	\underset{(0.037)}{0.78} & 
		\underset{(0.039)}{0.80} & 
		\underset{(0.058)}{0.78} 
		\end{bmatrix}'$
		
		&
		
		$\begin{bmatrix} 
		\underset{(0.037)}{0.78} & 
		\underset{(0.039)}{0.80} & 
		\underset{(0.057)}{0.76} 
		\end{bmatrix}'$
		
		\\   
\begin{tabular}{@{}c@{}}non-Gaussian  \\proxy weighting\end{tabular}    &

		$\begin{bmatrix}  
		\underset{(0.019)}{0.74} & 
		\underset{(0.020)}{0.74} & 
		\underset{(0.031)}{0.72} 
		\end{bmatrix}'$
		&
		
		$\begin{bmatrix}   
		\underset{(0.021)}{0.74} & 
		\underset{(0.027)}{0.45} & 
		\underset{(0.046)}{0.45} 
		\end{bmatrix}'$
		
		\\ \hline
		$T=800$&& 
		
		\\
non-Gaussian  &

		$\begin{bmatrix}      
		\underset{(0.016)}{0.69} & 
		\underset{(0.016)}{0.69} & 
		\underset{(0.025)}{0.67} 
		\end{bmatrix}'$
		
		&
		
		$\begin{bmatrix}  
			\underset{(0.016)}{0.68} & 
		\underset{(0.016)}{0.69} & 
		\underset{(0.025)}{0.71} 
		\end{bmatrix}'$
		
		\\
\begin{tabular}{@{}c@{}}non-Gaussian  \\proxy weighting\end{tabular}    &

		$\begin{bmatrix}  
		\underset{(0.010)}{0.67} & 
		\underset{(0.011)}{0.68} & 
		\underset{(0.016)}{0.69} 
		\end{bmatrix}'$
		&
		
		$\begin{bmatrix}    
		\underset{(0.011)}{0.72} & 
		\underset{(0.017)}{0.62} & 
		\underset{(0.027)}{0.64} 
		\end{bmatrix}'$
	\end{tabular}  
\end{table}

 \begin{table}
	\caption{Missing data proxy - Average point estimates and MSE for the impact of $\varepsilon_{\tau ,t}$ ($T=250$).     }
	\label{Table: Finite sample performance - missing} 
	\begin{tabular}{ c  |      c   c       }
		    & exogenous proxy   & endogenous  proxy  \\  
		 &$z_{\tau, t}= \varepsilon_{\tau, t}  +	\eta_t  $&$z_{\tau ,t}= \varepsilon_{\tau, t} -0.37 	\varepsilon_{y, t} +	\eta_t  $
		\\ \hline
		 
		proxy (frequentist)  &

		$\begin{bmatrix}
		\underset{(0.016)}{\hphantom{-}0.00}  &
		\underset{(0.017)}{-0.49} & 
		\underset{(0.041)}{\hphantom{-}0.99} 
		\end{bmatrix}'$
		&
		
		$\begin{bmatrix}  
		\underset{(0.015)}{\hphantom{-}0.00}  &
		\underset{(0.106)}{-0.81} & 
		\underset{(0.410)}{\hphantom{-}0.41} 
		\end{bmatrix}'$
		
		\\    
non-Gaussian  &

		$\begin{bmatrix}          
		\underset{(0.017)}{\hphantom{-}0.00}  &
		\underset{(0.020)}{-0.47} & 
		\underset{(0.047)}{\hphantom{-}0.97} 
		\end{bmatrix}'$
		
		&
		
		$\begin{bmatrix} 
		\underset{(0.017)}{\hphantom{-}0.00}  &
		\underset{(0.020)}{-0.47} & 
		\underset{(0.047)}{\hphantom{-}0.97} 
		\end{bmatrix}'$
		
		\\   
\begin{tabular}{@{}c@{}}non-Gaussian  \\proxy weighting\end{tabular}    & 
		
		$\begin{bmatrix} 
		\underset{(0.010)}{\hphantom{-}0.00}  &
		\underset{(0.012)}{-0.49} & 
		\underset{(0.028)}{\hphantom{-}1.00} 
		\end{bmatrix}'$
		&
		
		$\begin{bmatrix}   
		\underset{(0.011)}{\hphantom{-}0.00}  &
		\underset{(0.023)}{-0.58} & 
		\underset{(0.076)}{\hphantom{-}0.84} 
		\end{bmatrix}'$

	\end{tabular}  
\end{table}

\begin{table}
	\caption{MC with missing data proxy -  Coverage and average length of  $68$\% credible bands ($T=250$).}
	\label{Table: Finite sample performance coverage - missing} 
	\begin{tabular}{ c  |      c   c       }
		    & exogenous proxy   & endogenous  proxy  \\  
		 &$z_{\tau, t}= \varepsilon_{\tau, t}  +	\eta_t  $&$z_{\tau ,t}= \varepsilon_{\tau, t} -0.37 	\varepsilon_{y, t} +	\eta_t  $
		\\ \hline 
		
non-Gaussian  &

		$\begin{bmatrix}          
		\underset{(0.026)}{0.71} & 
		\underset{(0.027)}{0.70} & 
		\underset{(0.041)}{0.71} 
		\end{bmatrix}'$
		
		&
		
		$\begin{bmatrix} 
		\underset{(0.026)}{0.72} & 
		\underset{(0.027)}{0.71} & 
		\underset{(0.041)}{0.71} 
		\end{bmatrix}'$
		
		\\   
\begin{tabular}{@{}c@{}}non-Gaussian  \\proxy weighting\end{tabular}    &

		$\begin{bmatrix}  
		\underset{(0.019)}{0.68} & 
		\underset{(0.020)}{0.69} & 
		\underset{(0.031)}{0.68} 
		\end{bmatrix}'$
		&
		
		$\begin{bmatrix}   
		\underset{(0.020)}{0.70} & 
		\underset{(0.024)}{0.57} & 
		\underset{(0.039)}{0.55} 
		\end{bmatrix}'$

	\end{tabular}  
\end{table}

 \begin{table}
	\caption{Two proxies - Average point estimates and MSE for the impact of $\varepsilon_{\tau ,t}$ ($T=250$).     }
	\label{Table: Finite sample performance - 2proxy} 
	\begin{tabular}{ c  |      c   c       }
		    & exogenous proxy   & endogenous  proxy  \\  
		 &$z_{\tau, t}= \varepsilon_{\tau, t}  +	\eta_t  $&$z_{\tau ,t}= \varepsilon_{\tau, t} -0.37 	\varepsilon_{y, t} +	\eta_t  $ \\  
		 &$z_{y, t}= \varepsilon_{y, t}  +	\eta_t  $&$z_{y, t}= \varepsilon_{y, t}  +	\eta_t $
		\\ \hline
		 
		proxy (frequentist)  &

		$\begin{bmatrix}
		\underset{(0.009)}{\hphantom{-}0.00}  &
		\underset{(0.009)}{-0.50} & 
		\underset{(0.022)}{\hphantom{-}0.99} 
		\end{bmatrix}'$
		&
		
		$\begin{bmatrix}  
		\underset{(0.007)}{-0.01} & 
		\underset{(0.106)}{-0.81} & 
		\underset{(0.383)}{\hphantom{-}0.41} 
		\end{bmatrix}'$
		
		\\    
non-Gaussian  &

		$\begin{bmatrix}          
		\underset{(0.017)}{\hphantom{-}0.00}  &
		\underset{(0.020)}{-0.48} & 
		\underset{(0.047)}{\hphantom{-}0.95} 
		\end{bmatrix}'$
		
		&
		
		$\begin{bmatrix} 
		\underset{(0.018)}{\hphantom{-}0.00}  &
		\underset{(0.019)}{-0.48} & 
		\underset{(0.049)}{\hphantom{-}0.95} 
		\end{bmatrix}'$
		
		\\   
\begin{tabular}{@{}c@{}}non-Gaussian  \\proxy weighting\end{tabular}    & 
		
		$\begin{bmatrix} 
		\underset{(0.008)}{\hphantom{-}0.00}  &
		\underset{(0.005)}{-0.51} & 
		\underset{(0.012)}{\hphantom{-}1.01} 
		\end{bmatrix}'$
		&
		
		$\begin{bmatrix}   
		\underset{(0.008)}{\hphantom{-}0.00}  &
		\underset{(0.016)}{-0.57} & 
		\underset{(0.036)}{\hphantom{-}0.90} 
		\end{bmatrix}'$

	\end{tabular}  
\end{table}

\begin{table}
	\caption{Two proxies - Coverage and average length of  $68$\% credible bands ($T=250$).}
	\label{Table: Finite sample performance coverage - 2proxy} 
	\begin{tabular}{ c  |      c   c       }
		    & exogenous proxy   & endogenous  proxy  \\  
		 &$z_{\tau, t}= \varepsilon_{\tau, t}  +	\eta_t  $&$z_{\tau ,t}= \varepsilon_{\tau, t} -0.37 	\varepsilon_{y, t} +	\eta_t  $\\  
		 &$z_{y, t}= \varepsilon_{y, t}  +	\eta_t  $&$z_{y, t}= \varepsilon_{y, t}  +	\eta_t $
		\\ \hline 
		
non-Gaussian  &

		$\begin{bmatrix}          
		\underset{(0.025)}{0.71} & 
		\underset{(0.027)}{0.70} & 
		\underset{(0.041)}{0.70} 
		\end{bmatrix}'$
		
		&
		
		$\begin{bmatrix} 
		\underset{(0.026)}{0.71} & 
		\underset{(0.027)}{0.70} & 
		\underset{(0.041)}{0.70} 
		\end{bmatrix}'$
		
		\\   
\begin{tabular}{@{}c@{}}non-Gaussian  \\proxy weighting\end{tabular}    &

		$\begin{bmatrix}  
		\underset{(0.016)}{0.65} & 
		\underset{(0.015)}{0.73} & 
		\underset{(0.023)}{0.74} 
		\end{bmatrix}'$
		&
		
		$\begin{bmatrix}   
		\underset{(0.016)}{0.68} & 
		\underset{(0.021)}{0.61} & 
		\underset{(0.033)}{0.66} 
		\end{bmatrix}'$

	\end{tabular}  
\end{table}

\begin{table}
	\caption{Different choices for $\bm \Sigma_z$, T=800}.
	\label{TableSigma} 
	\begin{tabular}{ c  |      c   c       }
		    & exogenous proxy   & endogenous  proxy  \\  
		&$z_{\tau ,t}= \varepsilon_{\tau, t}  +	\eta_t$ &
  $z_{\tau ,t}= \varepsilon_{\tau, t} -0.37 	\varepsilon_{y, t} +	\eta_t$
		\\ \hline
\begin{tabular}{@{}c@{}}non-Gaussian  \\proxy weighting\end{tabular}    &

		$\begin{bmatrix}          
			\underset{(0.027)}{\hphantom{-}0.00}  &
		\underset{(0.031)}{-0.48} & 
		\underset{(0.066)}{\hphantom{-}0.96} 
		\end{bmatrix}'$
		
		&
		
		$\begin{bmatrix} 
		\underset{(0.032)}{\hphantom{-}0.01}  &
		\underset{(0.081)}{-0.67} & 
		\underset{(0.270)}{\hphantom{-}0.60} 
		\end{bmatrix}'$
		
		\\    
\begin{tabular}{@{}c@{}}non-Gaussian  \\proxy weighting \\ Sigma Est v2 \end{tabular}    & 
		
		$\begin{bmatrix}  
		\underset{(0.024)}{\hphantom{-}0.00}  &
		\underset{(0.030)}{-0.49} & 
		\underset{(0.074)}{\hphantom{-}0.95} 
		\end{bmatrix}'$
		&
		
		$\begin{bmatrix}   
			\underset{(0.033)}{\hphantom{-}0.00}  &
		\underset{(0.084)}{-0.67} & 
		\underset{(0.276)}{\hphantom{-}0.60} 
 
		\end{bmatrix}'$
		
				\\   
\begin{tabular}{@{}c@{}}non-Gaussian  \\proxy weighting \\ Sigma Est v3 \end{tabular}    & 
		
		$\begin{bmatrix}  
		\underset{(0.039)}{\hphantom{-}0.00}  &
		\underset{(0.041)}{-0.49} & 
		\underset{(0.099)}{\hphantom{-}0.94} 
		\end{bmatrix}'$
		&
		
		$\begin{bmatrix}   
		\underset{(0.040)}{\hphantom{-}0.01}  &
		\underset{(0.086)}{-0.64} & 
		\underset{(0.258)}{\hphantom{-}0.64} 
		\end{bmatrix}'$
        
	\end{tabular}  
\end{table}

 \begin{table}
	\caption{Alternative DGP - note average and mse *100 for readability - Average point estimates and MSE for the impact of $\varepsilon_{\tau ,t}$ ($T=250$).     }
	\label{TableMSErealstic} 
	\begin{tabular}{ c  |      c   c       }
		    & exogenous proxy   & endogenous  proxy   \\  
		 &$z_{\tau, t}= 0.0121 \varepsilon_{\tau, t}  +	0.0121 \eta_t  $
   &$z_{\tau ,t}= 0.0121 \varepsilon_{\tau, t} -0.006 	\varepsilon_{y, t} +0.0121	\eta_t  $
		\\ \hline 
		 
		proxy (frequentist)  &

		$\begin{bmatrix}
		\underset{(0.001)}{\hphantom{-}1.18}  &
		\underset{(0.000)}{\hphantom{-}0.13} & 
		\underset{(0.000)}{-0.41}  
		\end{bmatrix}'$
		&
		
		$\begin{bmatrix}  
		\underset{(0.014)}{\hphantom{-}0.08}  &
		\underset{(0.000)}{\hphantom{-}0.11} & 
		\underset{(0.001)}{-0.69} 
		\end{bmatrix}'$
		
		\\    
non-Gaussian  &

		$\begin{bmatrix}          
		\underset{(0.001)}{\hphantom{-}1.20}  &
		\underset{(0.001)}{\hphantom{-}0.12} & 
		\underset{(0.000)}{-0.41} 
		\end{bmatrix}'$
		
		&
		
		$\begin{bmatrix} 
		\underset{(0.001)}{\hphantom{-}1.20}  &
		\underset{(0.001)}{\hphantom{-}0.12} & 
		\underset{(0.000)}{-0.41} 
		\end{bmatrix}'$
		
		\\   
\begin{tabular}{@{}c@{}}non-Gaussian  \\proxy weighting\end{tabular}    & 
		
		$\begin{bmatrix} 
		\underset{(0.000)}{\hphantom{-}1.22}  &
		\underset{(0.000)}{\hphantom{-}0.13} & 
		\underset{(0.000)}{-0.42} 
		\end{bmatrix}'$
		&
		
		$\begin{bmatrix}   
		\underset{(0.001)}{\hphantom{-}1.08}  &
		\underset{(0.000)}{\hphantom{-}0.13} & 
		\underset{(0.000)}{-0.45} 
		\end{bmatrix}'$

	\end{tabular}  
 	\footnotesize{\textit{Note: The true impact of the shock $\varepsilon_{\tau, t}$ is $\begin{bmatrix}   
			0.0121  & 0.0014   & -0.0042
			\end{bmatrix}'*100$.  }}
\end{table}  

\begin{table}
	\caption{Alternative DGP - note length *100 for readability - Coverage and average length of  $68$\% credible bands ($T=250$).}
	\label{TableCoveragerealstic} 
	\begin{tabular}{ c  |      c   c       }
		    & exogenous proxy   & endogenous  proxy  \\  
		 &$z_{\tau, t}= 0.0121 \varepsilon_{\tau, t}  +	0.0121 \eta_t  $
   &$z_{\tau ,t}= 0.0121 \varepsilon_{\tau, t} -0.006 	\varepsilon_{y, t} +0.0121	\eta_t  $
		\\ \hline 
non-Gaussian  &

		$\begin{bmatrix}          
		\underset{(0.055)}{0.66} & 
		\underset{(0.055)}{0.68} & 
		\underset{(0.019)}{0.66} 
		\end{bmatrix}'$
		
		&
		
		$\begin{bmatrix} 
		\underset{(0.055)}{0.66} & 
		\underset{(0.055)}{0.67} & 
		\underset{(0.018)}{0.67} 
		\end{bmatrix}'$
		
		\\   
\begin{tabular}{@{}c@{}}non-Gaussian  \\proxy weighting\end{tabular}    & 
		
		$\begin{bmatrix}  
		\underset{(0.040)}{0.68} & 
		\underset{(0.038)}{0.69} & 
		\underset{(0.013)}{0.68} 
		\end{bmatrix}'$
		&
		
		$\begin{bmatrix}   
		\underset{(0.057)}{0.61} & 
		\underset{(0.042)}{0.68} & 
		\underset{(0.018)}{0.63} 
		\end{bmatrix}'$

	\end{tabular}  
\end{table}

\begin{table}
	\caption{Alternative DGP - note length *100 for readability - Coverage and average length of  $68$\% credible bands ($T=250$).}
	\label{TableCoveragerealstic} 
	\begin{tabular}{ c  |      c   c       }
		    & exogenous proxy   & endogenous  proxy  \\  
		 &$z_{\tau, t}= 0.0121 \varepsilon_{\tau, t}  +	0.0121 \eta_t  $
   &$z_{\tau ,t}= 0.0121 \varepsilon_{\tau, t} -0.006 	\varepsilon_{y, t} +0.0121	\eta_t  $
		\\ \hline 
non-Gaussian  &

		$\begin{bmatrix}          
		\underset{(0.055)}{0.66} & 
		\underset{(0.055)}{0.68} & 
		\underset{(0.019)}{0.66} 
		\end{bmatrix}'$
		
		&
		
		$\begin{bmatrix} 
		\underset{(0.055)}{0.66} & 
		\underset{(0.055)}{0.67} & 
		\underset{(0.018)}{0.67} 
		\end{bmatrix}'$
		
		\\   
\begin{tabular}{@{}c@{}}non-Gaussian  \\proxy weighting\end{tabular}    & 
		
		$\begin{bmatrix}  
		\underset{(0.040)}{0.68} & 
		\underset{(0.038)}{0.69} & 
		\underset{(0.013)}{0.68} 
		\end{bmatrix}'$
		&
		
		$\begin{bmatrix}   
		\underset{(0.057)}{0.61} & 
		\underset{(0.042)}{0.68} & 
		\underset{(0.018)}{0.63} 
		\end{bmatrix}'$

	\end{tabular}  
\end{table}

\clearpage
\newpage

\section{Estimated cyclical elasticities of tax revenues and government spending}

Figure \ref{fig:elasticities comp} reports the posterior of the cyclical elasticities of tax revenues ($\Theta_{y}$) and government spending ($\gamma_{y}$) respectively. As discussed in \citet{caldara2017analytics}, these two parameters crucially determine the size of the estimated multipliers. The tax multiplier increases in the size of the elasticity of tax revenues. In contrast, the smaller the elasticity of government spending, the larger the spending multiplier. 

The intuition for this relation can be summarized as follows. There exists a positive correlation between government spending and output in the data, which any identification approach decomposes into a fraction explained by government spending shocks and a fraction explained by the remaining shocks of the SVAR. The timing assumption imposed by \citet{mertens2014reconciliation} implies that government spending does not respond contemporaneously to any other shock. Therefore, all the positive contemporaneous relations in the data must be explained by the government spending shock. This leads to a spending multiplier of about one. If the systematic response of government spending increases, the remaining shocks explain a larger part of the positive correlation and the spending multiplier decreases. In contrast, if the systematic response decreases, that is, turns negative, the multiplier increases. A similar reasoning applies to the identification of tax shocks.

As shown in Figure \ref{fig:elasticities comp}, the different identification approaches provide different estimates for the elasticities of tax revenues and government spending. The fiscal proxy SVAR based on \cite{mertens2014reconciliation} leads to a tax revenue elasticity centered around three. 
The estimated elasticity is considerably smaller for the non-fiscal proxy SVAR and our non-Gaussian proxy weighting model. While the non-fiscal proxy approach results in an elasticity centered above two, our proposed non-Gaussian model implies an elasticity centered below two.\footnote{The value of the estimated elasticity of our non-Gaussian model is close to the one calculated by \citet{Follette2010} based on institutional details of tax revenues. In particular, \citet{Follette2010} estimate the elasticity of tax revenues with respect to output for the federal government, and obtain a value of 1.6 for the period 1986–2008 and 1.4 for 1960–85.} Given the positive relationship between the tax elasticity and the tax multiplier, this explains the large differences between the estimated tax multipliers across identification strategies. With a larger tax elasticity, the fiscal proxy SVAR produces the largest tax multiplier, whereas our data driven strategy leads to the smallest tax elasticity and tax multiplier. The picture is similar for the spending elasticity and the government spending multiplier. By adopting a zero restriction on the spending elasticity, the fiscal proxy SVAR results in the largest spending elasticity across identification strategies. In contrast, the non-fiscal proxy SVAR produces the smallest spending elasticity and thus the largest spending multiplier. Our non-Gaussian proxy weighting model implies a value in between the estimates of the other approaches and thus the estimated government spending multiplier is larger (smaller) than the one obtained by the fiscal proxy (non-fiscal proxy) approach.  

 \begin{figure}[h]
	\centering
		\caption{Posterior of tax revenue elasticity $\Theta_{y}$ and government spending elasticity $\gamma_{y}$  }\label{fig:elasticities comp} 
	\includegraphics[width=1\textwidth]{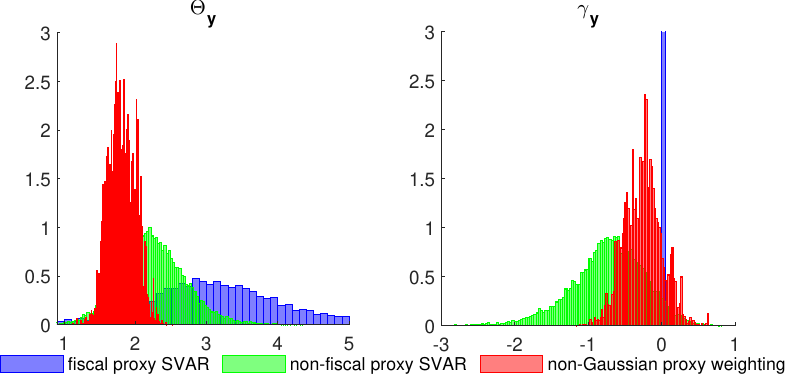} 
  \floatfoot{The figure compares the Posterior of tax revenue elasticity $\Theta_{y}$ and government spending elasticity $\gamma_{y}$ between our proxy shrinkage VAR with the non-fiscal proxy SVAR proposed by \citet{caldara2017analytics} as well as the fiscal proxy SVAR from \cite{mertens2014reconciliation}. }
\end{figure}

 \clearpage
 \newpage

\section{Empirical Robustness Checks}\label{Appendix:rob}

Our main finding that the government spending multiplier is larger than the tax multiplier is actually highly robust to modifications of the baseline empirical specification. Figure \ref{fig:rob_CKdata} shows that the responses are similar if we use the original data on output, tax revenues, and government spending from \citet{caldara2017analytics}. The main difference to the data of our baseline model is that \citet{caldara2017analytics} define government spending as the sum of federal and state and local spending whereas our baseline measure, following \citet{mertens2014reconciliation}, only includes federal spending. The central finding also remains when controlling for the monetary policy response to the fiscal stimulus by including consumer price inflation and the 3-month T-bill rate as additional endogenous variables in the VAR (see Figure \ref{fig:rob_inflation}). This five variable VAR is similar to the baseline model considered by \citet{caldara2017analytics}. We further show that the responses are fairly stable across different subsamples. Figures \ref{fig:rob_pre80} and \ref{fig:rob_post80} report the responses when splitting the sample into periods before and after 1980, respectively. While both multipliers are larger before 1980, the difference between the government spending and tax multiplier is positive in both periods. The main qualitative finding is also present when extending the sample until 2019 as shown in Figure \ref{fig:application_extendedsample}.\footnote{Because the tax proxy end in 2006, we follow previous literature, see, e.g. \cite{Paul2020} or \cite{Känzig2023}, and set the proxy to zero for the remaining periods.} Finally, we address the issue of anticipation by adding as a fourth endogenous variable, in turn: a measure of expected future taxes that is implied by tax exempt municipal bond yields and perfect arbitrage, constructed by \citet{Leeper2012}; and the defense spending news variable, which contains professional forecasters' projections of the path of future military spending, constructed by \citet{Ramey2011}. Figures \ref{fig:rob_anticipaiton_tax} and \ref{fig:rob_anticiaption_spending} present the respective responses.

Figure \ref{fig:benchmarkBcenter} displays the  output multipliers for the baseline specification used in the main text. In contrast to the results reported in the main text, the shocks are now labeled based on restricting the posterior draws to unique sign-permutation representatives centered around a first step estimator, see Section 2.2.
The estimated output multipliers are similar to those in the main text, indicating that the results are not driven by the labeling approach. 
For the first step estimator, we estimate the VAR with a maximum likelihood estimator assuming independent and skewed $t$-distributed shocks to estimate the simultaneous interaction under the restriction that government spending does not react contemporaneously
to tax and output shocks. The restrictions identify and label the government spending
shock. The remaining two shocks are labeled based on the tax and TFP proxies. That is
the shock with the highest correlation with the tax proxy is labeled as the tax shock and
the shock with the highest correlation with the TFP proxy is labeled as the output shock.
Importantly, the first step estimator is only used to determine the center of the space of
admissible $\bm B$ and our model can deviate from the government spending restrictions as well
as the “highest correlation” restrictions used to label the first step estimator. 

 Figure  \ref{fig:fiscalfreq} and Figure \ref{fig:nonfiscalfreq} demonstrate that our Gaussian proxy weighting approach for the fiscal and non-fiscal SVAR  yields similar results compared to a traditional frequentist moment-based proxy VAR estimation. In both figures, the dotted lines display the point estimators for the output multipliers obtained through the frequentist moment-based proxy VAR estimation with the proxies and zero restrictions employed in \cite{mertens2014reconciliation} and \cite{caldara2017analytics}, respectively.

Figure \ref{fig:nGfiscal}, Figure \ref{fig:nGnonfiscal}, and Figure \ref{fig:nGspending} display the output multipliers in the baseline specification used in the main text. These multipliers are estimated using a non-Gaussian proxy weighting SVAR approach. In Figure \ref{fig:nGfiscal}, the estimation employs only the Tax proxy, in Figure \ref{fig:nGnonfiscal} the non-Gaussian proxy weighting SVAR uses only the TFP proxy, and in Figure \ref{fig:nGfiscal} the non-Gaussian proxy weighting SVAR uses only the government spending proxy.

Figure \ref{fig:irf_zprocess} shows the results when the proxies are cleaned by previous values of its own past, the past of the other proxy variables, and the past of the macroeconomic variables (output, tax revenues and government spending).  The results turn out to be robust.  Yet another robustness check shows the results for different choices of $\bm \Sigma_z$, see section 2.4 in the main paper, in figure \ref{fig:shocks_diff_DD_sigmadz2} and figure \ref{fig:shocks_diff_DD_sigmadz3}. The results do not change across different choices for $\bm \Sigma_z$.

 \begin{figure}[h!] 
	\centering
		\caption{Estimated output multipliers for the non-Gaussian proxy weighting SVAR, using data from \citet{caldara2017analytics}}\label{fig:rob_CKdata}
	\includegraphics[width=1\textwidth]{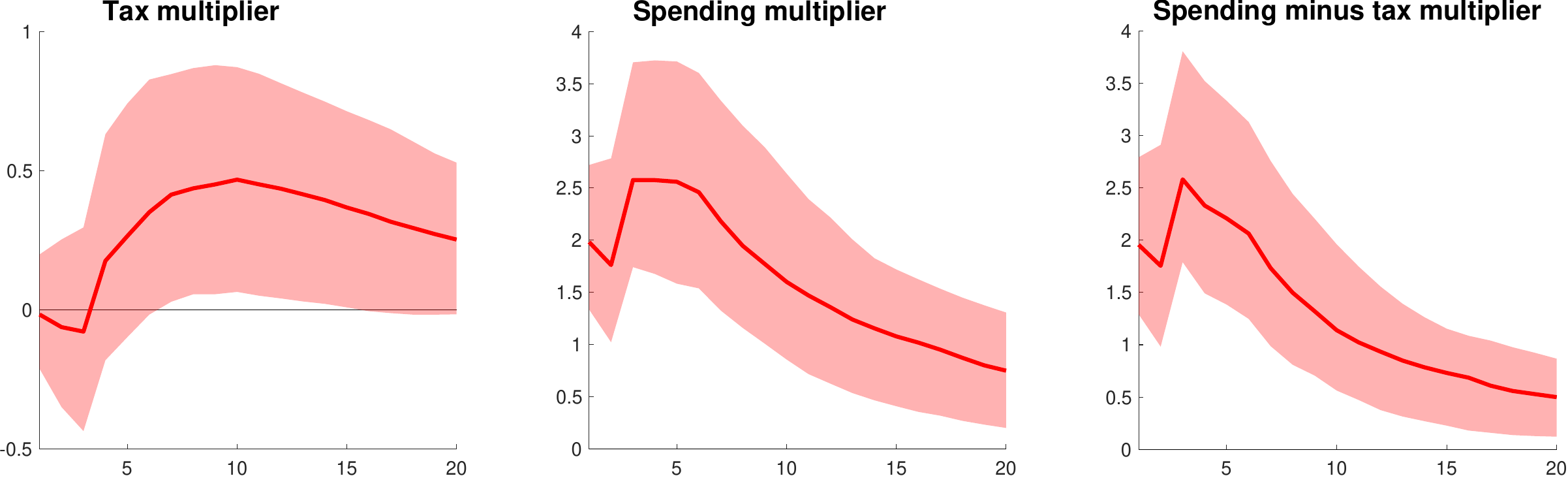}
  \floatfoot{The figure shows the posterior median as well as $68$\% credible bands for the impulse responses of output to a tax shock and a government spending shock. The right subfigure shows the posterior median as well as 68\% credible bands for the difference of the two output responses.}
\end{figure}   

\begin{figure} 
	\centering
		\caption{Estimated output multipliers for the non-Gaussian proxy weighting SVAR, controlling for inflation and interest rate}\label{fig:rob_inflation}
	\includegraphics[width=1\textwidth]{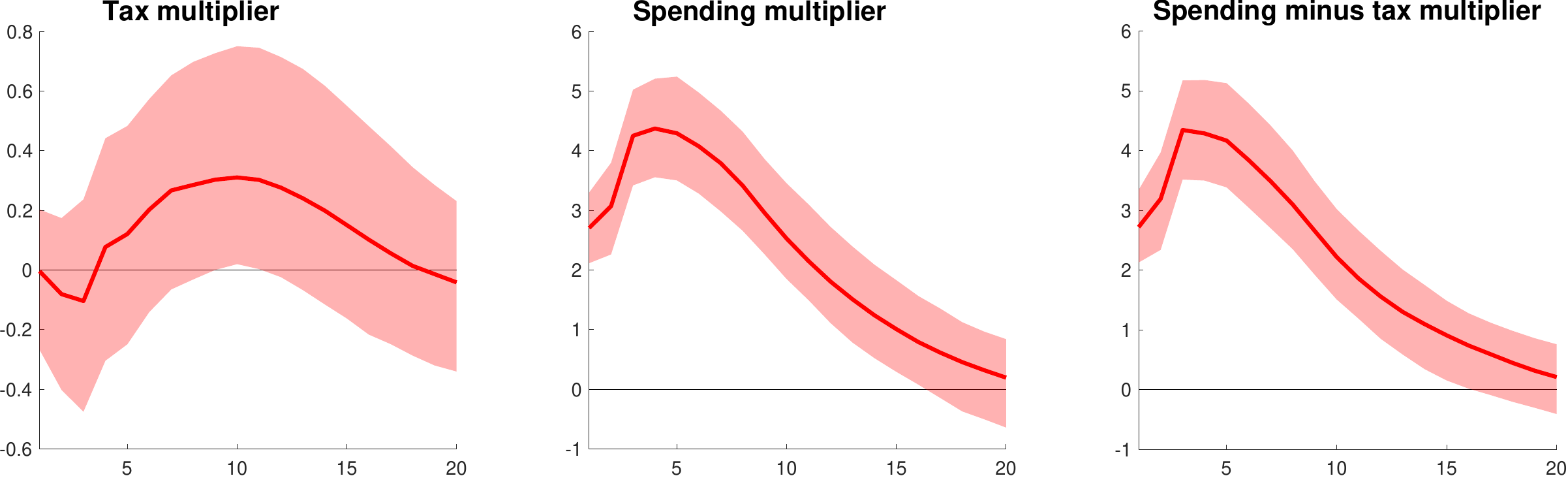}
  \floatfoot{The figure shows the posterior median as well as $68$\% credible bands for the impulse responses of output to a tax shock and a government spending shock. The right subfigure shows the posterior median as well as 68\% credible bands for the difference of the two output responses. }
\end{figure} 

 \begin{figure} 
	\centering
		\caption{Estimated output multipliers for the non-Gaussian proxy weighting SVAR, before 1980}\label{fig:rob_pre80}
	\includegraphics[width=1\textwidth]{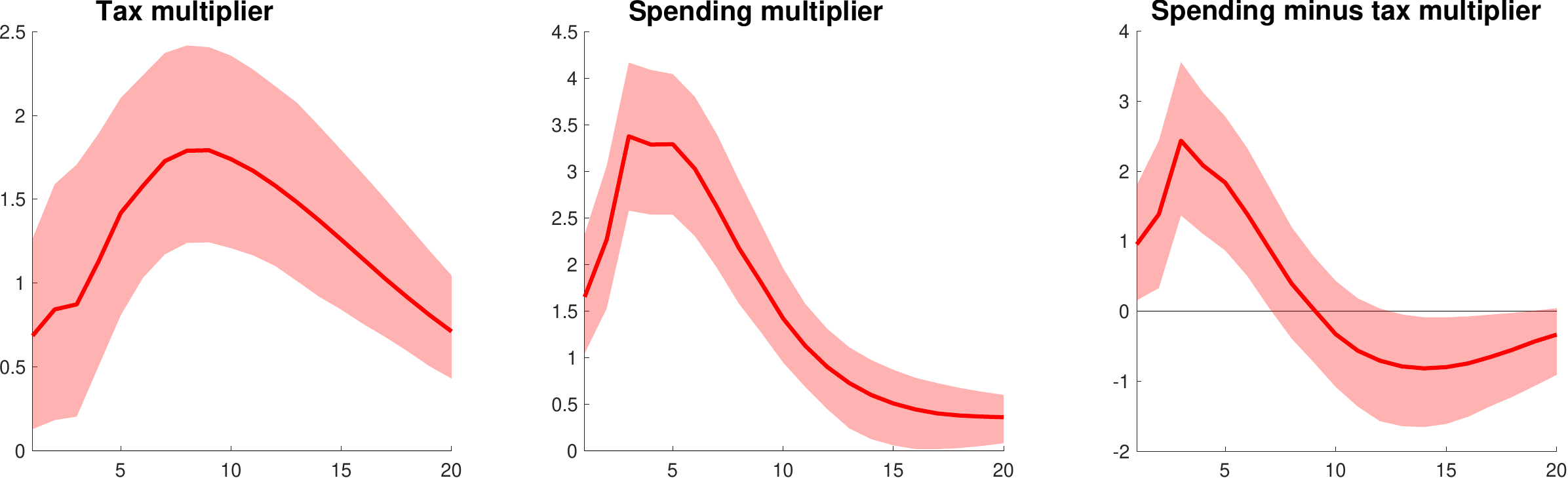}
  \floatfoot{The figure shows the posterior median as well as $68$\% credible bands for the impulse responses of output to a tax shock and a government spending shock. The right subfigure shows the posterior median as well as 68\% credible bands for the difference of the two output responses. }
\end{figure}   

 \begin{figure} 
	\centering
		\caption{Estimated output multipliers for the non-Gaussian proxy weighting SVAR, after 1980}\label{fig:rob_post80}
	\includegraphics[width=1\textwidth]{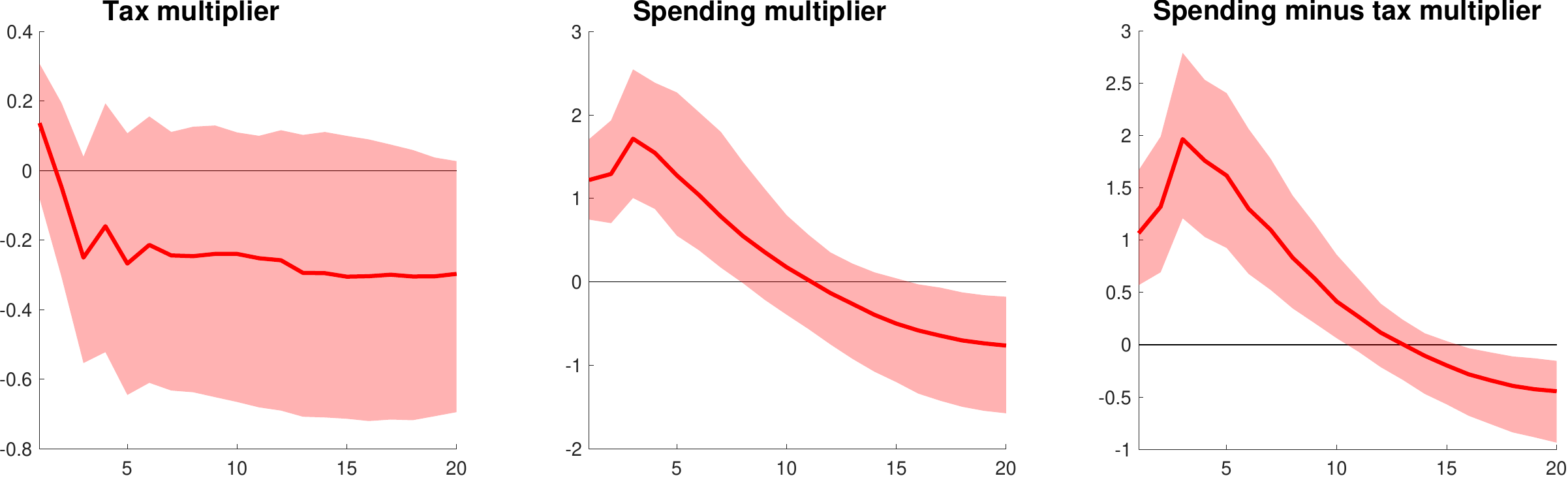}
  \floatfoot{The figure shows the posterior median as well as $68$\% credible bands for the impulse responses of output to a tax shock and a government spending shock. The right subfigure shows the posterior median as well as 68\% credible bands for the difference of the two output responses. }
\end{figure}   

\begin{figure}[h!] 
	\centering
		\caption{Estimated output multipliers for the non-Gaussian proxy weighting SVAR with data up to 2019}\label{fig:application_extendedsample}
	\includegraphics[width=1\textwidth]{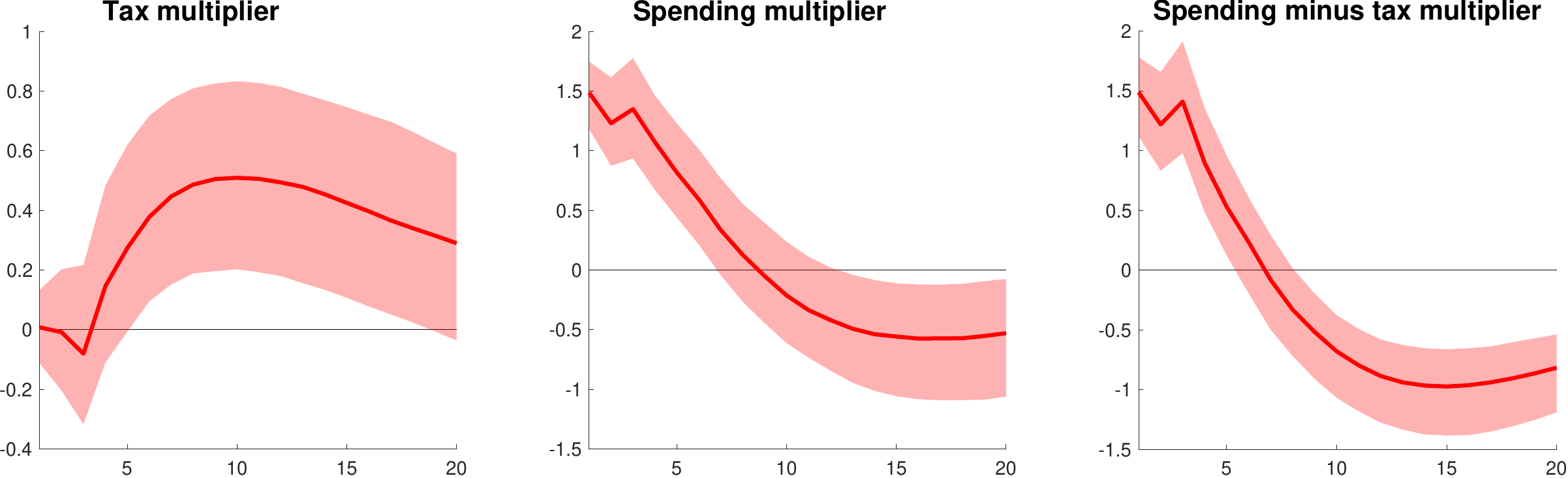}
  \floatfoot{The figure shows the posterior median as well as $68$\% credible bands for the impulse responses of output to a tax shock and a government spending shock. The right subfigure shows the posterior median as well as 68\% credible bands for the difference of the two output responses.}
\end{figure}

 \begin{figure} 
	\centering
		\caption{Estimated output multipliers for the non-Gaussian proxy weighting SVAR, controlling for implicit tax rate}\label{fig:rob_anticipaiton_tax}
	\includegraphics[width=1\textwidth]{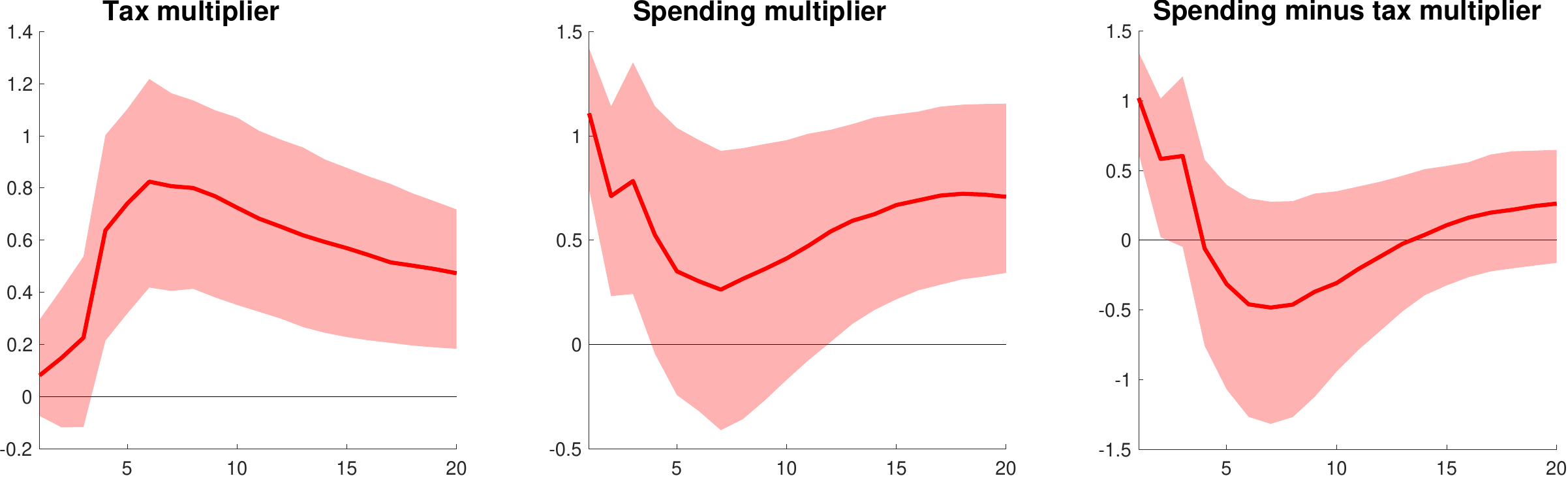}
  \floatfoot{The figure shows the posterior median as well as $68$\% credible bands for the impulse responses of output to a tax shock and a government spending shock. The right subfigure shows the posterior median as well as 68\% credible bands for the difference of the two output responses. }
\end{figure} 

 \begin{figure} 
	\centering
		\caption{Estimated output multipliers for the non-Gaussian proxy weighting SVAR, controlling for defense news}\label{fig:rob_anticiaption_spending}
	\includegraphics[width=1\textwidth]{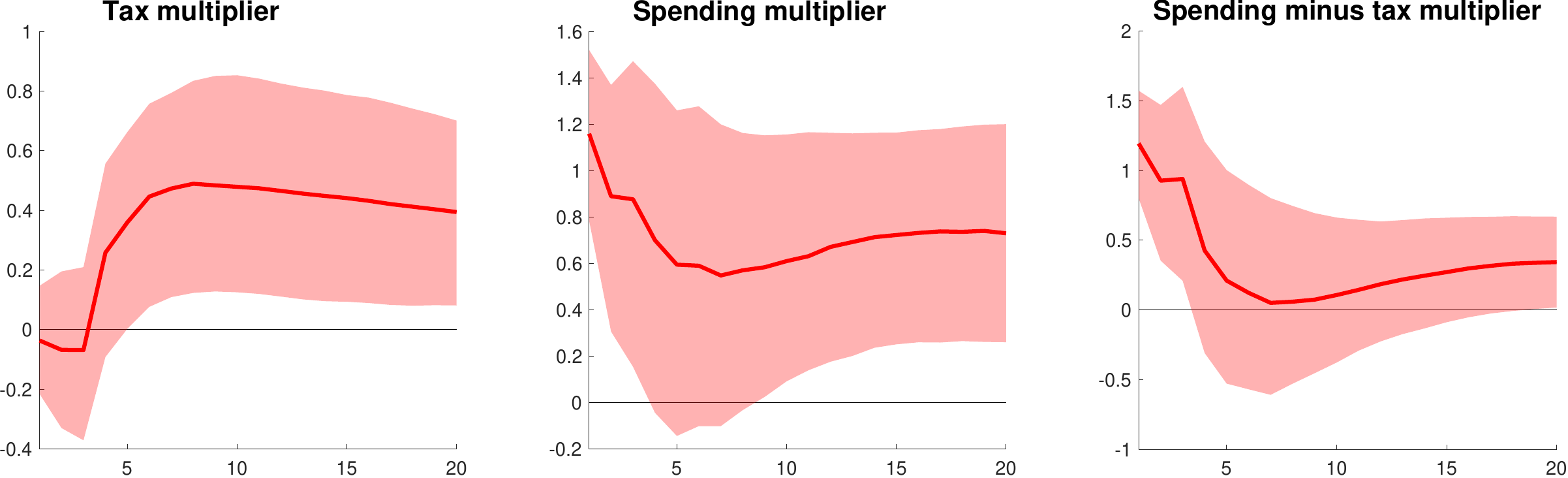}
  \floatfoot{The figure shows the posterior median as well as $68$\% credible bands for the impulse responses of output to a tax shock and a government spending shock. The right subfigure shows the posterior median as well as 68\% credible bands for the difference of the two output responses. }
\end{figure}

 \begin{figure}[h!] 
	\centering
		\caption{ Estimated output multipliers for the non-Gaussian proxy weighting SVAR using an alternative labeling approach} \label{fig:benchmarkBcenter}
	\includegraphics[width=1\textwidth]{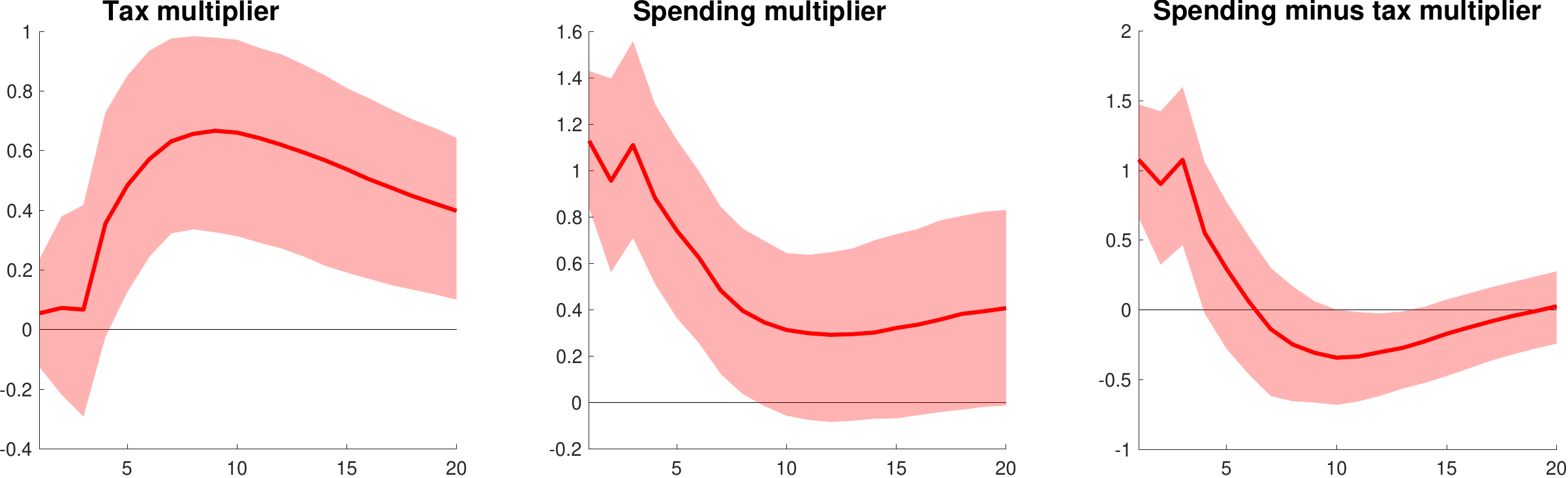}
  \floatfoot{The figure shows the posterior median as well as $68$\% credible bands for the impulse responses of output to a tax shock and a government spending shock. The right subfigure shows the posterior median as well as 68\% credible bands for the difference of the two output responses. 
  The shocks are now labeled based on restricting the posterior draws to unique sign-permutation representatives centered around a first step estimator, see Section 2.2.  For the first step
estimator, we estimate the VAR with a maximum likelihood estimator assuming independent and skewed t-distributed shocks to estimate the simultaneous interaction under the restriction that government spending does not react contemporaneously to tax and output shocks. The restrictions identify and label the government spending shock. The remaining two shocks are labeled based on the tax and TFP proxies. That is, the shock with the highest correlation with the tax proxy is labeled as the tax shock and the shock with the highest correlation with the TFP proxy is labeled as the output shock.   
  }
\end{figure}

 \begin{figure}[h!] 
	\centering
		\caption{Comparison of the estimated output multipliers  for the fiscal Gaussian proxy weighting SVAR and a frequentist fiscal proxy SVAR }\label{fig:fiscalfreq}
	\includegraphics[width=1\textwidth]{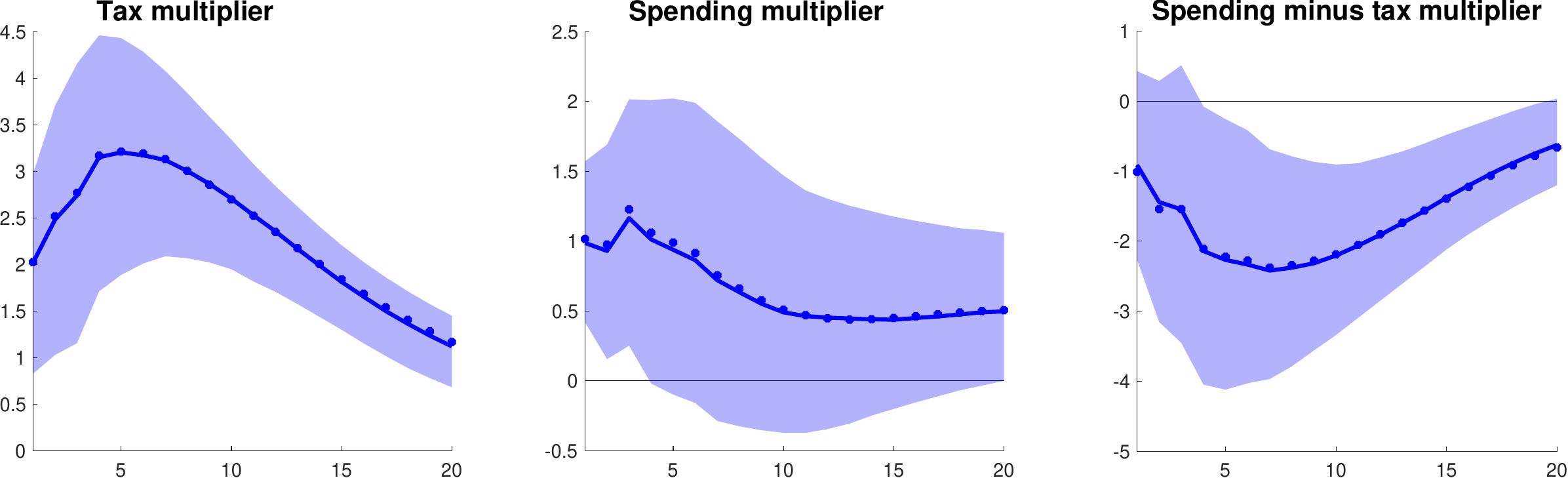}
  \floatfoot{The figure shows the posterior median as well as $68$\% credible bands for the impulse responses of output to a tax shock and a government spending shock. The right subfigure shows the posterior median as well as 68\% credible bands for the difference of the two output responses. The model is estimated using the fiscal Gaussian proxy weighting SVAR imposing the government spending restriction $b_{23}=0$  and   the tax proxy weighting moment conditions in Equation (22)  with $\mu_{\tau g}=\mu_{\tau y}=0$. 
  The   dotted line shows the point estimator using a frequentist moment-based proxy estimator for the fiscal proxy SVAR.
  }
\end{figure}   

 \begin{figure}[h!] 
	\centering
		\caption{Comparison of the estimated output multipliers  for the non-fiscal Gaussian proxy weighting SVAR and a frequentist non-fiscal proxy SVAR.}\label{fig:nonfiscalfreq}
	\includegraphics[width=1\textwidth]{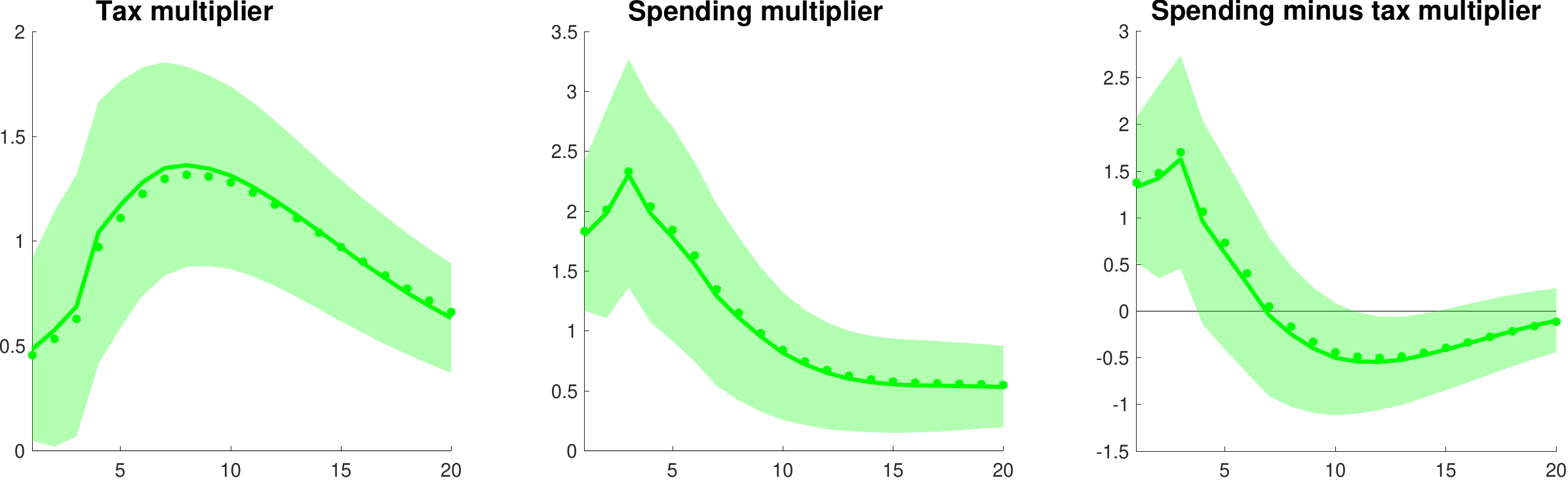}
  \floatfoot{The figure shows the posterior median as well as $68$\% credible bands for the impulse responses of output to a tax shock and a government spending shock. The right subfigure shows the posterior median as well as 68\% credible bands for the difference of the two output responses.  The model is estimated using the non-fiscal Gaussian proxy weighting SVAR imposing the government spending restriction $b_{13}=0$  and    the TFP proxy weighting moment conditions in Equation (23) with $\mu_{y \tau}=\mu_{y g}=0$.
  The   dotted line shows the point estimator using a frequentist moment-based proxy estimator for the non-fiscal proxy SVAR.}
\end{figure}

 \begin{figure}[h!] 
	\centering
		\caption{Estimated output multipliers for the non-Gaussian proxy weighting SVAR using only the tax proxy}\label{fig:nGfiscal}
	\includegraphics[width=1\textwidth]{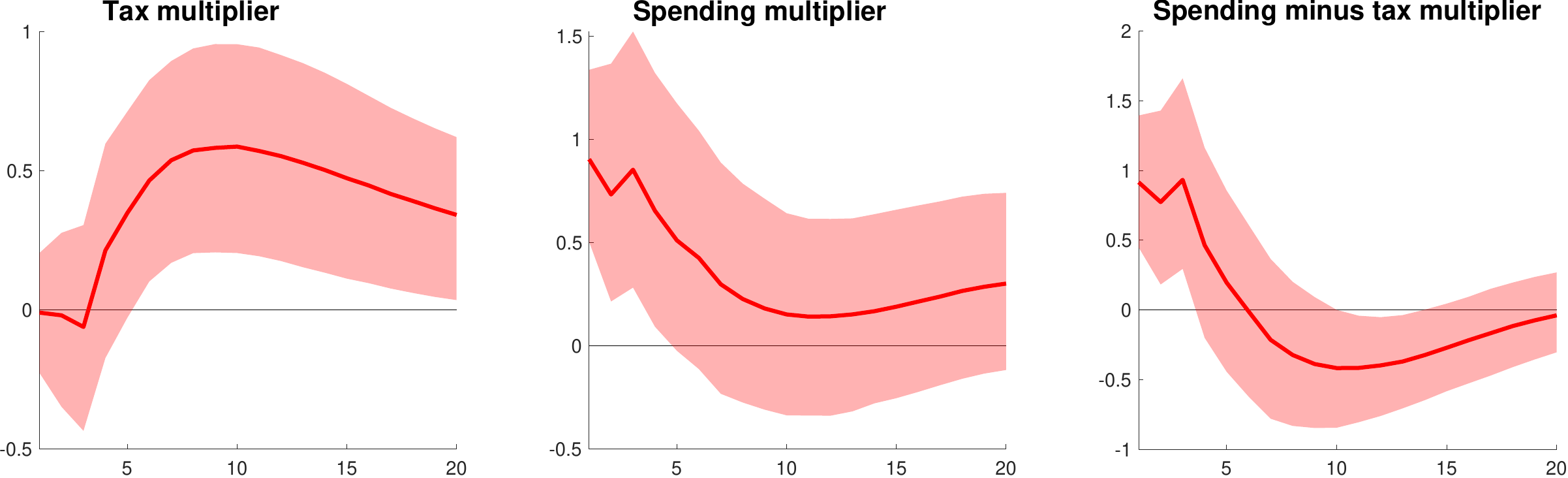}
  \floatfoot{The figure shows the posterior median as well as $68$\% credible bands for the impulse responses of output to a tax shock and a government spending shock. The right subfigure shows the posterior median as well as 68\% credible bands for the difference of the two output responses.}
\end{figure}   

 \begin{figure}[h!] 
	\centering
		\caption{Estimated output multipliers for the non-Gaussian proxy weighting SVAR using only the TFP proxy}\label{fig:nGnonfiscal}
	\includegraphics[width=1\textwidth]{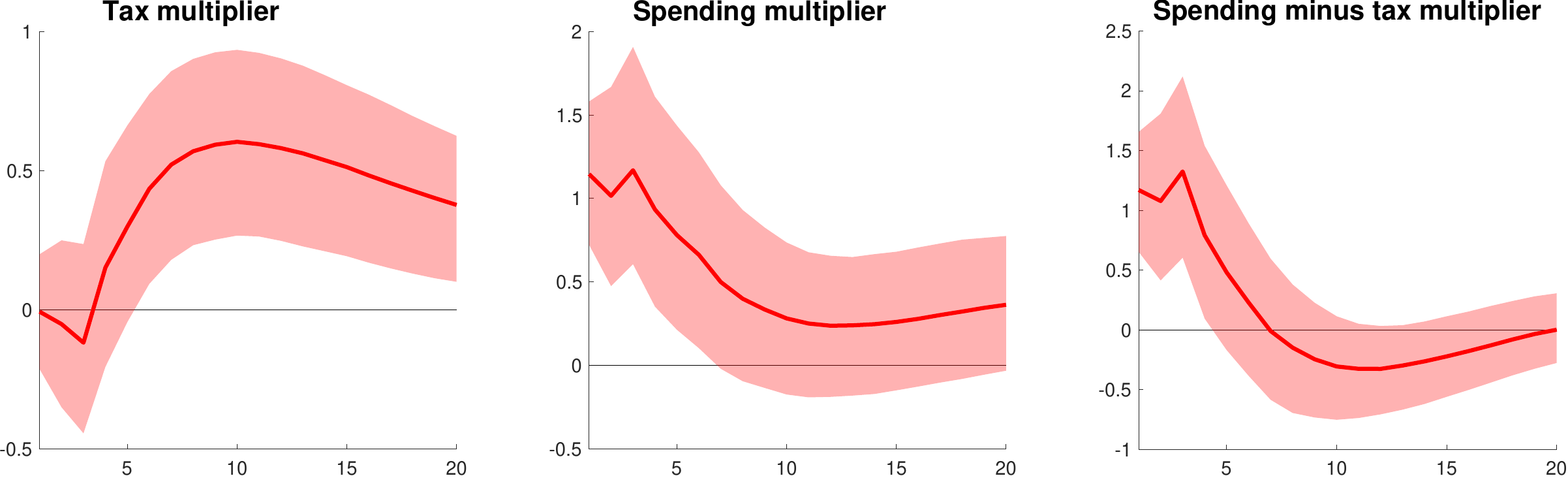}
  \floatfoot{The figure shows the posterior median as well as $68$\% credible bands for the impulse responses of output to a tax shock and a government spending shock. The right subfigure shows the posterior median as well as 68\% credible bands for the difference of the two output responses.}
\end{figure}   

 \begin{figure}[h!] 
	\centering
		\caption{Estimated output multipliers for the non-Gaussian proxy weighting SVAR using only the spending proxy}\label{fig:nGspending}
	\includegraphics[width=1\textwidth]{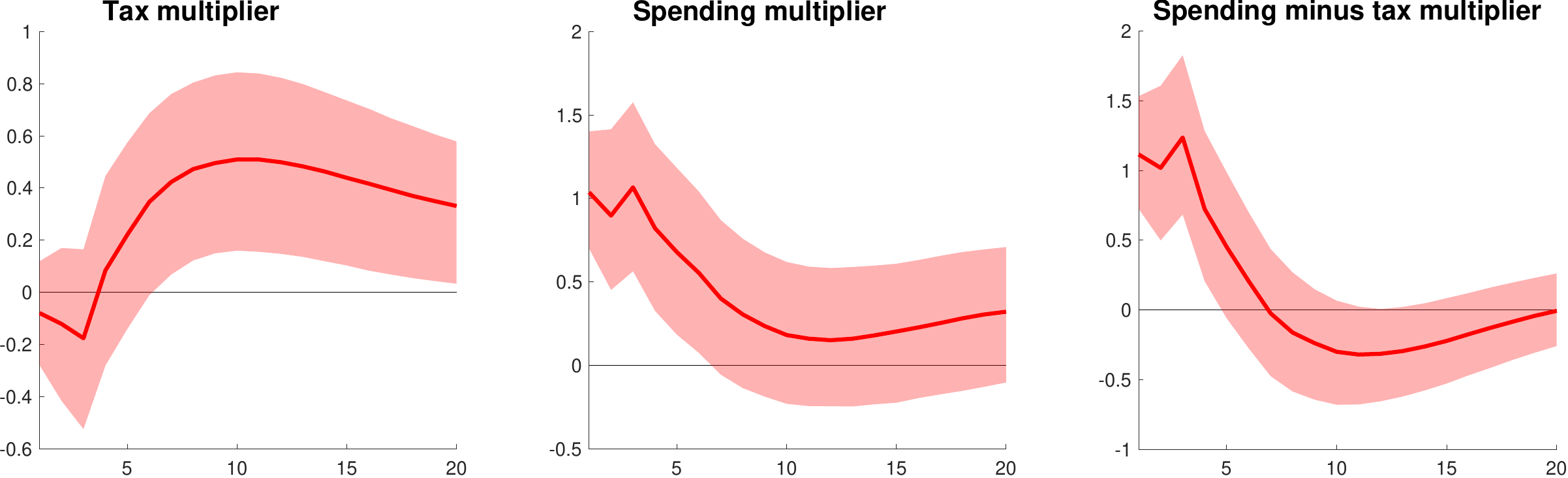}
  \floatfoot{The figure shows the posterior median as well as $68$\% credible bands for the impulse responses of output to a tax shock and a government spending shock. The right subfigure shows the posterior median as well as 68\% credible bands for the difference of the two output responses.}
\end{figure}

 \begin{figure}[h!] 
	\centering
		\caption{Estimated output multipliers for the non-Gaussian proxy weighting SVAR using the proxies after regression them on all macroeconomic variables and proxy variables.}\label{fig:irf_zprocess}
	\includegraphics[width=1\textwidth]{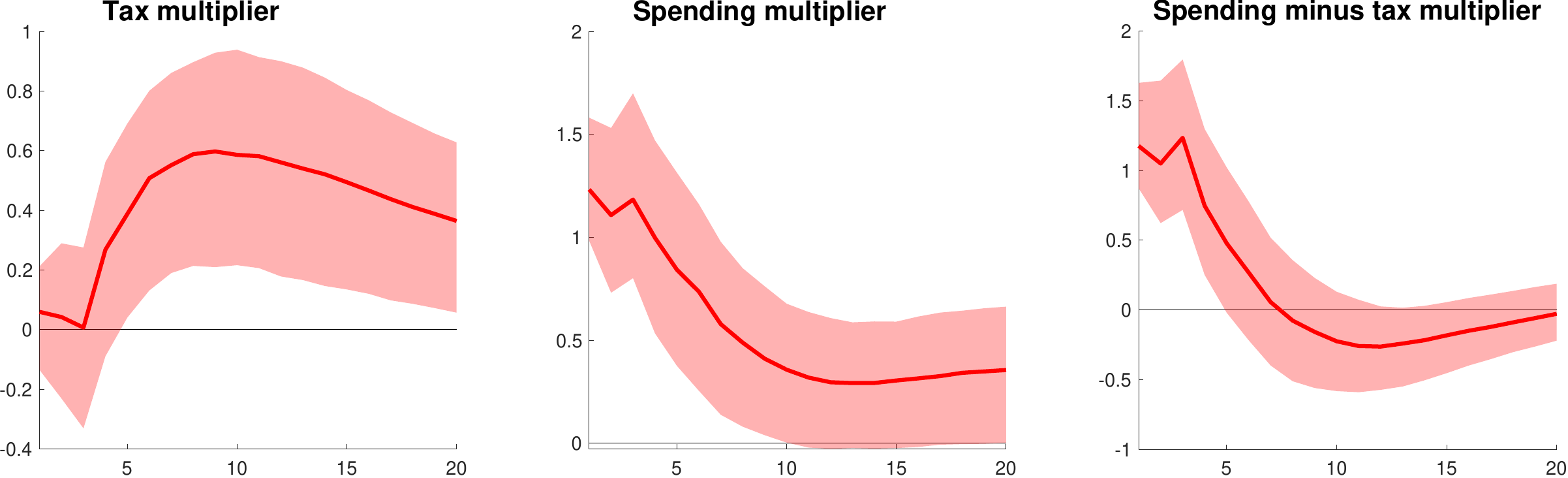}
  \floatfoot{The figure shows the posterior median as well as $68$\% credible bands for the impulse responses of output to a tax shock and a government spending shock. The right subfigure shows the posterior median as well as 68\% credible bands for the difference of the two output responses.}
\end{figure}

 \begin{figure}[h!] 
	\centering
		\caption{Estimated output multipliers for the non-Gaussian proxy weighting SVAR using an estimated $\bm \Sigma_z$ based on the empirical covariance matrix as described in section 2.4 in the main paper.}\label{fig:shocks_diff_DD_sigmadz2}
	\includegraphics[width=1\textwidth]{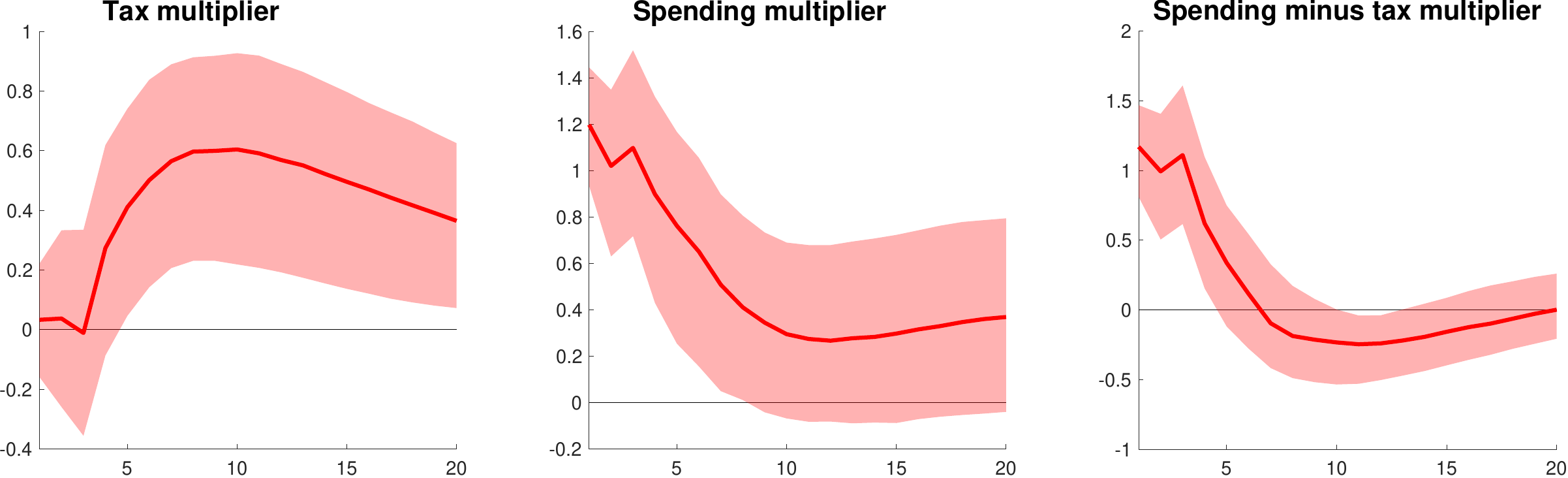}
  \floatfoot{The figure shows the posterior median as well as $68$\% credible bands for the impulse responses of output to a tax shock and a government spending shock. The right subfigure shows the posterior median as well as 68\% credible bands for the difference of the two output responses.
  We follow \cite{yin2009bayesian} and set  $\bm \Sigma_z$ to the empirical covariance matrix $ \bm \Sigma_z=  \frac{1}{T}\sum_{t=1}^T (z_t e_{2t}, \dots,z_t e_{nt})'\\(z_t e_{2t}, \dots,z_t e_{nt})-D(\bm z)D(\bm z)' $.  }
\end{figure}   

 \begin{figure}[h!] 
	\centering
		\caption{Estimated output multipliers for the non-Gaussian proxy weighting SVAR using an estimated $\bm \Sigma_z$ as described in section 2.4 in the main paper.}\label{fig:shocks_diff_DD_sigmadz3}
	\includegraphics[width=1\textwidth]{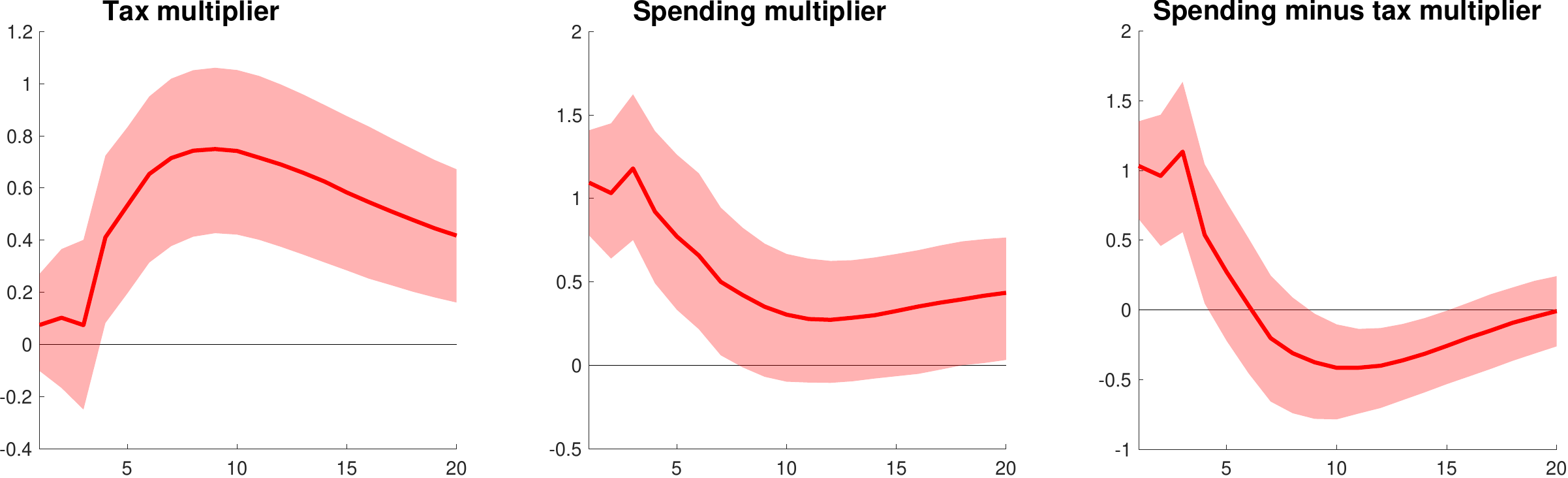}
  \floatfoot{The figure shows the posterior median as well as $68$\% credible bands for the impulse responses of output to a tax shock and a government spending shock. The right subfigure shows the posterior median as well as 68\% credible bands for the difference of the two output responses.
We estimate $\bm \Sigma_z$ and  shrink $\bm \Sigma_z$ to $\text{diag}(\sigma_z^2,..\\.,\sigma_z^2)$. In particular, we assume that $\bm \Sigma_z (i,k)\sim N(m_{i,k},\sigma_{i,k}^2)$ with $m_{i,k}=\sigma_z^2$ if $i=k$ and otherwise we set $m_{i,k}=0$ and $\sigma_{i,k}^ 2\sim  IG(c,d)$ with $c=d=0$.}
\end{figure}

\clearpage
\newpage

 \section{Impact on the estimated historical shocks}\label{Appendix:shocks}
To further highlight the implied differences across identification strategies, Figure \ref{fig:estimatedshocks} shows for selected periods the times series for the estimated structural shocks. The plots in the first row present the estimated tax shocks, the government spending shocks are shown in the second row, and the figures in the third row display the output shocks. While the estimated shocks share a similar pattern for most periods, there are some discrepancies worth mentioning. For example, during the beginning of the 1980s, the fiscal proxy SVAR  approach identifies a sequence of positive tax shocks, whereas our non-Gaussian proxy weighting model and the non-fiscal proxy model attributes most of the variation to negative output shocks. Notably, the US economy entered a recession in 1980Q1, a clear case for a negative output shock. Assigning a sizeable fraction of the fall in economic activity in the early 1980s to exogenous increases in tax revenues as done by the fiscal proxy approach should lead to a larger tax multiplier compared to our proposed approach which attributes it to negative output shocks, which seems to be much more in line with the historical narrative. Similarly, the fiscal proxy approach identifies a large positive tax shock and a relatively small negative output shock (in absolute terms) during the recession at the beginning of the 1990s. In contrast, our non-Gaussian proxy weighting model and the non-fiscal proxy approach attribute most of the output variation to negative output shocks. Again, interpreting this recessionary period as mainly driven by positive tax shocks leads to a larger estimated tax multiplier when applying the fiscal proxy approach compared to assigning most of the variation to negative output shocks as implied by our non-Gaussian proxy weighting approach. Further examples of events that our non-Gaussian proxy weighting model mainly interprets as negative output shocks but the fiscal proxy approach registers as positive tax shocks, and vice versa are:
\begin{itemize}
\item The economic expansion in the early 1970s that the \citet{mertens2014reconciliation} register as a large tax cut under conditions consistent with an expansionary output shock.
\item The economic expansion in the late 1970s which the \citet{mertens2014reconciliation} approach mainly attributes to a tax cut, to help explain a large output boom.
\end{itemize}

\begin{figure}[h]
	\centering
	\caption{Structural shocks of proxy VARs over time }\label{fig:estimatedshocks}
	\includegraphics[width=1.0\textwidth]{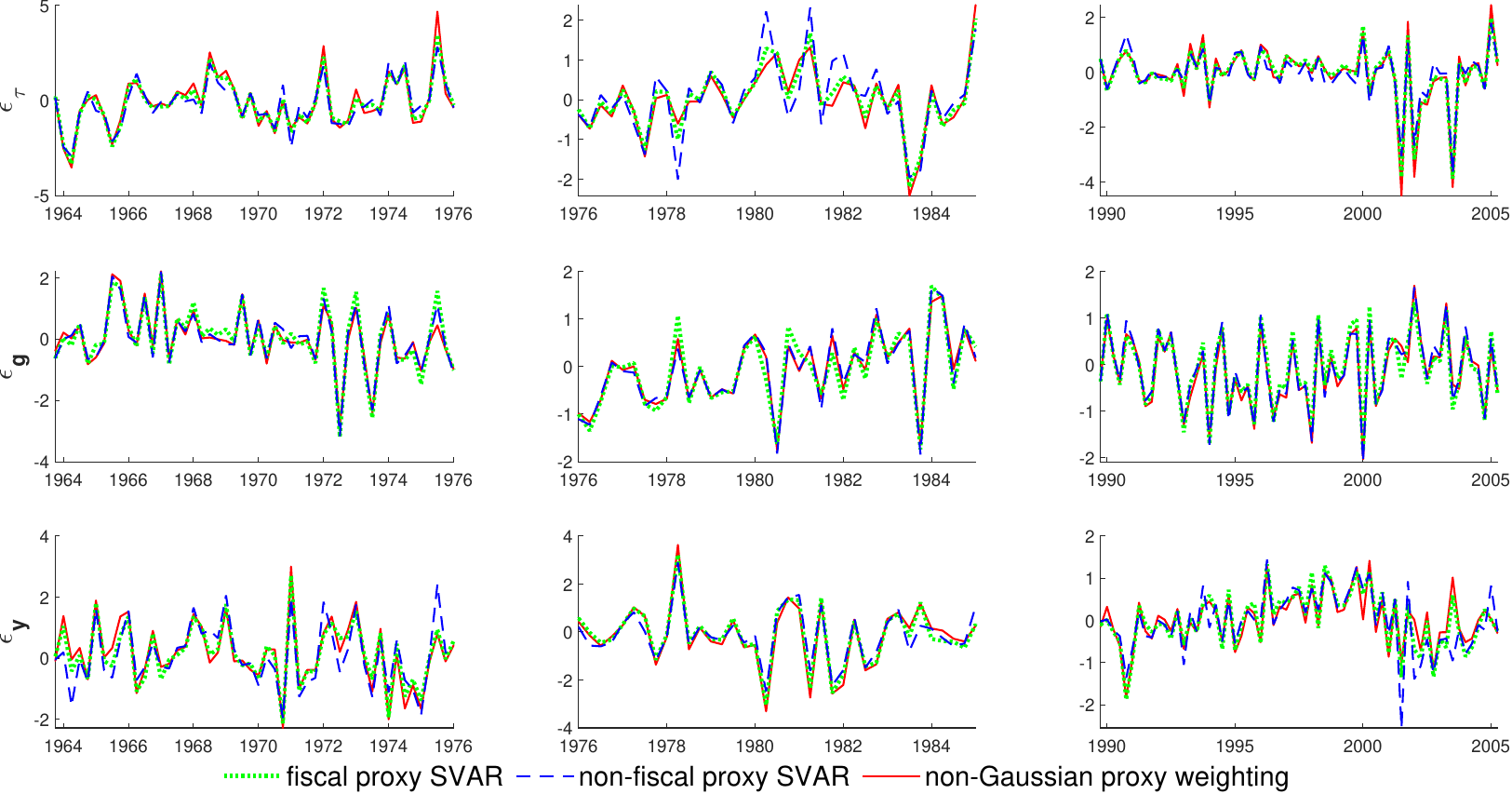} 
%  \floatfoot{Estimated structural shocks for the fiscal proxy SVAR are in blue, for the non-fiscal proxy are in green and for the proxy shrinkage SVAR are in red.}
\end{figure}

Taken together with the result of a negative correlation between the fiscal proxy and the structural output shock, these divergent patterns in the identified shocks between our proposed non-Gaussian proxy weighting model and the fiscal proxy model further reveal that the fiscal proxy VAR approach shows the tendency to interpret positive output shocks as negative tax shocks and vice versa. As a consequence, by taking this shortcoming of the fiscal proxy into account, our non-Gaussian proxy weighting approach leads to a much smaller tax multiplier compared to the standard fiscal proxy identification strategy. \citet{lewis2021identifying} finds a similar feature of the proxy-identified tax shocks proposed by \citet{mertens2014reconciliation}. In contrast to his approach, our non-Gaussian proxy weighting model shows that the potential enodogeneity of the tax proxy provides an explanation for the misclassified tax innovations.  

Regarding the non-fiscal proxy model, there are also some periods that deserve further attention. For example, the non-fiscal proxy model identifies a larger negative output shock in 1965Q3 compared to our non-Gaussian proxy weighting model without any clear evidence of the US economy entering a recession. In addition, the non-Gaussian proxy weighting model shows a larger positive government spending shock in that period, which is very much in line with the strong increase in military spending related to the Vietnam war (see also \citet{Ramey2018} who identify a strong unexpected expansion in military spending for the fiscal year of 1965).\footnote{A similar line of reasoning applies to 2002Q1, where our non-Gaussian proxy weighting model identifies a larger government spending shock compared to the non-fiscal proxy model which is in line with the strong increase in military spending that followed the terror attacks in 2001.} Another example is 1984Q1 and 1984Q2, where the non-Gaussian proxy weighting model identifies two positive output shocks, whereas the non-fiscal proxy model implies two negative output shocks. During that period the US economy showed robust growth rates close to 2\% and no indication of a strong growth decline. Thus, the identified shock sequence of our non-Gaussian proxy weighting model seems to be more in line with the macroeconomic history.\footnote{A similar pattern can be observed for 1992Q1 and 1992Q2 where the non-fiscal proxy model leads to two consecutive negative output shocks without any signs of an economic recession.} The differences are also visible for the period 2000Q1. Here, the output shock of the non-Gaussian proxy weighting model is smaller than the one of the non-fiscal model. In that period, the US economy experienced a decline in GDP growth and the official smoothed recession indicator increased from 0\% in 1999Q4 to 12\% in 2000Q1. Again, we think that the identified shock of our proposed non-Gaussian proxy weighting model fits better to the historical narrative.

Our result on the potential endogeneity of the TFP instrument, also aligns with several recent contributions on how demand shocks affect productivity. For example, \cite{Jorda2024} provide historical international evidence for long-run effects of monetary policy and propose a model with endogenous TFP to reconcile the empirical findings. Concerning the relationship between the fiscal transmission mechanism and endogenous productivity, \cite{Jorgensen2022,Klein2024,Alessandro2019} all find that an exogenous government spending shock significantly affects the Fernald utilization-adjusted TFP series, which questions the exogeneity of that measure. Our findings imply that not accounting for this endogeneity can potentially result in biased estimates on tax and government spending multipliers.

\subsection{Correcting the Proxies}\label{sec: new proxies}

Given the evidence of a correlation between both the fiscal and the non-fiscal proxies with non-target shocks, we now construct new proxies that are orthogonal to the disturbances of the model. Thus, these new proxies arguably fulfill the exogeneity assumption and therefore can be used in a standard proxy variable VAR approach. To get new proxy measures, we proceed as follows. For the narrative tax proxy, we regress the proxy on a constant and the median structural government spending and output shocks obtained from applying the non-Gaussian proxy weighting approach. Similarly, for the TFP measure, we regress the proxy on a constant and the median structural tax and government spending shocks. The residuals of these regressions capture movements in the original proxies that are not related to non-target shocks. One should keep in mind that the construction of the new proxies is model-specific. In particular, for our application, we use the simple three variables baseline VAR, which is a common model used in the fiscal policy literature. In addition, using the constructed proxies as observable data might lead to a generated regressor problem.

%Importantly, even if the reader is not convinced by our non-Gaussian identification approach she may think about the corrections in the proxies coming from narrative arguments, see the discussion below. If she believes in the narrative arguments it is reassuring that our statistical identification approach and the corrected proxies lead to the same conclusion.

Figure \ref{fig:taxproxy} presents our new tax proxy measure and compares it with the original series. For most episodes, both measures move in the same direction and show only marginal differences. However, for some key dates there are sizeable discrepancies. The original tax proxy shows a tendency to indicate an exogenous tax increase during periods of economic recession and an exogenous tax decrease during episodes of economic expansion. For example, in 1971Q1 the original measure shows a relatively strong tax cut, although the actual change to the tax code included only modest adjustments to depreciation rules. In general, this period better fits the narrative of an expansionary output shock. In addition, with the US economy entering a recession in the early 1980s, the original measure shows two pronounced tax changes, a tax increase in 1980Q2 and a tax cut in 1981Q3. However, the macroeconomic narrative seems to be more consistent with a series of contractionary output shocks. Our new measure, which accounts for confounding innovations in the tax proxy measure, indicates a tax increase in the early 1970s economic expansion and almost no change in the tax code during the early 1980s recession.

\begin{figure}
\caption{New proxy variables}
\centering
\begin{subfigure}%{\textwidth}
  \centering
  \includegraphics[width=.47\linewidth]{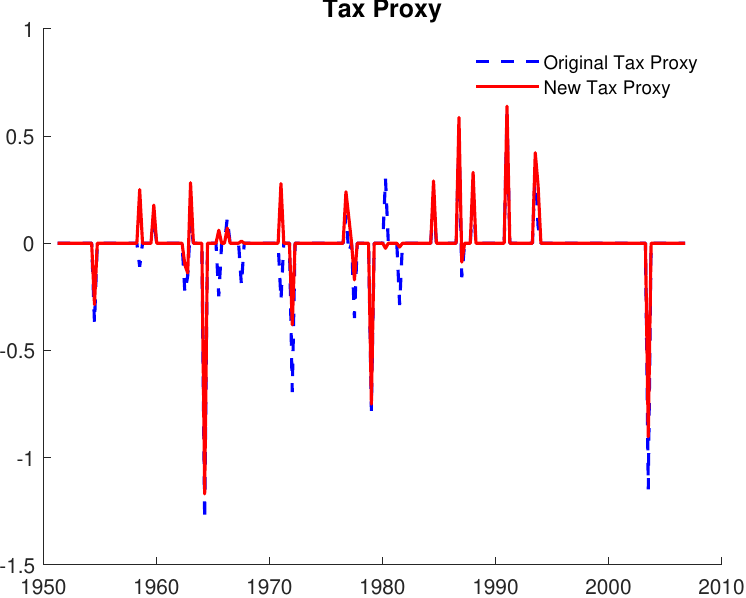}
  %\caption{With }
  \label{fig:taxproxy}
\end{subfigure}%
\begin{subfigure}%{\textwidth}
  \centering
  \includegraphics[width=.47\linewidth]{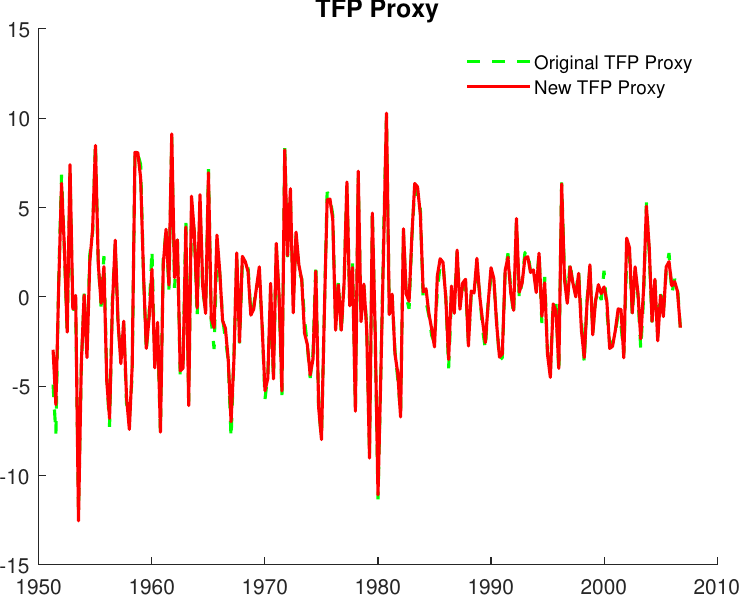}
  %\caption{With original updated fiscal and non-fiscal proxies}
  \label{fig:tfpproxy}
\end{subfigure}
\label{fig:newproxies} 
\floatfoot{The new tax proxy is residual of the regression 
	$z_{\tau,t} = \beta_0 + \beta_1 \varepsilon_{g,t} + \beta_2 \varepsilon_{y,t} + u_t$, meaning the variation of the tax proxy unexplained by the government spending and output shock.The new TFP proxy is residual of the regression 
$z_{TFP,t} = \beta_0 + \beta_1 \varepsilon_{\tau,t} + \beta_2 \varepsilon_{g,t} + u_t$, meaning the variation of the TFP proxy unexplained by the tax and government spending shock.}
\end{figure}

We also calculate a new TFP proxy as shown in Figure \ref{fig:tfpproxy}. The differences between the original and the new TFP proxy, which takes potential endogeneity concerns into account are relatively small. However, differences can be observed for the periods like 1965Q3, 1984Q1, 1984Q2, 2002Q1. In the appendix, we discuss in detail why we think our new TFP proxy measure better aligns with the historical narrative.

Given that our new proxy measures remove endogenous variation in the original tax and TFP series, we can use them as instruments in a standard proxy VAR setting that rests on the idea of an available instrument fulfilling the exogeneity assumption. Figure \ref{fig:newmultipliers} presents the results of the fiscal and non-fiscal models, respectively, when relying on the newly constructed proxies and compares the obtained estimates to our non-Gaussian proxy weighting model. The differences in estimated multipliers across models are much smaller. As such, the fiscal proxy model now leads to an estimated tax multiplier that is much smaller compared to the case when using the original (potentially endogenous) tax proxy. Both the fiscal and non-fiscal model induce estimated tax multipliers that mainly lie within the credible bands of our non-Gaussian model. The same applies to the estimated government spending multiplier, which takes on values close to unity across all models. Importantly, when endogenous variation in the proxies is removed, all models lead to the conclusion that the government spending multiplier is larger than the tax multiplier. Thus, our evidence suggests that endogeneity in the tax and TFP proxies is responsible for the large differences in estimated multipliers across estimation strategies. Because the fiscal and non-fiscal approach have to assume that the proxies are indeed exogenous, this finding can only be obtained by a strategy that evaluates the exogeneity of the proxies like our proposed non-Gaussian proxy weighting model.

 \begin{figure}
	\centering
	\caption{Comparison of estimated output multipliers between the different models with new proxies}\label{fig:newmultipliers} 
	\includegraphics[width=1\textwidth]{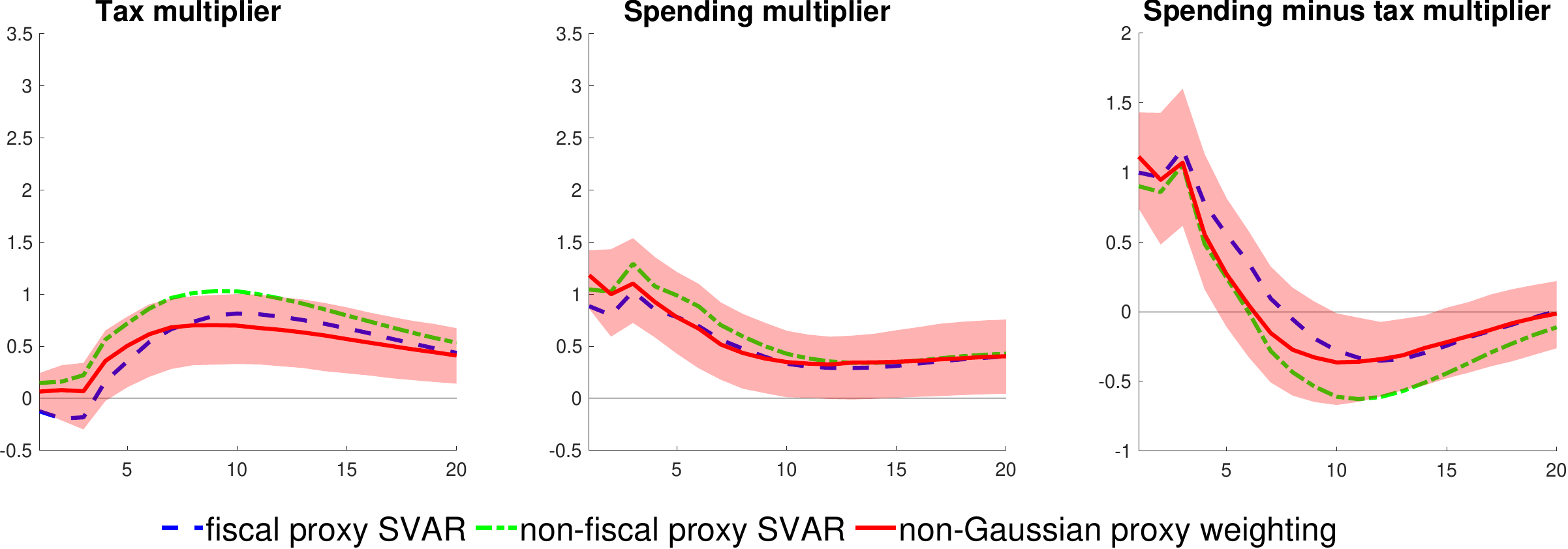} 
  \floatfoot{The figure compares the output responses between our non-Gaussian proxy weighting VAR with $68$\% credible bands to the median responses in the non-fiscal proxy SVAR proposed by \citet{caldara2017analytics} as well as the fiscal proxy SVAR from \cite{mertens2014reconciliation}. But we replace the old proxy variables with the new proxy variables. }
\end{figure}  

 \small{
\addcontentsline{toc}{section}{References} 
\bibliography{literatur}}\clearpage

\end{document}